\documentclass[pdflatex,sn-mathphys-num]{sn-jnl}
\usepackage{graphicx}%
\usepackage{multirow}%
\usepackage{amsmath,amssymb,amsfonts}%
\usepackage{amsthm}%
\usepackage{mathrsfs}%
\usepackage[title]{appendix}%
\usepackage{xcolor}%
\usepackage{textcomp}%
\usepackage{manyfoot}%
\usepackage{booktabs}%
\usepackage{algorithm}%
\usepackage{algorithmicx}%
\usepackage{algpseudocode}%
\usepackage{listings}%
\usepackage{comment}
\usepackage{gensymb}

\usepackage{longtable}
\usepackage{array}
\usepackage{caption}
\newcolumntype{N}{>{\centering\arraybackslash}m{0.05\textwidth}} 
\newcolumntype{L}{>{\raggedright\arraybackslash}m{.46\textwidth}} 
\newcolumntype{R}{>{\raggedright\arraybackslash}m{.46\textwidth}} 

\theoremstyle{thmstyleone}%
\theoremstyle{thmstyletwo}%
\theoremstyle{thmstylethree}%
\begin{document}

\title[Article Title]{A quantitative analysis of Galilei's observations of Jupiter satellites from the Sidereus Nuncius}


\author*[1]{\fnm{Andrea} \sur{Longhin}}\email{andrea.longhin@unipd.it}
\affil*[1]{\orgdiv{Department of Physics and Astronomy}, \orgname{University of Padova and INFN-Padova}, \orgaddress{\street{via Marzolo 8}, \city{Padova}, \postcode{35132}, \country{Italy}}}

\abstract{We present a new careful and comprehensive analysis the observations of the satellites of Jupiter from the Sidereus Nuncius that extends and complements previous similar studies. Each observation is compared to the predictions obtained using a modern sky simulator, verifying and trying to understand them individually. The work considers both the information that can be extracted from the sketches and the angular measurements reported by Galilei in the text. The consistency and relation between these two data-sets is explored. The angular measurements also allow assessing the absolute accuracy of the angular measurements in relation to modern ephemerides. Exploiting the simulations we use the data to evaluate the performances of the telescope in terms of separation power of close-by satellites and the inefficiency in the detection connected to the proximity to the disk. A sinusoidal fit of the data obtained from the available sketches, allows measuring the relative major semi-axes of the satellites' orbits and their periods with a statistical precision of 2-4\% and 0.1-0.3\% respectively. The posterior fit error is used to estimate the accuracy in the reconstruction of the elongations. We show that with this data one can infer in a convincing way the third law of Kepler for the Jupiter system. The 1:2 and 1:4 orbital resonance between the periods of Io and Europa/Ganymede can be determined with \% precision. In order to obtain these results it is important to separate the four datasets. This operation was an extremely difficult task for Galilei as the analysis will evidence. Nevertheless we show how some indication on the periods emerge from the  using the modern Lomb-Scargle technique without having to separate the four data-sets. We briefly extend the use of the simulator to verify the accuracy in the seven observations of the Moon and the performance in reproducing the Pleiades, the Orion belt, the Orion head and the Beehive cluster. Finally we present some results on the achievable imaging of Jupiter and the Moon obtained with a replica of the instrument. This exercise highlights the challenges connected with these observations and confirms the excellence underlying this amazing set of early scientific data.}


\maketitle
\tableofcontents

\clearpage
\section{Introduction}
\label{intro}

The night of January 7, 1610 was a special one for Galilei and, in hindsight, for the history of science. As he writes in the Sidereus Nuncius, he had already pointed to Jupiter with the earlier version of the instrument but no significant feature had emerged. The one that he had advertised to the Doge of Venice in August 1609 had a magnification of about 8$\times$. After making the famous inkwashes of the Moon in November, December 1609, with the improved version of his cannocchiale, that now achieves a magnification of about 30$\times$, Galilei points again to the major planet. He sees three peculiar brilliant ``stars'' very close to the planet and strangely well aligned to a quite peculiar direction, the one of the Ecliptic plane. Two are to the left and one to the right (first entry of Tab.~\ref{tab:1}). He writes he had not paid too much attention to the relative distances as he expected them to be fixed stars but, \textit{guided by I know not what fate}, (\textit{nescio quo fato ducto}) he came back to observe it the night after. It is hard to believe he did not already had an intuition of the exceptional discovery that was behind the corner. Which is the probability of having by change three aligned stars so closely packed to Jupiter? And in that very meaningful direction? 

On January 8, Jupiter lies to the East of the three stars (second entry of Tab.~\ref{tab:1}). In reality the easternmost ``star'' (Callisto) was still there, to the East side, but Galilei does not notice it, this time. This might mean that indeed he was still not completely aware of what was going on and a still a bit less focused on these observations. As we will see he almost never missed a ``detectable'' satellite by mistake again thereafter. In particular Callisto will always be observed in the following two months, unless when it was too close to the disk. The three stars were then in reality not the same three satellites: Io had passed from the left to right side and Europa had become visible to the right after being too close to Jupiter the day before. 

\begin{figure}
    \centering
\includegraphics[width=\linewidth]{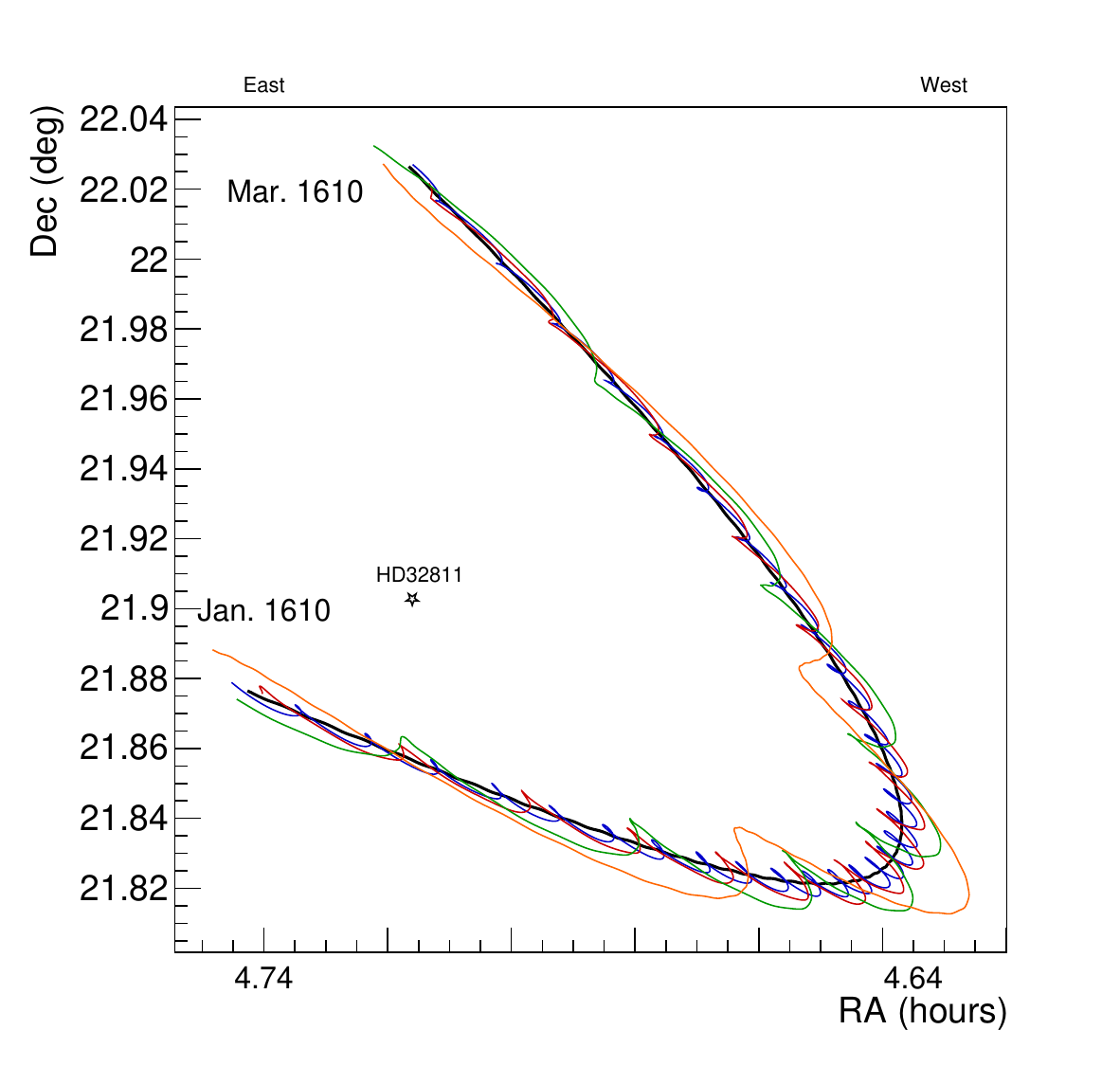}\\
    \caption{Motion of Jupiter (black) and of its satellites (Io in blue, Europa in red, Ganymede in green and Callisto in Orange) across the sky from Jan. 7 1610 to March 2 1610. The marker points the position of the star HD32811 that was recorded by Galileo in observations 61, 62 and 64 (see Tab.~\ref{fig:1}.)}
    \label{fig:path}
\end{figure}

At this point, two intriguing observations was making Galilei eager to observe the situation in the following days: first the relative separations of the ``stars'' appear a little different; second, one had to assume that Jupiter was moving eastward (i.e. in retrograde motion) while in reality, as the opposition had already passed in 8 Dec. 1609, it was expected to be moving westward. The path of the Jupiter system in those days is shown in Fig.~\ref{fig:path}(data from \cite{PDSRings_Ephem3Jup}). The combined motion is quite complex. Galilei cross checks the motion of the planet with respect to a ``truly'' fixed star (HD32811) only much after with the last observations of February and early March. It is interesting to notice that the same star had appeared much closer to Jupiter as the planet was overcoming it from East to West. After the second day anyway the baseline, unexciting hypothesis that these were the same three fixed stars was still in place but a bit shaky already. This amazing story has been covered in the literature by historians of Physics and Astronomy in great detail. Of particular interest is the paper of 2011 by Gingerich and van Helden\cite{GingerichVanHelden2011} where they elaborate in detail on how Galileo became convinced that the four points of light were satellites and not stars.

But coming back to those days, the cloudy sky of Padua on the 9$^{\rm{th}}$ creates some ``suspense'' but the scene of the 10$^{\rm{th}}$ is a real breakthrough as Jupiter is now West of the two ``stars'' instead of even farther to the East: this is inconsistent with Jupiter ``overcoming'' the two fixed stars by moving East. Moreover one of the three stars had disappeared indicating that something more interesting was going on. 
A thrilling adventure that will accompany him uninterruptedly for each clear night for the following two months was just about to start. It is remarkable how many clear days there were in Padua (/Venice) during those winter months: just ten days were cloudy out of about sixty. On average nowadays the fraction of cloudy days is 40-50\% according to \cite{weatherPD}. This detailed account of the dance of the satellites of Jupiter is an history of scientific rigor and excitement: it lead him to observe until late at night in several occasions. 

Earlier in the Sidereus Galilei encourages astronomers in determining the periods of the satellites as this was not possible to him due to the restrictions in time. Still, despite this consideration it is amazing how accurately Galilei decided to document what he saw. It is also interesting to notice how early Galilei wanted to focus on the determination of the periods, an activity that will keep him busy for years. Despite being insufficient to estimate the periods, according to Galilei, the set of observations is extremely rich and well professionally documented. It is kind of amazing considering that this was just meant as a quick first release in support of the discovery. 

In fact these observations were so exceptional and groundbreaking to deserve such a solid and complete basis of evidence despite this might seem redundant to our eyes. In \cite{BucciantiniCamerotaGiudice} we find a good account of the trepidation with which Galilei was awaiting the reactions of his contemporaries colleagues and especially the one of Kepler.

It should also be not forgotten that Galilei might have felt the pressure of releasing the book as soon as possible due to the risk of being anticipated by other scientists. This risk was indeed quite well founded given the large spread of low magnification/low quality telescopes that had interested Europe already since 1608\cite{VanHelden1977,BucciantiniCamerotaGiudice}
and the actual presence of competitors (Mayr, Harriot \cite{Roche1982}).  Despite this he deemed necessary to report such detailed data until the very close to the publication deadlines. The dedication to Cosimo de Medici is dated 12 Mar. 1610 and on Mar. 13 the book was in press of Tommaso Baglioni in Venice. The last observation dates back to just ten days earlier. This choice tells a lot about of the mindset of Galilei and this emphasis on data cannot but induce a certain degree of admiration in modern experimentalists.

In this work we present a new, detailed analysis of the observations of the satellites of Jupiter reported by Galileo Galilei in the Sidereus Nuncius \cite{SN} during the period from January 7 to March 3, 1610 besides some checks of other observations found in the same book.

\subsection{Previous results}

The idea of comparing this exciting data-set to modern ephemeris is not new. This approach was pioneered in 1962-1964, by the Belgian astronomer Jean Meeus that re-examined Galileo’s observations of Jupiter’s moons~\cite{Meeus1962,Meeus1964}, showing that all four were visible on 7 Jan. 1610. He first showed how it was not possible to resolve satellites being too close to each other or closer than about three Jovian radii from the planet’s limb. Using his newly computed satellite tables, Meeus reconstructed their configurations from 1600 to 2200.

Later, the historian Stillman Drake extended Meeus’ work by analysing all of Galileo’s Jupiter observations\cite{Drake1976,Drake1979}. 

In 1982 John Roche\cite{Roche1982} analyzed the observed elongations and compared them against modern ephemerides. He discusses how the measurements of Galilei were later used by Harriot and combined with his measurements to calculate the periods of the satellites. In doing so this he performs a careful evaluation of a subset of the data of the Sidereus showing something already noticed by North \cite{North1960} i.e. that beside a random error there is a systematic component that makes the angular measurements overestimate the real elongations by large factors. Roche interestingly mentions (pag. 25) that this is also observed in the data of Mayr and Harriot. He reports an overestimation of 1.4 for Galilei, 1.3 for Harriot and 1.35 for Mayr. This point is also discussed in the book by Stillman Drake of 1983\cite{Drake1983}. We will come back to this interesting point and the hypotheses of Roche later during the discussion of our results that fully support this observation with a much larger data-sample. We shows that it is a real scale factor that is consistent for the four satellites . 

In his 1983 commentary on Sidereus Nuncius, Drake compared Galilei’s notes with modern ephemerides and concluded that Galilei could generally resolve two satellites only if they were separated by more than $\sim 10^{^{\prime\prime}}$~\cite{Drake1983}. 

In 1997 Levi, F.A., Levi-Donati also presented a verification of the observations of the Sidereus~\cite{Levi1997} using several programs (Planet 4.0, Voyager 1.0, Microprojects Astro 1.3 and 1.4, StarryNight Mac 1.0.1). In 2012 Bernieri published a work\cite{Bernieri2012} where he revisits Galileo’s observations of Jupiter’s moons, using a modern planetarium software (TheSky6) to reconstruct the actual angular separations of Jupiter’s satellites as Galileo would have seen them. He carefully accounts for uncertainties in Galileo’s timing ($\pm$15~min) and the software’s positional precision ($\sim 1^{\prime\prime}$). This analysis reveals that the telescope
could only resolve satellite–satellite separations greater than approximately 19–20$^{\prime\prime}$ (much larger than earlier estimates of $\sim$10$^{\prime\prime}$ by Drake) and that satellite–planet separations needed to exceed roughly 50$^{\prime\prime}$ for detection, due to glare and optical aberrations. Additionally, Bernieri calculates that the mean error in Galileo’s recorded angular measurements is about 57$^{\prime\prime}$, matching Galileo’s own estimation in Sidereus. Notably, these measurement errors show a systematic positive bias and correlate with larger separations—indicating an effect that we will discuss in more detail in our analysis. 

A study involving a sinusoidal fit of the orbits by Alessandro Bettini is presented in \cite{Bettini2016Gravitation} (2016). In Fig. 4.20. Bettini considers later data taken in 1611 and performs a sinusoidal fit of the elongations of Callisto and Ganymede. He observes that Galilei obtained a very good agreement in the periods (1.76, 3.53, 7.16, 16.3 days to be compared with modern values of 1.77 3.55, 7.1, 16.75). Concerning the elongations the knowledge evolved as shown in Tab.~\ref{tab:Bettini}.
\begin{table}[h!]
\centering
\caption{Angular radii of the orbits of Jupiter’s satellites (in units of Jupiter radii). From \cite{Bettini2016Gravitation}.}
\begin{tabular}{lcccc}
\hline
Epoch   & Io   & Europa & Ganymede & Callisto \\
\hline
1610    & 3.5  & 5.7    & 8.8      & 15.3 \\
1611    & 3.8  & 6.2    & 8.4      & 15 \\
1611?   & 4.0  & 7.0    & 10.0     & 15 \\
1612    & 5.7  & 8.6    & 14.0     & ``almost 25'' \\
Modern  & 5.58 & 8.88   & 14.16    & 24.90 \\
\hline
\end{tabular}
\label{tab:Bettini}
\end{table}
Bettini also shows how these data allow to verify the third Kepler on the Jovian system and test the universality of the gravitational constant for the Jovian and the solar system.

Among the earlier work on the subject it is worth mentioning a very nice web page developed at the same time of Bernieri's work by Ernie Wright \cite{WrightSidereusNuncius} which provides animations of the satellites and a direct comparison for several configurations.

In this work we extend significantly the analysis of the Sidereus Nuncius dataset: 
\begin{itemize}
    \item We  present sinusoidal fits of the elongations vs time. From the fit parameters we show a comparison of orbital parameters with modern expectations. Data have been corrected from Italian to modern hours (Sec.~\ref{modernh}). The residuals of the fit are used to estimate the resolution of the measurements.
    \item We present a detailed comparison of two datasets (Sec.~\ref{dataset},~\ref{digit}): the one coming from an analysis of the drawings in the published Sidereus and the angular data given by Galilei in the text, starting from Jan. 12. Using this dataset, we can, in principle, characterize in absolute terms the accuracy of the angular measurements by comparing with the ephemeris providing the expected elongations in degrees, and not anymore in terms of the Jupiter disk whose definition is affected by the systematics on its determination with the telescope. To check this hypothesis, we test if the angular data support the evidence of the decrease in apparent angle of the Jupiter system as it moved away from the Earth from January to March (a 15\% effect). In addition this dataset is closer to the real measurements of Galilei as it does not pass through the additional layer of complexity introduced by the printing process.
    \item For the first time we show how a frequency analysis of sparse data (Lomb-Scargle periodograms) allows getting a fairly solid evaluation of the orbital periods of Callisto and Ganymede without having to disentangle the individual satellites;
    \item We extend the study proposed by Bettini to this dataset (test of the third Kepler law);
    \item We introduce a check of the 1:2:4 resonance of the periods of the innermost satellites;
    \item We evaluate the correctness of all the configurations and discuss the only known anomaly of February 12. Each configuration drawn by Galilei is set aside to the prediction from the software allowing to evaluate in a direct and intuitive way the reasons for missing observations;
    \item We reevaluate the resolving power of pairs of satellites and the inefficiency introduced by the proximity to Jupiter's disk by analyzing the distribution of the positions of the missed satellites or missed pairs at ``truth-level'' using the simulator;
    \item We extend the comparison with the simulator for other observations reported in the Sidereus (the Pleiades, the Beehive cluster, the Orion belt and head) in Sec.~\ref{otherobservations}. Here we also present some comparisons between the original notes and the Sidereus and a cross check of the absolute angular measurements of Galilei for a known stellar pattern;
    \item We then consider the seven inkwashes of the Moon that were the basis for the engravings of the Sidereus and, using  the generally accepted dates of these observations, that are widely reported in the previous literature, we accurately compare them to modern expectations of the Moon appearance with the support of the simulator. 
    \item In the last part of the paper we describe the technical difficulties involved in collecting these data by the direct experience obtained by observing Jupiter and the Moon with a replica of one of the telescopes of Galilei that we developed for this purpose; We give a particular emphasis on the difficulty of the observations introduced by the narrowness of the field of view, a critical aspect for the success of these observations, that nevertheless is generally not particularly considered.
\end{itemize}

\section{Analysed data}
\label{dataset}
This analysis focuses on the first observations presented in the Sidereus Nuncius without considering those performed in later years by Galileo himself. We do not consider here the information provided by the handwritten notes of Galilei that are carefully analysed in the literature:~\cite{GalileiOpere,GingerichVanHelden2003}. We anyway compared systematically the handwritten notes sketches with the diagrams in the Sidereus in Appendix \ref{appendixE}.

\begin{figure}
\begin{minipage}[t]{0.48\linewidth}
\begin{tabular}{|l|}\hline
1)\includegraphics[height=8mm]{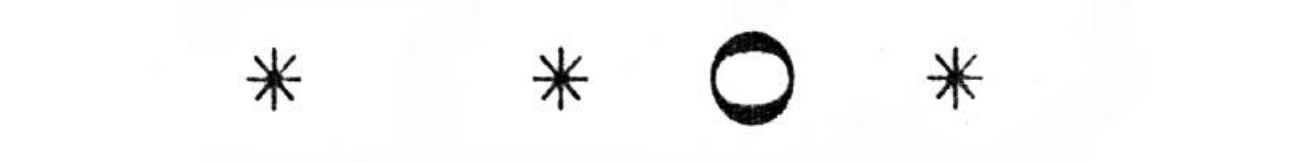}\\\hline
2)\includegraphics[height=8mm]{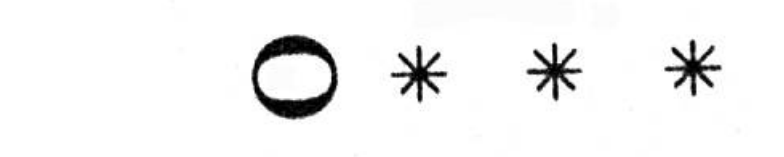}\\\hline
3)\includegraphics[height=8mm]{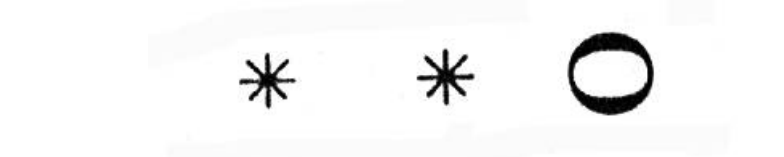}\\\hline
4)\includegraphics[height=8mm]{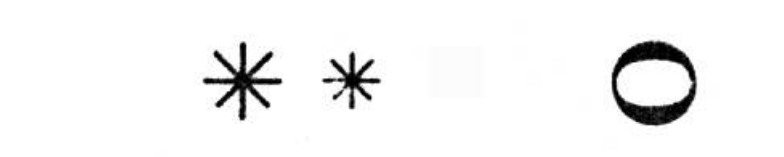}\\\hline
5)\includegraphics[height=8mm]{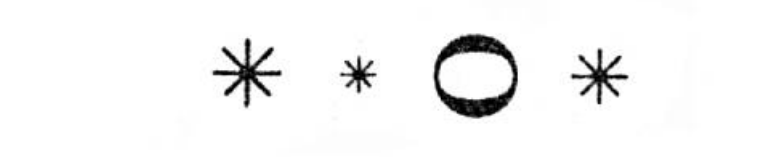}\\\hline
6)\includegraphics[height=8mm]{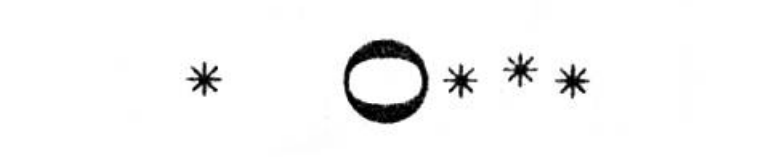}\\\hline
7)\includegraphics[height=8mm]{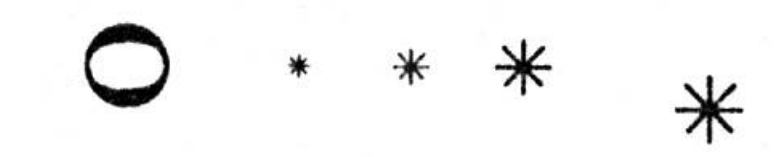}\\\hline
8)\includegraphics[height=8mm]{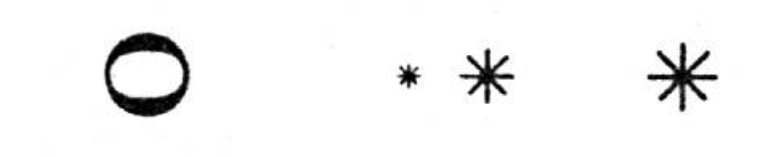}\\\hline
9)\includegraphics[height=8mm]{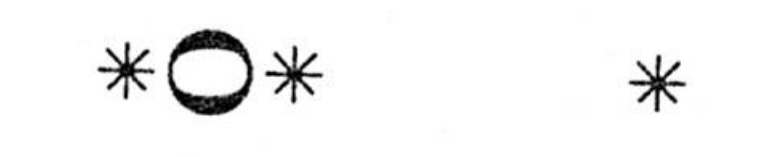}\\\hline
10)\includegraphics[height=8mm]{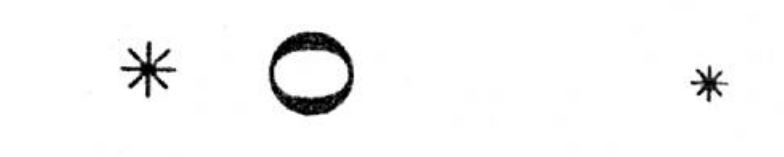}\\\hline
11)\includegraphics[height=8mm]{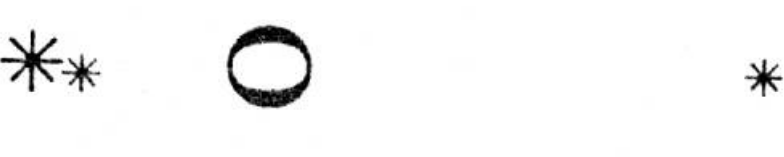}\\\hline
12)\includegraphics[height=8mm]{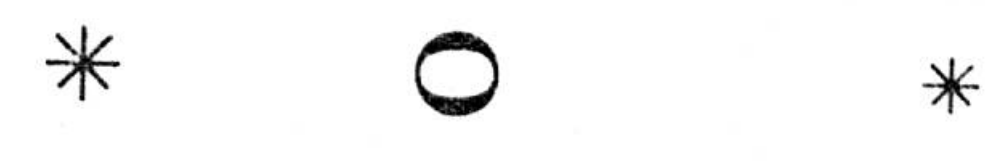}\\\hline
13)\includegraphics[height=8mm]{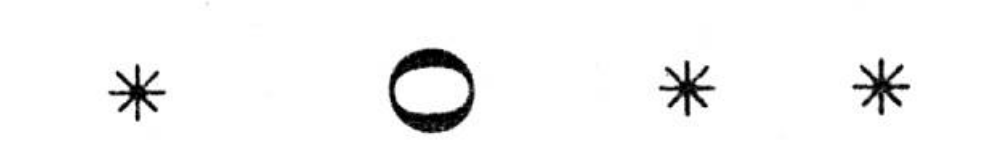}\\\hline
14)\includegraphics[height=8mm]{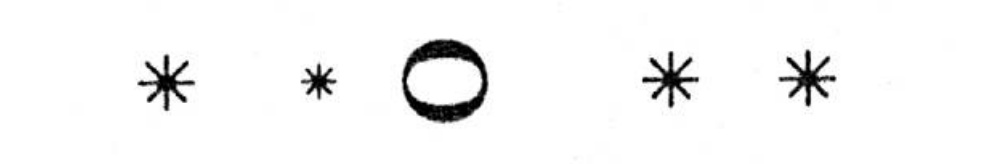}\\\hline
15)\includegraphics[height=8mm]{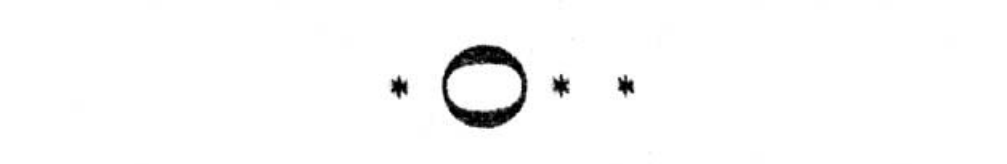}\\\hline
16)\includegraphics[height=8mm]{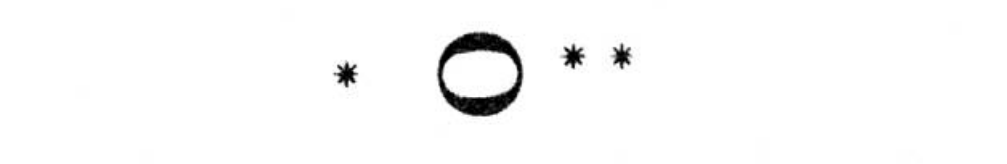}\\\hline
\end{tabular}
\end{minipage}%
\begin{minipage}[t]{0.48\linewidth}
\begin{tabular}{|l|}\hline
17)\includegraphics[height=8mm]{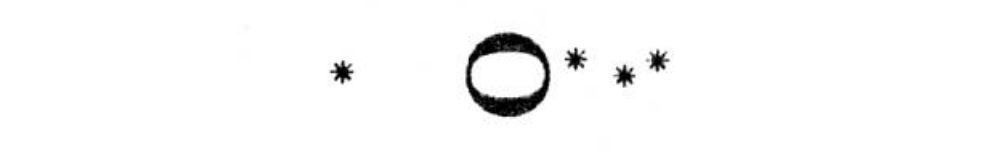}\\\hline
18)\includegraphics[height=8mm]{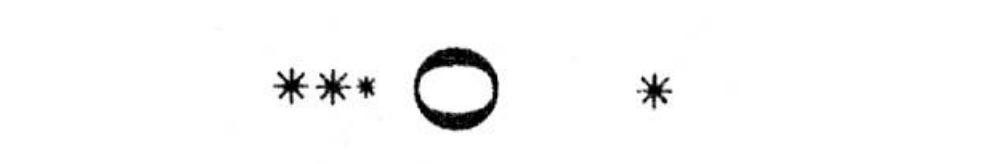}\\\hline
19)\includegraphics[height=8mm]{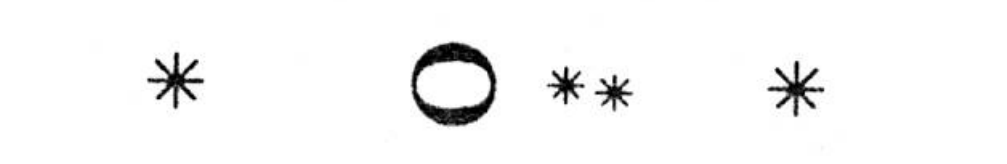}\\\hline
20)\includegraphics[height=8mm]{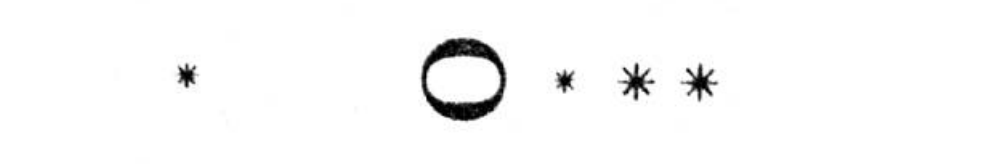}\\\hline
21)\includegraphics[height=8mm]{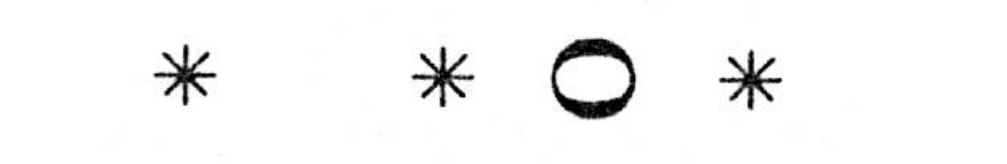}\\\hline
22)\includegraphics[height=8mm]{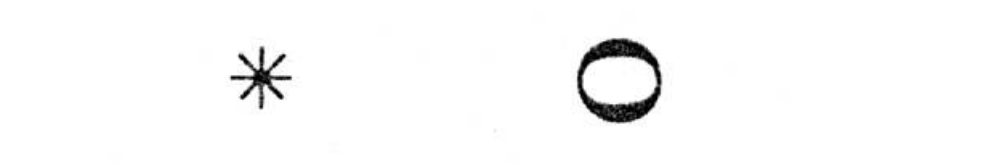}\\\hline
23)\includegraphics[height=8mm]{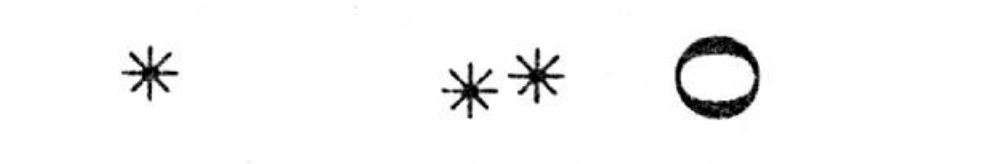}\\\hline
24)\includegraphics[height=8mm]{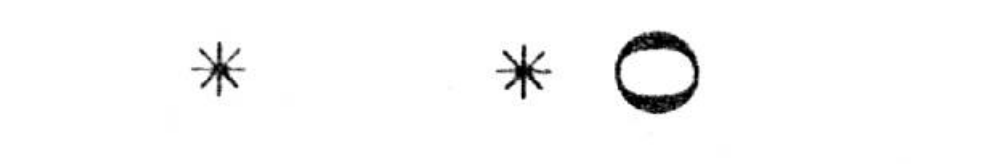}\\\hline
25)\includegraphics[height=8mm]{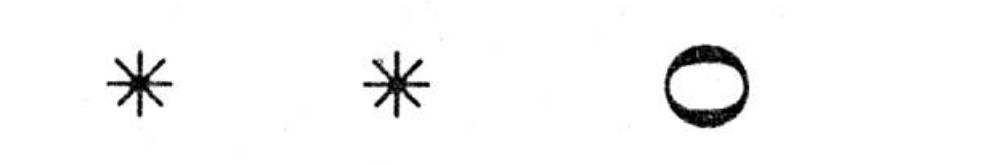}\\\hline
26)\includegraphics[height=8mm]{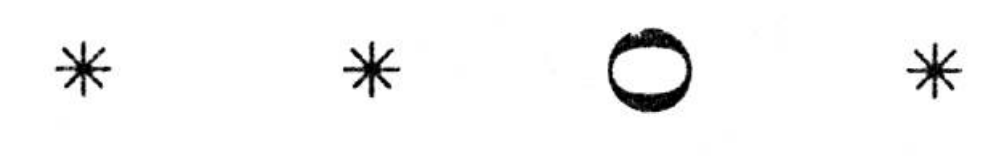}\\\hline
27)\includegraphics[height=8mm]{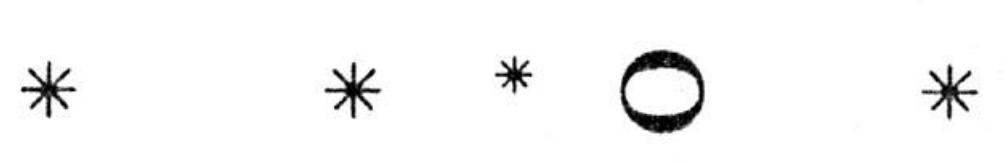}\\\hline
28)\includegraphics[height=8mm]{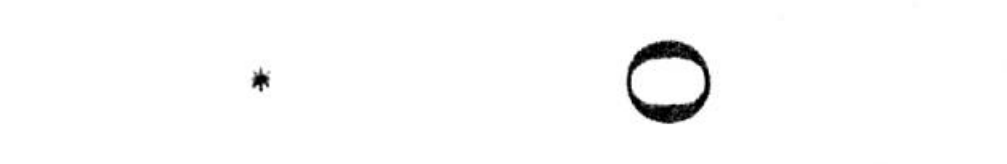}\\\hline
29)\includegraphics[height=8mm]{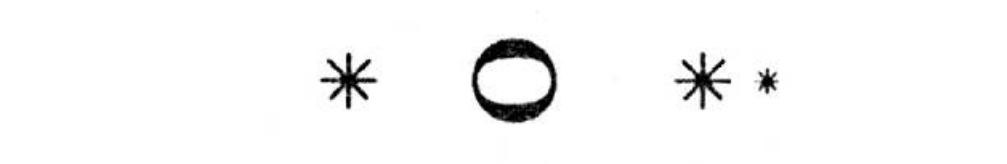}\\\hline
30)\includegraphics[height=8mm]{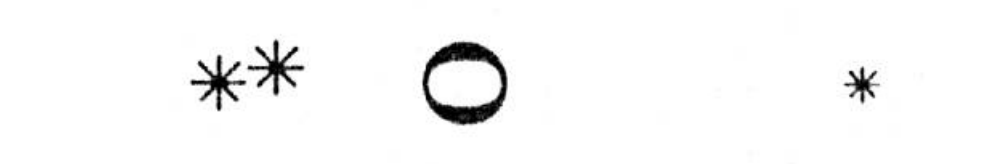}\\\hline
31)\includegraphics[height=8mm]{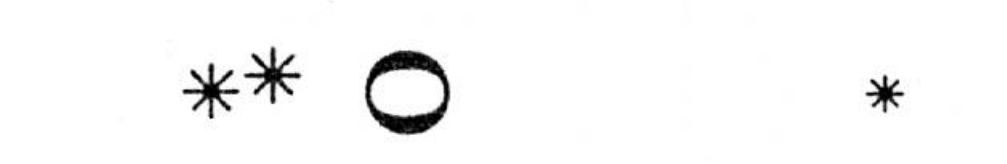}\\\hline
32)\includegraphics[height=8mm]{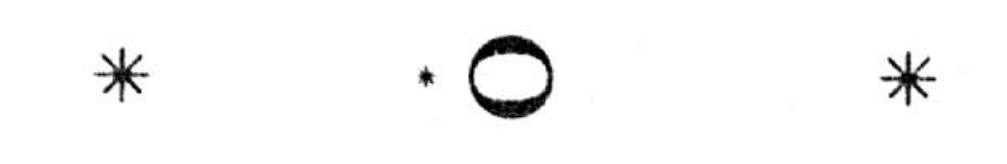}\\\hline
\end{tabular}
\end{minipage}
\caption{\label{fig:0} Sidereus Nuncius sketches A, from January 7 to February 1st.}
\end{figure}

\begin{figure}
\begin{minipage}[t]{0.48\linewidth}
\begin{tabular}{|l|}\hline
33)\includegraphics[height=8mm]{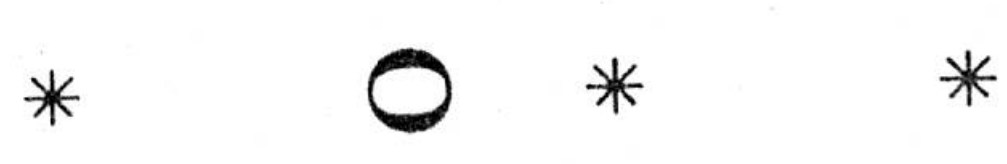}\\\hline
34)\includegraphics[height=8mm]{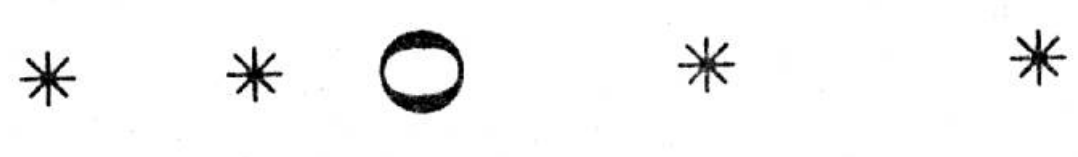}\\\hline
35)\includegraphics[height=8mm]{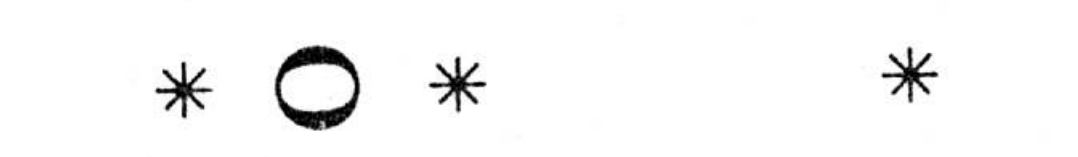}\\\hline
36)\includegraphics[height=8mm]{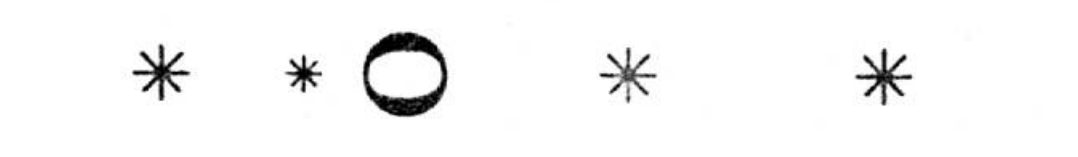}\\\hline
37)\includegraphics[height=8mm]{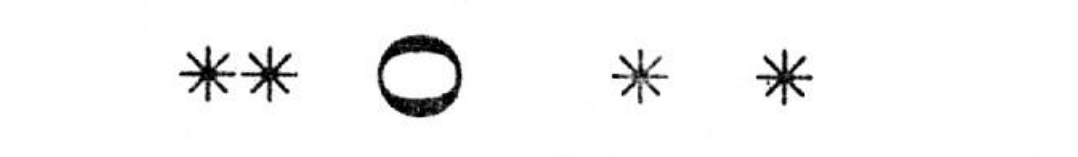}\\\hline
38)\includegraphics[height=8mm]{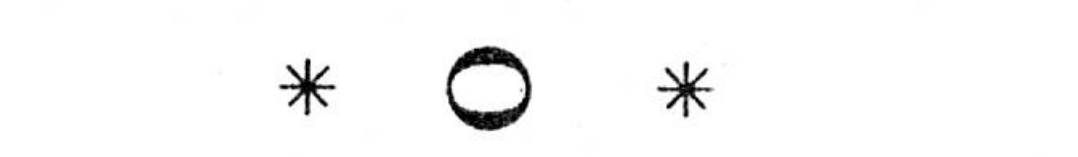}\\\hline
39)\includegraphics[height=8mm]{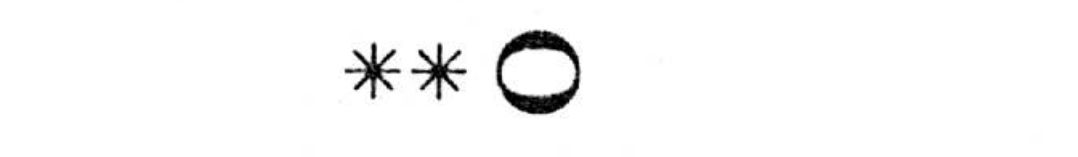}\\\hline
40)\includegraphics[height=8mm]{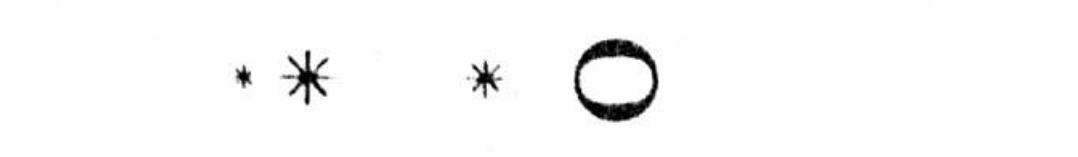}\\\hline
41)\includegraphics[height=8mm]{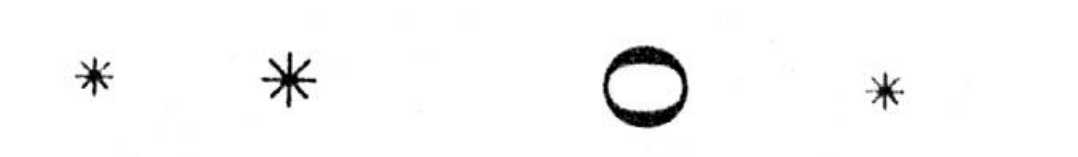}\\\hline
42)\includegraphics[height=8mm]{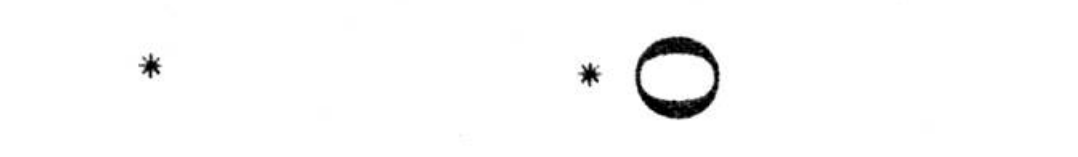}\\\hline
43)\includegraphics[height=8mm]{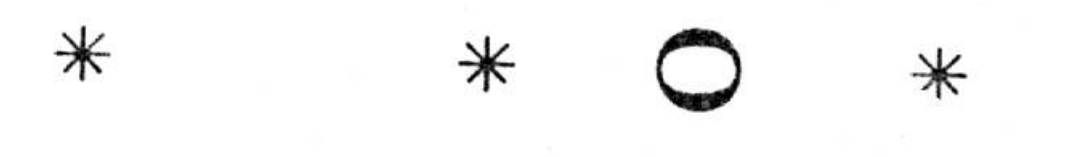}\\\hline
44)\includegraphics[height=8mm]{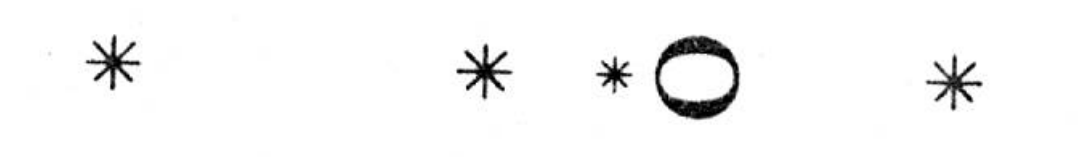}\\\hline
45)\includegraphics[height=8mm]{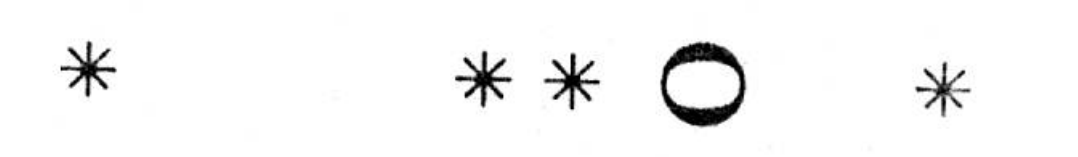}\\\hline
46)\includegraphics[height=8mm]{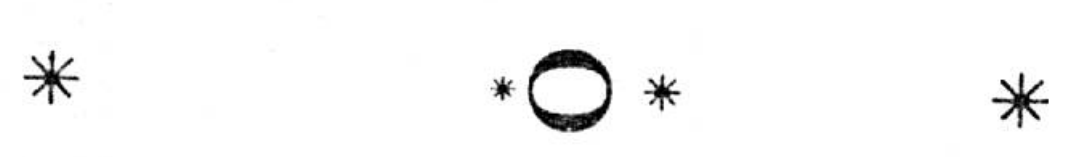}\\\hline
47)\includegraphics[height=8mm]{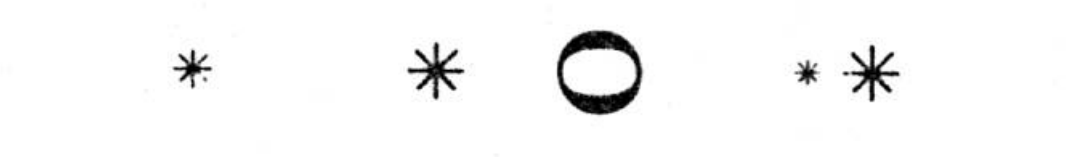}\\\hline
48)\includegraphics[height=8mm]{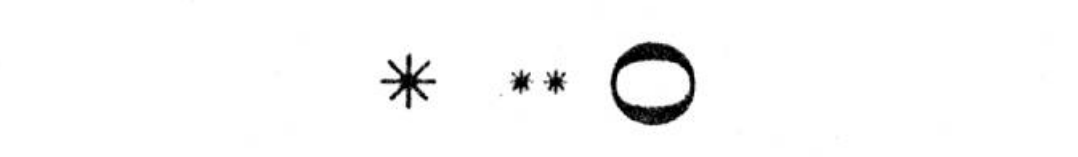}\\\hline
\end{tabular}
\end{minipage}%
\begin{minipage}[t]{0.48\linewidth}
\begin{tabular}{|l|}\hline
49)\includegraphics[height=8mm]{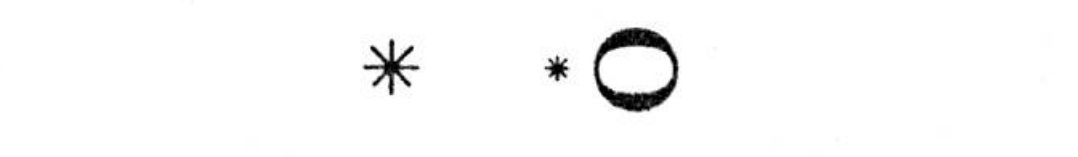}\\\hline
50)\includegraphics[height=8mm]{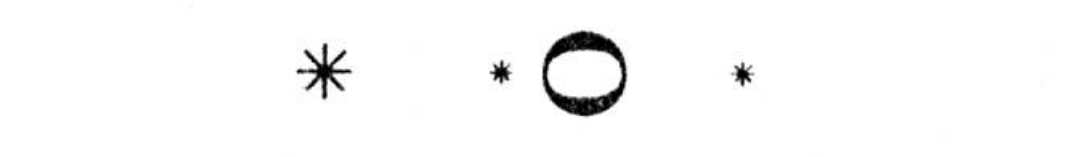}\\\hline
51)\includegraphics[height=8mm]{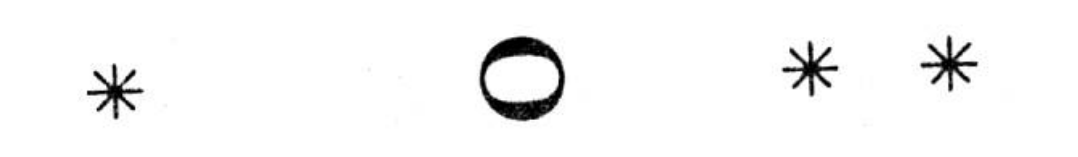}\\\hline
52)\includegraphics[height=8mm]{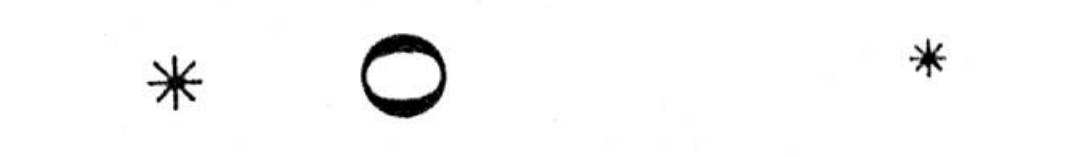}\\\hline
53)\includegraphics[height=8mm]{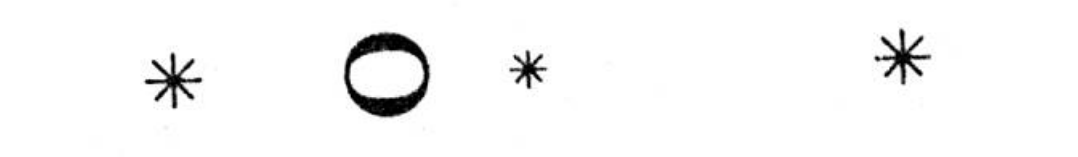}\\\hline
54)\includegraphics[height=8mm]{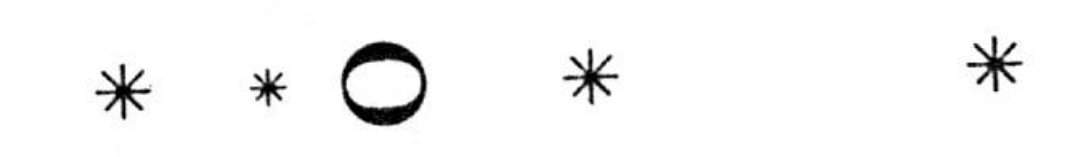}\\\hline
55)\includegraphics[height=8mm]{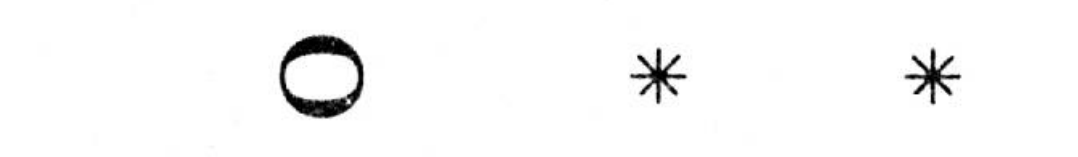}\\\hline
56)\includegraphics[height=8mm]{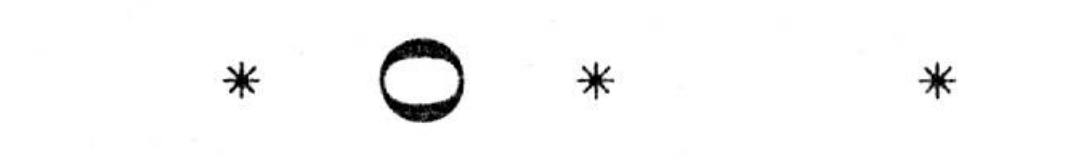}\\\hline
57)\includegraphics[height=8mm]{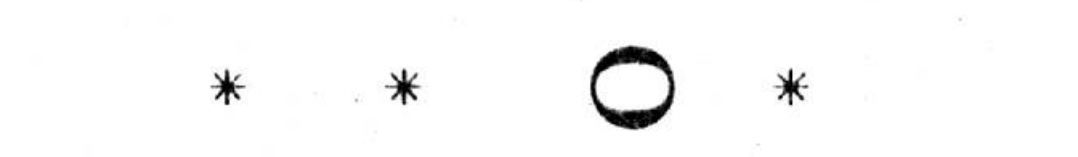}\\\hline
58)\includegraphics[height=8mm]{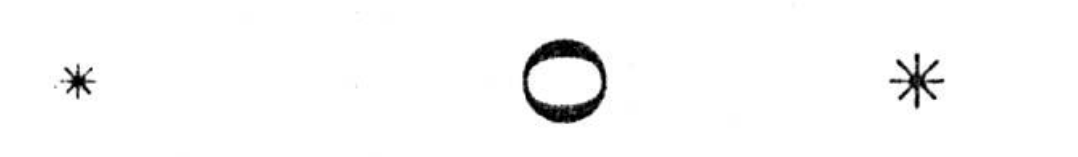}\\\hline
59)\includegraphics[height=8mm]{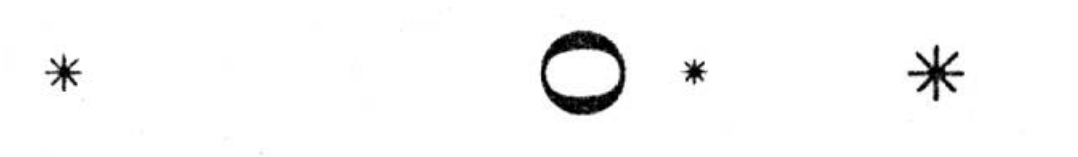}\\\hline
60)\includegraphics[height=8mm]{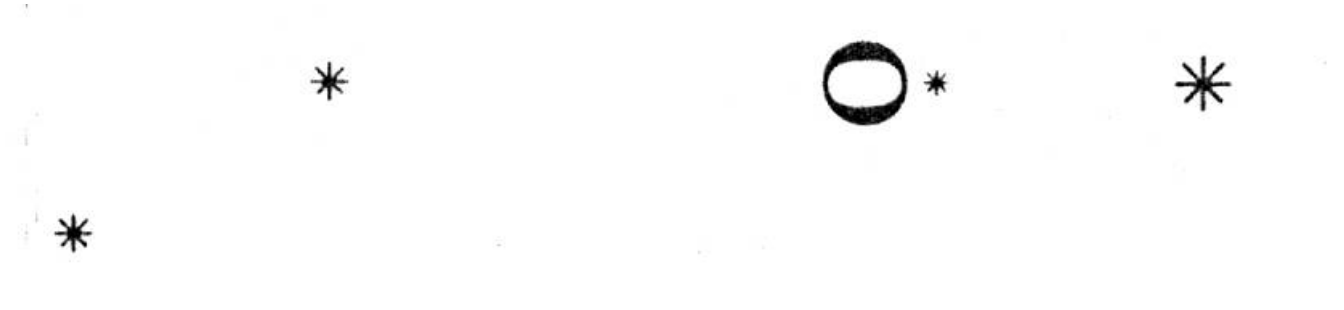}\\\hline
61)\includegraphics[height=8mm]{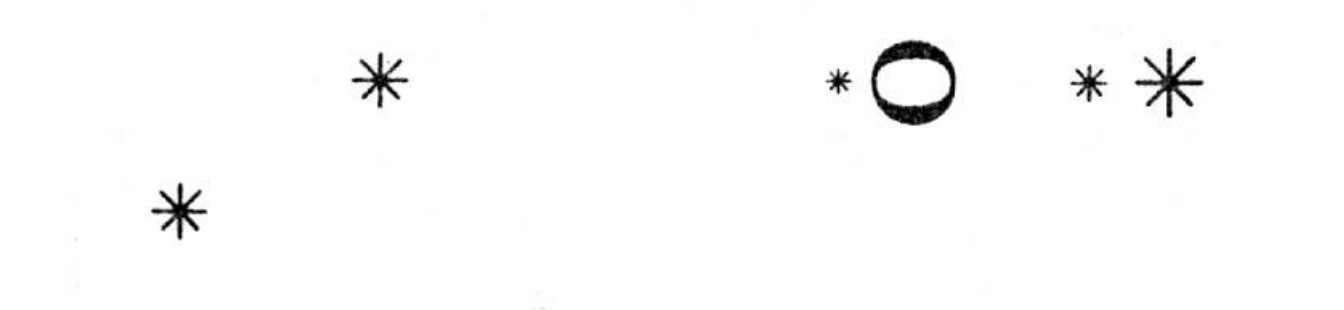}\\\hline
62)\includegraphics[height=8mm]{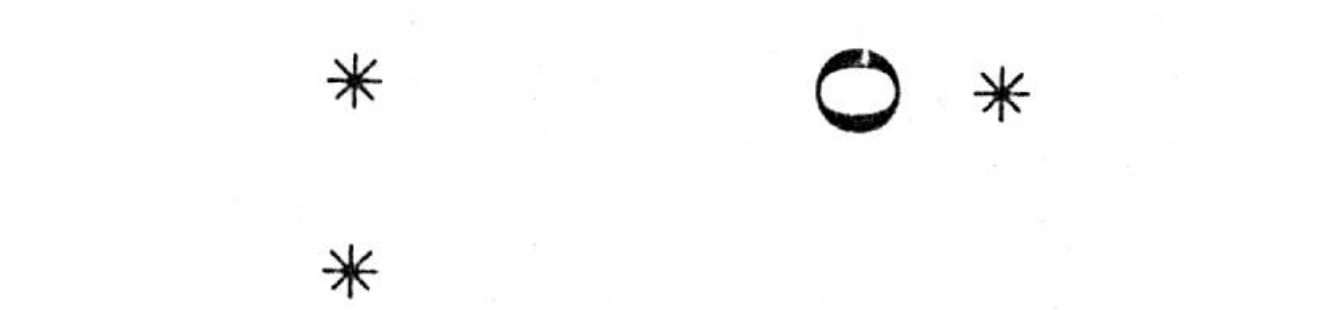}\\\hline
63)\includegraphics[height=8mm]{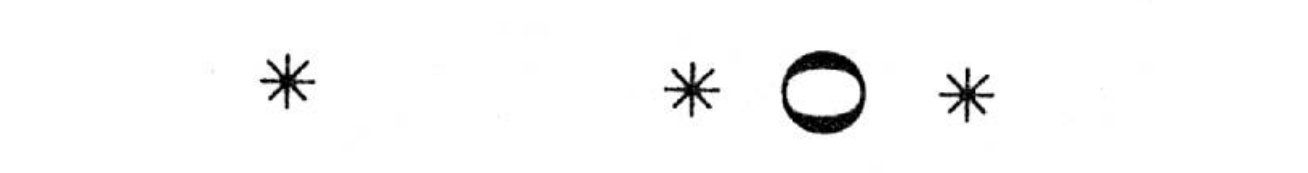}\\\hline
64)\includegraphics[height=8mm]{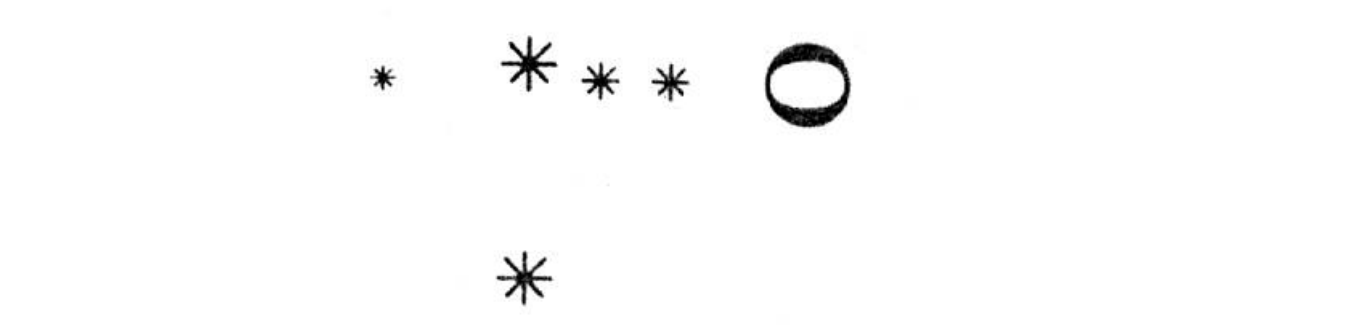}\\\hline
\end{tabular}
\end{minipage}
\caption{\label{fig:1} Sidereus Nuncius sketches B, from February 2 to March 1st.}
\end{figure}

The images were extracted from the Sidereus Nuncius dititally available at \cite{SN}. We show the collection of the sketches in chronological order with a progressive numeric labeling in Figs.~\ref{fig:0}-\ref{fig:1}. 

In total Galileo presents 65
sketches in 1610 from January 7$^{\rm{th}}$ up to March 1$^{\rm{st}}$ basically inspecting Jupiter whenever the meteorological conditions in Padua or Venice allowed him. Here we consider only 64 sketches because one is a duplicate of a Jupiter configuration with the addition of the above-mentioned fixed reference star.
As it became evident that the change in the
configurations had visible changes on a scale of a few
hours\footnote{The period of Io is just about 1.8 days.} Galileo
decided to observed the system at different times during the same
night. Such multiple observations occurred for example on Jan. 15 (id
7, 8), Jan. 17 (id 10, 11), Jan. 19 (id 13, 14), Jan. 20 (id 15, 16,
17), Jan. 22 (id 19, 20), Jan. 23 (id 21, 22), Jan. 24 (id 23, 24),
Jan. 26 (id 26, 27), Jan. 31 (id 30, 31), Feb. 2 (id 33, 34), Feb. 4
(id 36, 37), Feb. 11 (id 43, 44, 45), Feb. 15 (id 48, 49, 50), Feb. 18
(id 53, 54), Feb. 26 (id 58, 59), Feb. 28 (id 62, 63). The range of
these measurements taken during the same night spans up to about six
hours. On Jan. 20 for example the first observation was done 1h15'
after sunset and the last 7h after sunset so at about 6 p.m. and 1
a.m..  Generally Galilei reported the configuration of the satellites
only except in four occasions (observation 60, 61, 62, 64) when he
also noted down the position of a nearby star. This star is named HD32811 or SAO 76962m or HIP 23784 in modern catalogues. It has a visual magnitude of 7.16, a distance of 872.09 light years, type B8/9V and is located at 4h 43m 18.2s +21$^\circ$54$^\prime$9.9$^{\prime\prime}$. These data were obtained with the online tool \textsc{Stellarium web} that we will use extensively in the following\cite{stellarium,Stellarium_v25_2,Zotti2021Stellarium}. The observation of HD32811 was intended to clarify the motion of Jupiter and of its satellites with respect to the ``really-fixed'' stars and came at end of the campaign.  The satellites are marked as asterisks and the size of the symbol is a proxy for the relative brightness. 
Galilei always takes care of specifying if the satellites appeared as aligned or not. See for example observations 6, 7, 11, 16, 17, 19, 23, 27, 30, 31 in Appendix \ref{fullcomp}. As we will see in the following the recordings of these displacements from the ecliptic plane are remarkably accurate and recorded with full awareness.

\begin{figure}
    \centering
    \includegraphics[width=0.9\linewidth]{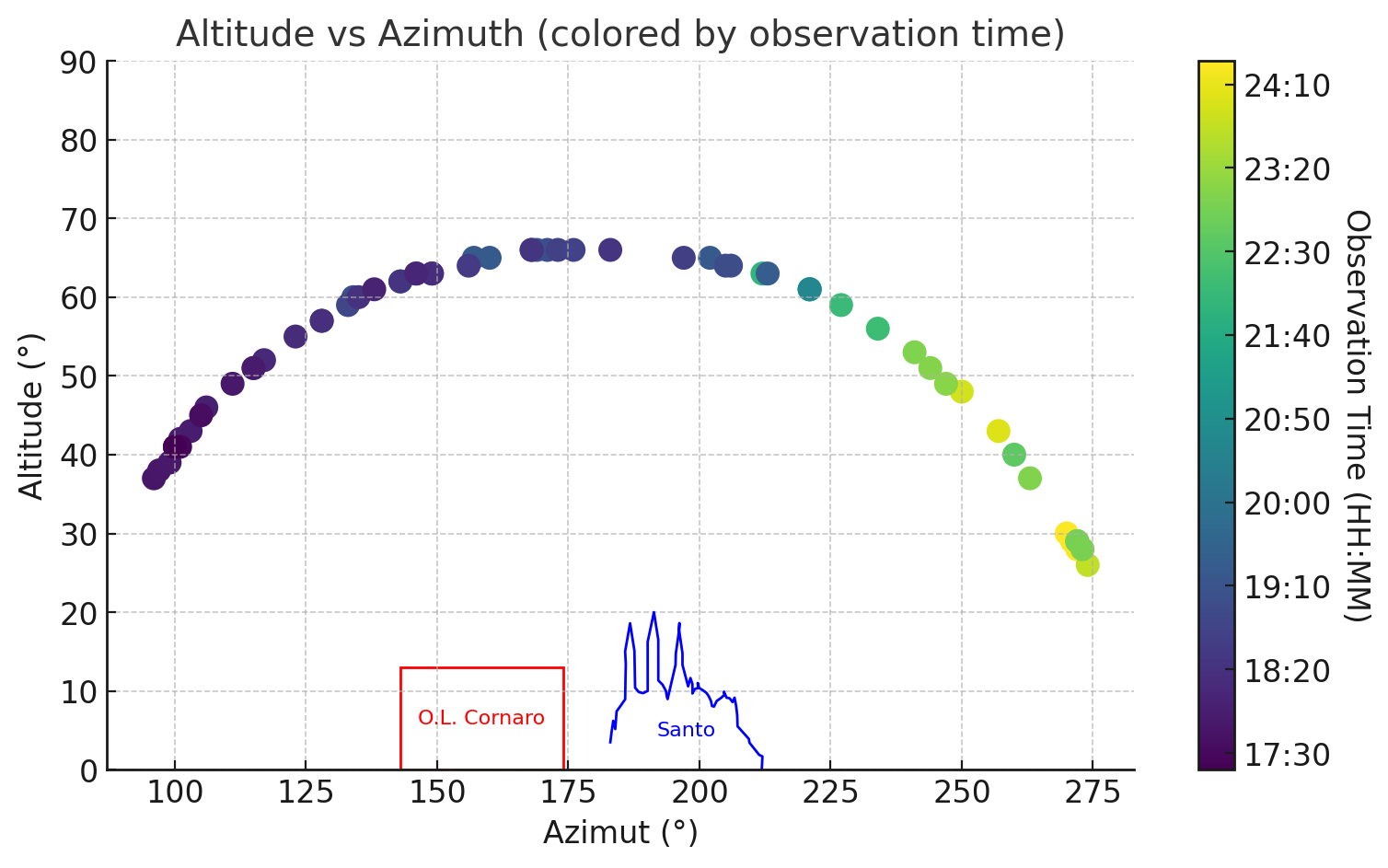}
    \caption{Distribution of the observations in azimuth and altitude for the supposed location of the observations in Padua (45.4032, 11.8818, currently at via Galileo Galilei 18). The color code indicates CET time of the observation.}
    \label{fig:altaz}
\end{figure}

The observations are generally thought to have taken place in the garden of Galilei's house in Padua\cite{Benucci2013casa}, located just north of the Basilica del Santo and the Odeo Cornaro at 45.4032, 11.8818. From Jan.~30 to Feb.~12 we know from the correspondence that Galilei was in Venice instead\cite{Bernieri2012}. The distribution of the position of Jupiter in alt-azimuthal coordinates is shown in Fig.~\ref{fig:altaz} with color codes indicating the observation time during the night. It is apparent how Galilei had a rather clear horizon as the observations are almost uniformly distributed. The asymmetry towards the East is due to the fact that the observations could not have been made during daytime. In general Galilei used to observe immediately after the sunset. The late observations at about midnight are limited by the fact that Jupiter started to be low on the horizon. The altitude angle for the telescope ranged from about 30 to 70$^\circ$. It is interesting to notice that the highest point of the Basilica del Santo is 72~m at a distance of about 200~m has hence an altitude of about 20$^\circ$, it covers an azimuth between roughly 183 and 212$^\circ$. The Odeo and Loggia Cornaro were at about 50~m with an height of about 12~m thus subtending an angle of 13$^\circ$ in an azimuth window between 174 and 143$^\circ$. The house itself is oriented almost east-west so the late observations might have been limited by the house itself as they happened about 5$^\circ$ north of the West direction. This setting is hence perfectly consistent with the observations\footnote{An early hypothesis that the tower of Porta Molino, along the medieval walls of Padua, was also the setting of the observation has been recently dismissed}.

\subsection{Dataset from the digitization of the sketches}
\label{digit}
Tables \ref{tab:0} and \ref{tab:1} summarize the measurements; the
first table refers to the observations performed in January while the
second to those of February and March.  Data have been obtained
from the analysis of digitized images using cursors that were manually centered on the satellites. For each image the size of the planet disk was also recorded to normalize the elongations in terms of this diameter. The diameter has always been found to be constant between the sketches with few exceptions (60, 61, 62, 64) where also the nearby star is drawn.  
The first column is a
sequential identifier that correspond to the one of Fig.~\ref{fig:0}.  The
following three columns give the time and hour of the observation as
reported by Galilei (in italic hours). The 5$^{ \rm{th}}$ column
reports the number of visible satellites. The last columns show the
position of the satellites ($x_i$, $y_i$) where $x$ runs horizontally
from left to right and $y$ vertically from bottom to top. The coordinates start here at the center of the disk and are normalized to the diameter of the disk in the drawings. In the
sketches the East is left and West right. This is the most natural
choice but is also motivated by the fact that Galilei's telescope was not inverting the image being composed of a semi-convex or convex objective and bi-concave or plano-concave eyepiece.

\begin{table}[]
\begin{scriptsize}
\centering
\begin{tabular}{ r r r r r r r r r }
\toprule
id & m & day & h:min & $n$ & $(x_0, y_0)$ & $(x_1, y_1)$ & $(x_2, y_2)$ & $(x_3, y_3)$\\
\midrule
1 & 1 & 7 & 1:00 & 3 & (-5.72, 0) & (-2.28, 0) & (2.42, 0)&\\
2 & 1 & 8 & / & 3 & (1.53, 0) & (3.14, 0) & (4.78, 0)&\\
3 & 1 & 10 & / & 2 & (-4.09, 0) & (-1.98, 0)&&\\
4 & 1 & 11 & / & 2 & (-4.95, 0) & (-3.64, 0)&&\\
5 & 1 & 12 & 3:00 & 3 & (-2.73, 0) & (-1.36, 0) & (1.50, 0)&\\
6 & 1 & 13 & / & 4 & (-2.16, 0) & (0.95, 0) & (1.64, 0.11) & (2.25, 0)\\
7 & 1 & 15 & 3:00 & 4 & (2.05, 0) & (3.41, 0) & (4.77, 0) & (7.05, -0.52)\\
8 & 1 & 15 & 7:00 & 3 & (3.14, 0) & (4.09, 0) & (6.43, 0)&\\
9 & 1 & 16 & 1:00 & 3 & (-0.98, 0) & (1.02, 0) & (5.41, 0)&\\
10 & 1 & 17 & 0:30 & 2 & (-1.95, 0) & (4.77, 0)&&\\
11 & 1 & 17 & 5:00 & 3 & (-2.82, 0.05) & (-2.25, -0.05) & (5.95, -0.14)&\\
12 & 1 & 18 & 0:20 & 2 & (-4.50, 0.09) & (5.95, -0.18)&&\\
13 & 1 & 19 & 2:00 & 3 & (-3.50, 0) & (3.11, 0) & (5.43, 0)&\\
14 & 1 & 19 & 5:00 & 4 & (-3.32, 0) & (-1.48, 0) & (2.75, 0) & (4.39, 0)\\
15 & 1 & 20 & 1:15 & 3 & (-1.00, 0) & (0.95, 0) & (1.75, 0)&\\
16 & 1 & 20 & 6:00 & 3 & (-1.59, 0) & (1.14, 0.18) & (1.73, 0.18)&\\
17 & 1 & 20 & 7:00 & 4 & (-2.00, 0) & (0.82, 0.14) & (1.41, 0) & (1.82, 0.14)\\
18 & 1 & 21 & 0:30 & 4 & (-1.95, 0) & (-1.45, 0) & (-1.09, 0) & (2.39, 0)\\
19 & 1 & 22 & 2:00 & 4 & (-3.36, 0) & (1.32, 0) & (1.93, -0.09) & (4.11, -0.09)\\
20 & 1 & 22 & 6:00 & 4 & (-3.32, 0) & (1.23, 0) & (2.09, 0) & (2.84, 0)\\
21 & 1 & 23 & 0:40 & 3 & (-4.86, 0) & (-1.82, 0) & (1.89, 0)&\\
22 & 1 & 23 & 5:00 & 1 & (-4.27, 0)&&&\\
23 & 1 & 24 & / & 3 & (-6.82, 0) & (-2.95, -0.14) & (-2.14, 0)&\\
24 & 1 & 24 & 6:00 & 2 & (-5.27, 0) & (-1.59, 0)&&\\
25 & 1 & 25 & 1:40 & 2 & (-6.82, 0) & (-3.68, 0)&&\\
26 & 1 & 26 & 0:40 & 3 & (-6.77, 0) & (-3.36, 0) & (3.43, 0)&\\
27 & 1 & 26 & 5:00 & 4 & (-7.32, 0) & (-3.70, 0) & (-1.75, 0.18) & (3.45, 0)\\
28 & 1 & 27 & 1:00 & 1 & (-4.91, 0)&&&\\
29 & 1 & 30 & 1:00 & 3 & (-1.95, 0) & (2.30, 0) & (3.07, 0)&\\
30 & 1 & 31 & 2:00 & 3 & (-2.93, 0) & (-2.23, 0.18) & (4.80, 0)&\\
31 & 1 & 31 & 4:00 & 3 & (-2.32, 0) & (-1.59, 0.20) & (5.75, 0)&\\
\bottomrule
\end{tabular}
\caption{Dataset 1 (January). The last four columns contain the position of the satellites from left to right in units of the Jupiter equatorial diameter, as in the sketches of Galilei.}
\label{tab:0}
\end{scriptsize}
\end{table}

\begin{table}[]
\centering
\begin{scriptsize}
\begin{tabular}{ r r r r r r r r r }
\toprule
id & m & day & h:min & $n$ & $(x_0, y_0)$ & $(x_1, y_1)$ & $(x_2, y_2)$ & $(x_3, y_3)$\\
\midrule
32 & 2 & 1 & 2:00 & 3 & (-4.66, 0) & (-1.00, 0) & (4.77, 0)&\\
33 & 2 & 2 & / & 3 & (-4.30, 0) & (2.48, 0) & (6.73, 0.14)&\\
34 & 2 & 2 & 7:00 & 4 & (-4.39, -0.05) & (-1.95, 0) & (3.36, 0.05) & (7.27, 0.14)\\
35 & 2 & 3 & 7:00 & 3 & (-1.55, 0) & (1.68, 0) & (7.00, 0.18)&\\
36 & 2 & 4 & 2:00 & 4 & (-2.86, 0) & (-1.18, 0) & (2.64, 0) & (5.64, 0)\\
37 & 2 & 4 & 7:00 & 4 & (-2.50, 0) & (-1.77, 0) & (2.59, 0) & (4.32, 0)\\
38 & 2 & 6 & / & 2 & (-2.14, 0) & (2.34, 0)&&\\
39 & 2 & 7 & / & 2 & (-1.95, 0) & (-1.18, 0)&&\\
40 & 2 & 8 & 1:00 & 3 & (-4.36, 0) & (-3.64, 0) & (-1.55, 0)&\\
41 & 2 & 9 & 0:30 & 3 & (-6.45, 0) & (-4.25, 0) & (2.86, 0)&\\
42 & 2 & 10 & 1:30 & 2 & (-6.18, 0.14) & (-1.02, 0)&&\\
43 & 2 & 11 & 1:00 & 3 & (-7.25, 0.14) & (-2.50, 0) & (2.84, 0)&\\
44 & 2 & 11 & 3:00 & 4 & (-6.89, 0.14) & (-2.50, 0.05) & (-0.95, 0) & (3.07, 0)\\
45 & 2 & 11 & 5:30 & 4 & (-7.23, 0.14) & (-2.59, 0) & (-1.55, 0) & (2.84, 0)\\
46 & 2 & 12 & 0:40 & 4 & (-6.09, 0.14) & (-0.82, 0) & (1.09, 0) & (5.32, -0.14)\\
47 & 2 & 13 & 0:30 & 4 & (-4.82, 0) & (-1.95, 0) & (2.45, 0) & (3.20, 0)\\
48 & 2 & 15 & / & 3 & (-2.86, 0) & (-1.55, 0) & (-1.14, 0)&\\
49 & 2 & 15 & 5:00 & 2 & (-2.89, 0) & (-0.91, 0)&&\\
50 & 2 & 15 & 6:00 & 3 & (-3.07, 0) & (-0.98, 0) & (1.86, 0)&\\
51 & 2 & 16 & 6:00 & 3 & (-4.77, -0.16) & (3.45, 0.14) & (5.09, 0.18)&\\
52 & 2 & 17 & 1:00 & 2 & (-2.68, 0) & (6.18, 0.14)&&\\
53 & 2 & 18 & / & 3 & (-2.50, 0) & (1.70, 0.09) & (6.11, 0.23)&\\
54 & 2 & 18 & 6:00 & 4 & (-3.07, 0) & (-1.36, 0) & (2.48, 0) & (7.27, 0.23)\\
55 & 2 & 19 & 0:40 & 2 & (4.00, 0) & (7.23, 0)&&\\
56 & 2 & 21 & 1:30 & 3 & (-2.14, 0) & (2.07, 0) & (6.09, 0)&\\
57 & 2 & 25 & 1:30 & 3 & (-4.77, 0) & (-2.68, 0) & (1.89, 0)&\\
58 & 2 & 26 & 0:30 & 2 & (-5.73, 0) & (4.16, 0)&&\\
59 & 2 & 26 & 5:00 & 3 & (-6.09, 0) & (1.36, 0) & (4.20, 0)&\\
60 & 2 & 26 & 5:00 & 4 & (-6.53, 0) & (0.94, 0) & (4.18, 0)&\\
61 & 2 & 27 & 1:04 & 4 & (-5.05, 0) & (-0.70, 0) & (1.68, 0) & (2.43, 0)\\
62 & 2 & 28 & 1:00 & 2 & (1.74, 0)&&&\\
63 & 2 & 28 & 5:00 & 3 & (-6.364, 0) & (-1.545, 0) & (1.682, 0)&\\
64 & 3 & 1 & / & 4 & (-5.18, 0) & (-3.41, 0.24) & (-2.53, 0) & (-1.71, 0)\\\bottomrule
\end{tabular}
\caption{Dataset 2 (February-March). The last four columns contain the position of the satellites from left to right in units of the Jupiter equatorial diameter, as in the sketches of Galilei.}
\label{tab:1}
\end{scriptsize}
\end{table}

\subsection{Conversion of the observation times}
\label{modernh}
Galilei expressed the observation times in terms of hours since sunset
following italic hours that was used at those times. The \textit{first hour}, for example, refers to one hour after sunset. To convert these data into modern hours (CET) we considered the time of sunset in Padua in the beginning of the year 
and sum the hour reported by Galilei to this value assuming that each hour
has the same length.  The time of sunset was approximated linearly as:
$h_{sunset}=16.58+0.02353~d$ where hours are expressed in decimal
fractions and $d$ is the ordinal day since the beginning of the year. The result of the conversion is shown in Tab.~\ref{tab:times} in Appendix where all the observations are associated to a time expressed in CET. The first hour corresponded to about 17:45 CET on Jan. 7 and at about 19:00 at the end of February. The latest observations (7$^{th}$ hour) corresponded instead to about 23:45, at the beginning of January, and 1:00 AM, at the beginning of March. Note that the setting time of Jupiter was about 5.30 and 2.00 at the beginning and end of the considered period.

It should be noted that for 11 observations Galilei does not mention the time. They are marked with a back-slash in the fourth column of Tab.~\ref{tab:0} and \ref{tab:1}. These are observations 2, 3, 4, 6, 23, 33, 38, 39, 48, 53 and 64. In this case for the subsequent analysis we assumed that the observation time was the first hour, a frequent choice. This introduces for these observations an error up to potentially up to 6 hours in the worst case. This indetermination on the time introduces a significant angular bias especially for satellites close to Jupiter when their angular velocities are of about 20, 15, 12, 8$^{\prime\prime}$/h for Io, Europa Ganymede and Callisto, respectively. Given the angular diameter of Jupiter (45-38$^{\prime\prime}$ in those days) it means that a satellite (especially Io) can easily be visible on different sides of the planet in a six hour time span. In the first column of Tab.~\ref{tab:stellarium} in Appendix we have clearly identified these cases with a $\star$. Looking at these eleven configurations it does not seem that this approximation poses specific problems in the interpretation but in some cases it might be argued that some satellites might have been missed because at the real observation hour the satellite might have been closer to Jupiter than at the assumed time (i.e. for obs. 23 Io was not seen even if it seems far away enough for having been seen; in obs. 38 Callisto is not separated from Jupiter even if it seems far enough; similar consideration for Io in obs. 39, Callisto in obs. 48). These cases where hence marked separately in the analysis of the efficiency of the detection near the disk that will be discussed later.

\subsection{Dataset from the  angular readings of Galileo}

In the Sidereus, starting from Jan. 12, Galilei begins to report the positions of the satellites in numbers by providing their angular displacements besides the sketches. He states: 
{\emph{
It was therefore established by me, and concluded beyond all doubt, that there are in the heavens three wandering stars revolving around Jupiter, similar to Venus and Mercury around the Sun; which was finally observed more clearly than in broad daylight through many subsequent observations: and not only three, but four were found to be the wandering stars performing their revolutions around Jupiter ... And I also measured the intervals between them with the method explained earlier.
}}
He also says the he noted down the time as he realized how fast were the changes (but still they are missing in 7 cases). Another step in awareness of the impact of the observations happened on Jan. 15, late night, when he started to take notes directly in Latin instead of Italian in view of a publication.

The readings are given in primes (pr.) and seconds of arc (sec). The smallest granularity used is 10$^{\prime\prime}$ but typically distances are expressed in round primes or 1/2 or 1/3 or a prime. 
Vertical displacements are reported but not quantified.
The distant satellites are often characterized by their distance from Jupiter while the other by their relative separations. For a modern reader it is interesting to notice how it was more natural in those days to use an accurate and descriptive account in words  with a not fixed pattern rather than choosing to provide a table. Thanks to the precise wording it is anyway easy to do the trivial math and come out with a table where the distances are expressed symmetrically with respect to Jupiter with eastern satellites having negative elongations. The collection of all the measurements is in Tab.~\ref{tab:angulardist0} (January) and Tab.~\ref{tab:angulardist1} (February-March).
\begin{table}[htbp]
\centering
\scriptsize
\setlength{\tabcolsep}{4pt}
\begin{tabular}{r r r r r r r r r r r}
\toprule
$id$ & y & m & d & h & min & $n_{sat}$ & $x_1$ ($^\prime$)& $x_2$  ($^\prime$)& $x_3$  ($^\prime$) & $x_4$ ($^\prime$) \\
\midrule
1 & 1610 & 1 & 7 & 1 & 0 & 0 & / & / & / & / \\
2 & 1610 & 1 & 8 & -1 & -1 & 0 & / & / & / & / \\
3 & 1610 & 1 & 10 & -1 & -1 & 0 & / & / & / & / \\
4 & 1610 & 1 & 11 & -1 & -1 & 0 & / & / & / & / \\
5 & 1610 & 1 & 12 & 3 & 0 & 0 & / & / & / & / \\
5.1 & 1610 & 1 & 12 & 1 & 0 & 2 & -2.00 & 2.00 & / & / \\
6 & 1610 & 1 & 13 & -1 & -1 & 4 & -2.00 & 1.00 & 2.00 & 3.00 \\
7 & 1610 & 1 & 15 & 3 & 0 & 4 & 2.00 & 4.00 & 6.00 & 11.00 \\
8 & 1610 & 1 & 15 & 7 & 0 & 3 & 3.00 & 4.50 & 9.00 & / \\
9 & 1610 & 1 & 16 & 1 & 0 & 3 & -0.67 & 0.67 & 8.00 & / \\
10 & 1610 & 1 & 17 & 0 & 30 & 2 & -3.00 & 11.00 & / & / \\
11 & 1610 & 1 & 17 & 5 & 0 & 3 & -3.00 & -2.67 & 11.00 & / \\
12 & 1610 & 1 & 18 & 0 & 20 & 2 & -8.00 & 10.00 & / & / \\
13 & 1610 & 1 & 19 & 2 & 0 & 3 & -6.00 & 5.00 & 9.00 & / \\
14 & 1610 & 1 & 19 & 5 & 0 & 4 & -6.00 & -3.00 & 5.00 & 9.00 \\
15 & 1610 & 1 & 20 & 1 & 15 & 3 & -1.00 & 1.00 & 2.00 & / \\
16 & 1610 & 1 & 20 & 6 & 0 & 3 & -2.00 & 0.67 & 1.00 & / \\
17 & 1610 & 1 & 20 & 7 & 0 & 3 & -2.00 & 0.33 & 0.83 & 1.00 \\
18 & 1610 & 1 & 21 & 0 & 30 & 4 & -2.50 & -1.67 & -0.83 & 4.00 \\
19 & 1610 & 1 & 22 & 2 & 0 & 4 & -5.00 & 1.00 & 1.67 & 7.00 \\
20 & 1610 & 1 & 22 & 6 & 0 & 4 & -5.00 & 1.33 & 2.67 & 4.00 \\
21 & 1610 & 1 & 23 & 0 & 40 & 3 & -7.00 & -1.33 & 3.33 & / \\
22 & 1610 & 1 & 23 & 5 & 0 & 0 & / & / & / & / \\
23 & 1610 & 1 & 24 & -1 & -1 & 3 & -11.50 & -2.50 & -2.00 & / \\
24 & 1610 & 1 & 24 & 6 & 0 & 2 & -8.00 & -3.00 & / & / \\
25 & 1610 & 1 & 25 & 1 & 40 & 2 & -11.00 & -6.00 & / & / \\
26 & 1610 & 1 & 26 & 0 & 40 & 3 & -11.33 & -5.33 & 5.00 & / \\
27 & 1610 & 1 & 26 & 5 & 0 & 4 & -11.33 & -5.33 & -0.50 & 5.00 \\
28 & 1610 & 1 & 27 & 1 & 0 & 1 & -7.00 & / & / & / \\
29 & 1610 & 1 & 30 & 1 & 0 & 3 & -2.50 & 3.00 & 4.00 & / \\
30 & 1610 & 1 & 31 & 2 & 0 & 3 & -2.83 & -2.33 & 10.00 & / \\
31 & 1610 & 1 & 31 & 4 & 0 & 3 & -2.67 & -2.33 & 10.00 & / \\
32 & 1610 & 2 & 1 & 2 & 0 & 3 & -6.00 & -0.33 & 8.00 & / \\
\bottomrule
\end{tabular}
\caption{Angular displacements data as reported in the text by Galilei starting from Jan. 12 1610. January data.}
\label{tab:angulardist0}
\end{table}

\begin{table}[htbp]
\centering
\scriptsize
\setlength{\tabcolsep}{4pt}
\begin{tabular}{r r r r r r r r r r r}
\toprule
$id$ & y & m & d & h & min & $n_{sat}$ & $x_1$ ($^\prime$)& $x_2$  ($^\prime$)& $x_3$  ($^\prime$) & $x_4$ ($^\prime$) \\
\midrule
33 & 1610 & 2 & 2 & -1 & -1 & 3 & -6.00 & 4.00 & 12.00 & / \\
34 & 1610 & 2 & 2 & 7 & 0 & 4 & -5.67 & -1.67 & 6.00 & 14.00 \\
35 & 1610 & 2 & 3 & 7 & 0 & 3 & -1.50 & 2.00 & 12.00 & / \\
36 & 1610 & 2 & 4 & 2 & 0 & 4 & -3.67 & -0.67 & 4.00 & 10.00 \\
37 & 1610 & 2 & 4 & 7 & 0 & 4 & -2.50 & -2.00 & 4.00 & 7.00 \\
38 & 1610 & 2 & 6 & -1 & -1 & 2 & -2.00 & 5.00 & / & / \\
39 & 1610 & 2 & 7 & -1 & -1 & 2 & -2.00 & -1.00 & / & / \\
40 & 1610 & 2 & 8 & 1 & 0 & 4 & -5.67 & -5.33 & -1.50 & -1.33 \\
40.1 & 1610 & 2 & 8 & 3 & 0 & 4 & -6.33 & -6.00 & -1.50 & -0.17 \\
40.2 & 1610 & 2 & 8 & 4 & 0 & 3 & -6.33 & -6.00 & -1.50 & / \\
41 & 1610 & 2 & 9 & 0 & 30 & 3 & -11.00 & -7.00 & 4.00 & / \\
42 & 1610 & 2 & 10 & 1 & 30 & 2 & -10.00 & -0.33 & / & / \\
42.1 & 1610 & 2 & 10 & 4 & 0 & 1 & -12.00 & / & / & / \\
43 & 1610 & 2 & 11 & 1 & 0 & 3 & -12.00 & -4.00 & 4.00 & / \\
44 & 1610 & 2 & 11 & 3 & 0 & 4 & -12.00 & -4.00 & -0.50 & 4.00 \\
45 & 1610 & 2 & 11 & 5 & 30 & 4 & -12.00 & -4.00 & -2.00 & 4.00 \\
46 & 1610 & 2 & 12 & 0 & 40 & 4 & -10.00 & -0.67 & 1.00 & 8.00 \\
46.1 & 1610 & 2 & 12 & 4 & 0 & 3 & -10.00 & 1.00 & 8.00 & / \\
47 & 1610 & 2 & 13 & 0 & 30 & 4 & -6.00 & -2.00 & 3.50 & 4.00 \\
48 & 1610 & 2 & 15 & -1 & -1 & 3 & -3.17 & -1.17 & -0.83 & / \\
49 & 1610 & 2 & 15 & 5 & 0 & 2 & -4.00 & -0.50 & / & / \\
50 & 1610 & 2 & 15 & 6 & 0 & 3 & -4.00 & -0.50 & 2.00 & / \\
51 & 1610 & 2 & 16 & 6 & 0 & 3 & -7.00 & 5.00 & 8.00 & / \\
52 & 1610 & 2 & 17 & 1 & 0 & 2 & -3.00 & 10.00 & / & / \\
52.1 & 1610 & 2 & 17 & 6 & 0 & 2 & -0.83 & 12.00 & / & / \\
53 & 1610 & 2 & 18 & 1 & 0 & 3 & -3.00 & 2.00 & 10.00 & / \\
53.1 & 1610 & 2 & 18 & 2 & 0 & 3 & -3.00 & 3.00 & 10.00 & / \\
54 & 1610 & 2 & 18 & 6 & 0 & 4 & -4.83 & -1.83 & 3.00 & 10.00 \\
55 & 1610 & 2 & 19 & 0 & 40 & 2 & 7.00 & 13.00 & / & / \\
56 & 1610 & 2 & 21 & 1 & 30 & 3 & -2.00 & 3.00 & 10.00 & / \\
57 & 1610 & 2 & 25 & 1 & 30 & 3 & -8.00 & -4.00 & 2.00 & / \\
58 & 1610 & 2 & 26 & 0 & 30 & 2 & -10.00 & 6.00 & / & / \\
59 & 1610 & 2 & 26 & 5 & 1 & 3 & -11.00 & 1.00 & 6.00 & / \\
60 & 1610 & 2 & 26 & 5 & 2 & 0 & / & / & / & / \\
61 & 1610 & 2 & 27 & 1 & 4 & 4 & -10.00 & -0.50 & 2.50 & 3.50 \\
61.1 & 1610 & 2 & 27 & 5 & 0 & 4 & -10.00 & -1.00 & 2.50 & 3.50 \\
62 & 1610 & 2 & 28 & 1 & 0 & 2 & -9.00 & 2.00 & / & / \\
63 & 1610 & 2 & 28 & 5 & 0 & 3 & -9.00 & -2.00 & 2.00 & / \\
64 & 1610 & 3 & 1 & 0 & 40 & 4 & -7.33 & -3.33 & -3.00 & -2.00 \\
64.1 & 1610 & 3 & 2 & 0 & 30 & 3 & -7.00 & -6.50 & 2.00 & / \\
\bottomrule
\end{tabular}
\caption{Angular displacements data as reported in the text by Galilei starting from Jan 12 1610. February-March data.}
\label{tab:angulardist1}
\end{table}

The first column of Tabs.~\ref{tab:angulardist0} and \ref{tab:angulardist1} are the measurement identifiers already introduced in Tabs.~\ref{tab:0} and \ref{tab:1}. In this case we have added 9 measurements that are reported in the text but do not have a corresponding sketch. We have denoted them as decimals (5.1,
40.1, 40.2, 42.1, 46.1, 52.1, 53.1, 61.1, 64.1).
It is important to notice that the measurements that we are considering are in total 72 (64 with sketches+ 9  with only quoted measurements in the text). 

\subsection{Comparison of the two datasets}

\begin{figure}
    \centering
\includegraphics[width=\linewidth]{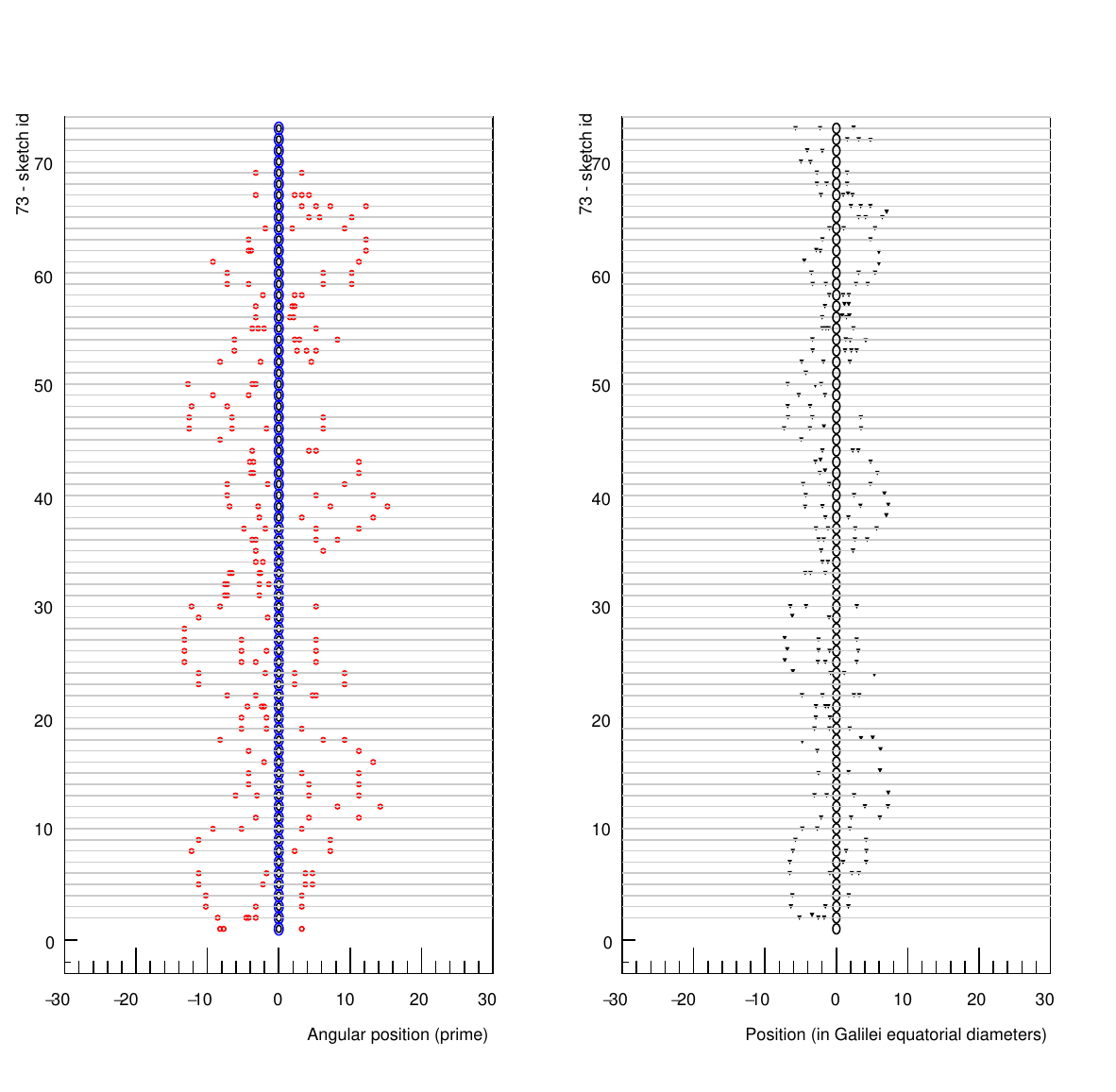}\\
    \caption{
    Left: reconstructed positions of Jupiter satellites from the angular data of Galilei (in primes). Right: reconstructed positions from the digitization of the sketches in units of the planet's equatorial diameters in the sketches themselves. Times increases from top to bottom.}
    \label{fig:pos}
\end{figure}

A graphical summary of the positions of the satellites is
shown in Fig.~\ref{fig:pos}. The time flows from top to
bottom. Comparing this plot with Figs.~\ref{fig:0} and \ref{fig:1}
ensures the absence of mistakes in the digitization process.

The left plot refers to the dataset with angular information (elongations in primes) while the one on the right comes from the digitization of the sketches and here the measurement are normalized to the diameter of Jupiter as in the drawings of the Sidereus.

\begin{figure}
    \centering
\includegraphics[width=\linewidth]{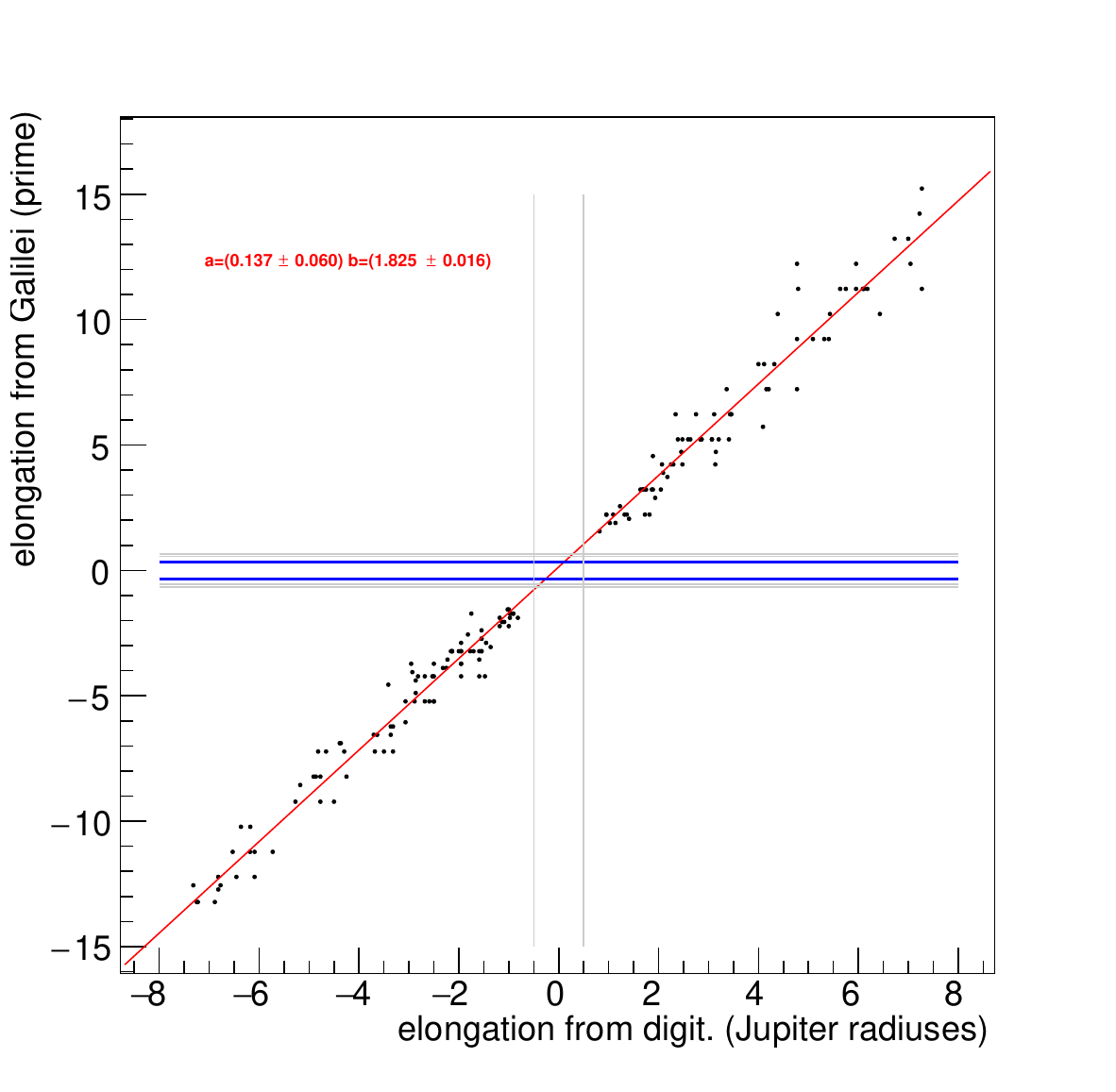}
    \caption{Scatter plot of the elongations with dataset-1 (digitization) and dataset-2 (angular measurements by Galilei) after the correction of Eq.~\ref{eq:corr}.}
    \label{fig:correlation}
\end{figure}

\begin{figure}
    \centering
\includegraphics[width=\linewidth]{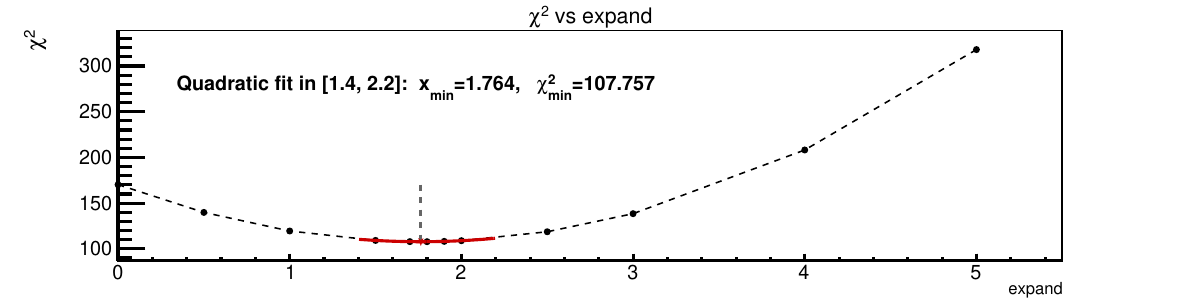}
    \caption{Variation of the $\chi^2$ of a linear fit to the data of Fig.~\ref{fig:correlation} as a function of $\epsilon$ in Eq.\ref{eq:corr}, here denoted as ``expand''.}
    \label{fig:chi2corr}
\end{figure}

In Fig.~\ref{fig:correlation} we show a scatter plot of the elongations with dataset-1 (digitization) and dataset-2 (angular measurements by Galilei). Initially we had interpreted the angular measurements as referred to the center of Jupiter while they should be clearly referred to the limb of Jupiter. In the first, wrong assumption, a clear offset between the positive and negative arms is evident. We have hence corrected the data by adding a half equatorial diameter of Jupiter with a scaling factor that was determined from the data:
\begin{equation}
\theta_i^{corr} = \theta_i + \rm{sign}(\theta_i)\epsilon\frac{\Delta_{avg}}{2}
\label{eq:corr}
\end{equation}
For the diameter of Jupiter ($\Delta_{avg}$) we have taken the average of the two extreme angular diameters of 45 and 38$^{\prime\prime}$. We have hence varied $\epsilon$ until the minimum $\chi^2$ was reached when performing a linear fit of all the points. The best match is achieved for $\epsilon=1.76$ (Fig.~\ref{fig:chi2corr}). This seems to suggest that the diameter of Jupiter was perceived with an expansion factor of this size i.e. of about 79-67$^{\prime\prime}$.
The correlation plot using the corrected angular elongations is shown in Fig.~\ref{fig:correlation}. The horizontal blue lines indicate the
real diameter of Jupiter (38-45$^{\prime\prime}$). The vertical gray lines the disk of Jupiter in the sketches that by definition has a width of one Jupiter diameter. From this figure it become clear how close the satellite recording fall from the disk of Jupiter. The linear fit yields the following relation between the datasets:
\begin{equation}
\theta_i^{corr}=(1.825 \pm 0.06)^\prime\times x_i+(0.137 \pm 0.060)^\prime
\label{eq:scaling}
\end{equation}
where we denoted with $x_i$ the elongations normalized to the diameter of Jupiter as reported in the drawings (dataset-1) and with $\theta_i^{corr}$ the angular recordings of Galilei corrected from the limb of Jupiter disk after the tuning done from the data.

\begin{figure}
    \centering
\includegraphics[width=\linewidth]{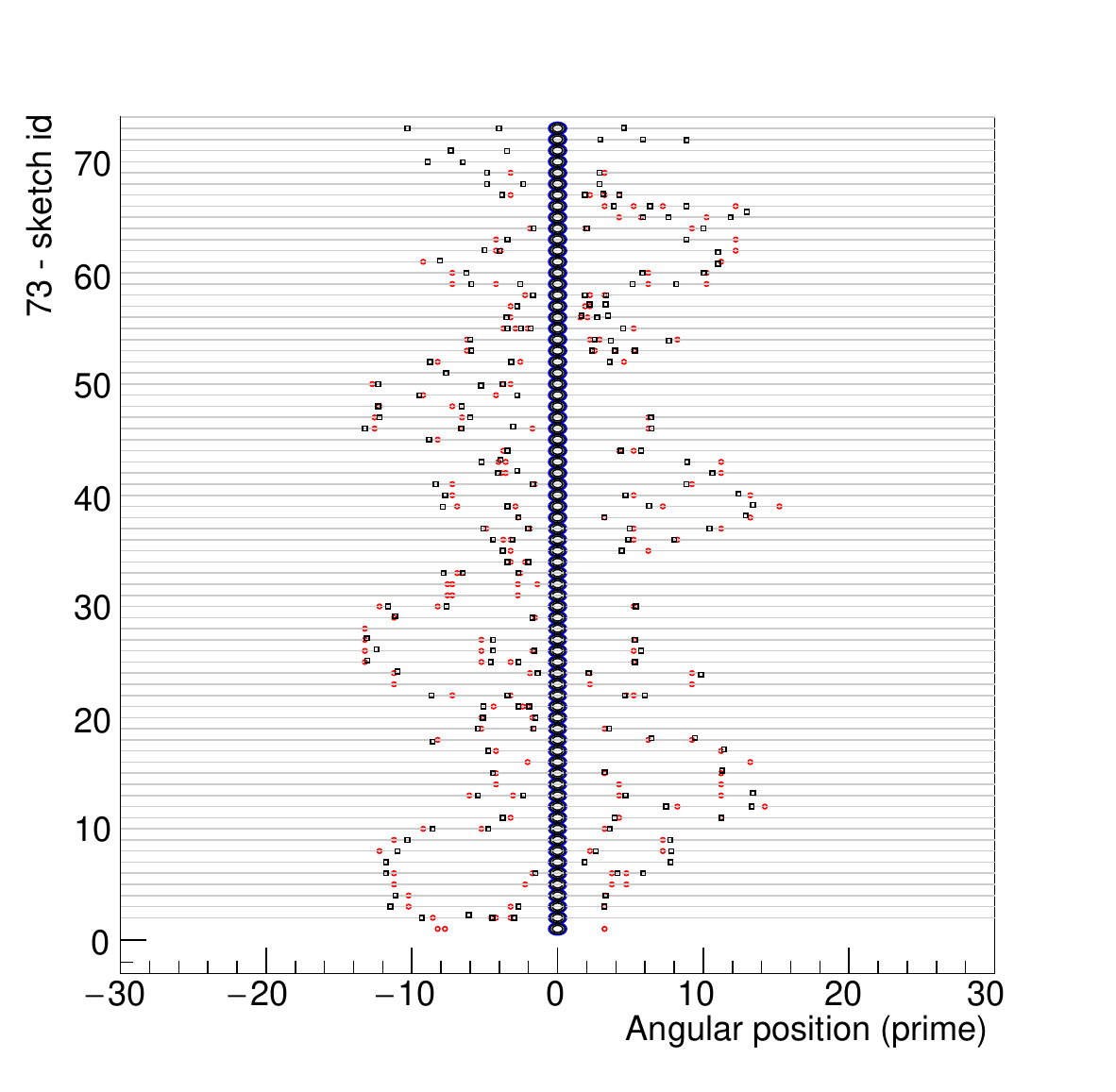}\\
    \caption{Dataset-1 (black markers) and Dataset-2 (red markers) superimposed after a proper scaling. The entries where only the dataset-1 data appear are those for which Galileo on report the angular data but no additional sketches. The dataset from digitization was expressed in terms of angular displacements using Eq.~\ref{eq:scaling}.}
    \label{fig:super}
\end{figure}

In Fig.~\ref{fig:super} the dataset-1 and dataset-2 have been superimposed with a proper scaling of dataset-1 given by the the linear fit parameters shown in Fig.~\ref{fig:correlation} and Eq.
\ref{eq:scaling}. The entries where only the dataset-1 data appear are those for which Galileo on report the angular data but no additional sketches.

\subsection{Overall distribution and four harmonic oscillators model}

\begin{figure}
    \centering
\includegraphics[width=0.5\linewidth]{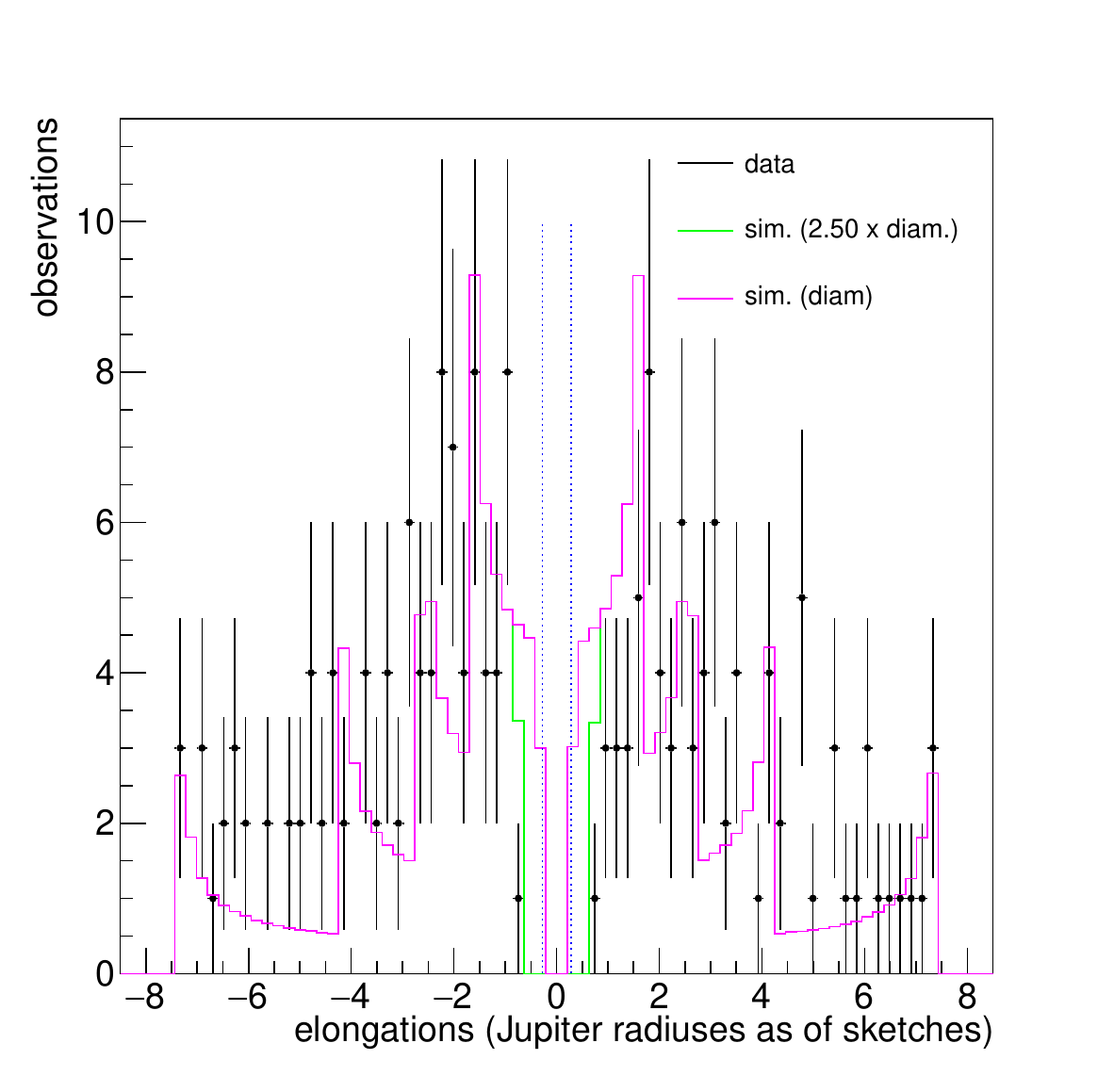}%
\includegraphics[width=0.5\linewidth]{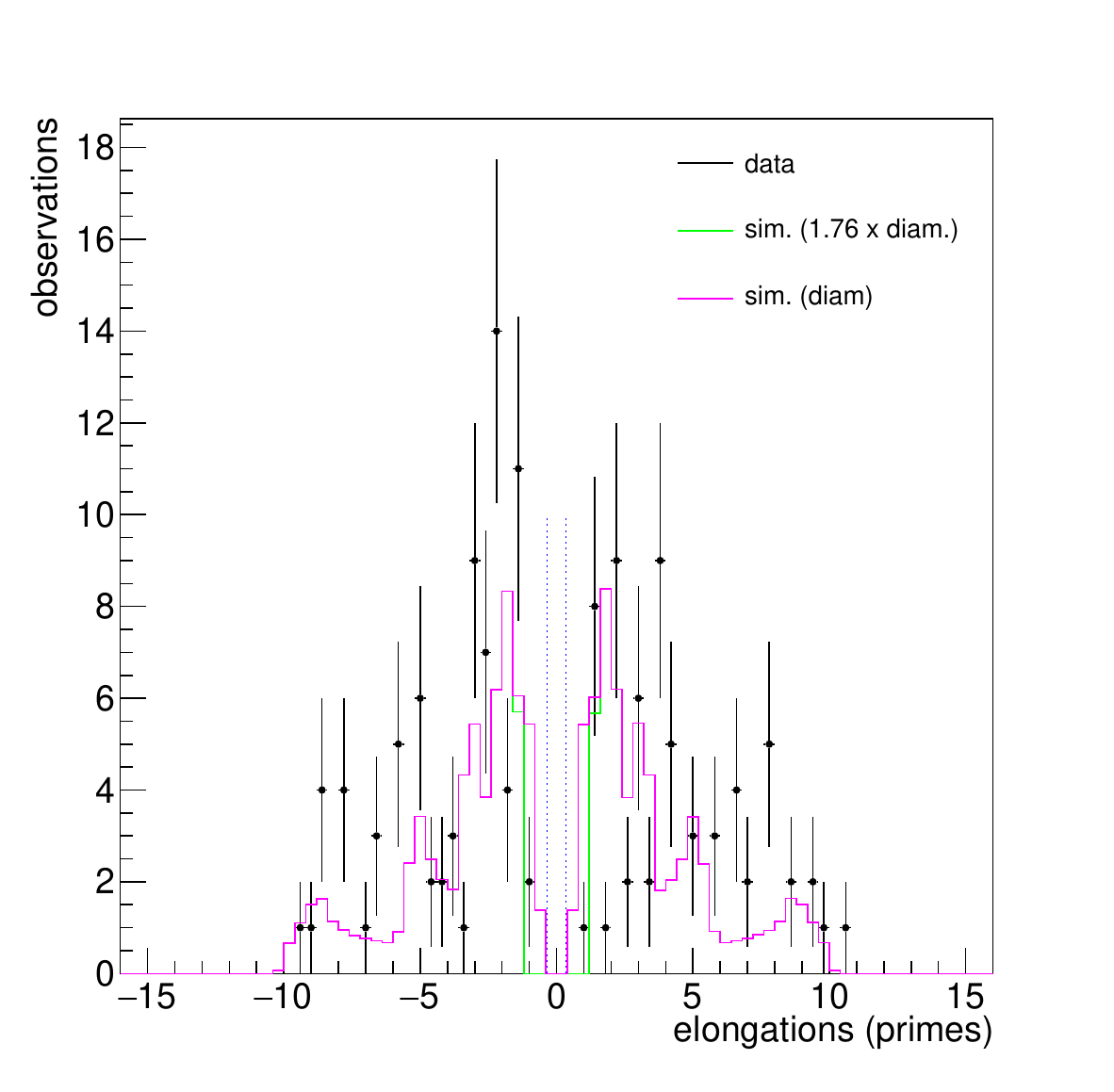}
    \caption{Distribution of the elongation of the satellites for data (bullets) and a simulation assuming harmonic motions with (green) and without (magenta) the inefficiency effect in the data related to the proximity to Jupiter's disk. Left plot refers to dataset-1, right plot to dataset-2.}
    \label{fig:hist}
\end{figure}

We now check if the distribution of the elongations is consistent with the sum of four harmonic oscillators corresponding to four distinct almost circular orbits. 
The distribution of the horizontal displacements ($x_i$ and $\theta_i^{corr}$) is shown in Fig.~\ref{fig:hist} (left and right plot respectively) for the data (bullets) and the expectation in the hypothesis that the motion is harmonic and neglecting the inclination of the orbits. Here we also implicitly assume that the sampling in time is uniform while in reality we know that observations appeared in clusters during the same night. The expected distribution in the case of the $\theta_i^{corr}$ is more smeared than for the $x_i$ because we are taking into account in the simulation the variation of the angular diameter of Jupiter from 45 to 38$^{\prime\prime}$. In the diameter normalized dataset naturally this effect is canceled out in the ratio. As expected the probability of observing the satellite when it is at the maximal elongation is higher as its projected velocity is smaller.

The expectation in the solid magenta line assumes that satellites are
only invisible if they are behind or in front of the disk of
Jupiter. The expectation is solid green line shows instead the
distribution assuming that the inefficient angular region close to
Jupiter corresponds to 2.5 and 1.76 times the real dimension of the disk.  These hypotheses are quite consistent with the data. The factor 1.76 for the angular data derives from the analysis presented before. Further discussion of this inefficiency near the disk is presented in Sec.~\ref{cannocchiale}. 

For dataset-1 the superposition of the distribution (that are each normalized to each
other) was done by eye by scaling the $x$-axis to match the most
extreme observations of Callisto. 
For dataset-2 the data had to be scaled down by a factor 0.7 to match with modern expectations on the satellites' angular elongations during that observational period. This systematic effect will be discussed in more detail in the context of the sinusoidal fits later in Sec.~\ref{results}.

Despite the mentioned approximations, the observation are semi-qualitatively
consistent with what we expect, also considering the large statistical uncertainties.  We will perform fully quantitative considerations on the dataset where
the information of individual satellites is extracted (Sec.~\ref{results}).

\section{Satellite-untagged analysis}

Before trying to associate the observations to individual satellites,
we perform a global analysis considering all the measurements
together. The disentangling of the contributions of single satellites was a rather hard task as the data were taken sparsely and often satellites were too close and unresolved. This was the subject of a significant effort by Galilei until 1619\cite{Drake1978} and several other astronomers (i.e. mainly Pereisc, Harriot, Mayr but also Kepler, Magini, Sizzi, Agucchi) as reported in \cite{Roche1982,Bettini2016Gravitation}. It was soon understood how important this could be as a tool for determining the longitude on Earth. The ``atlantic labour'' of Galilei is analysed in a publication by Stillman Drake\cite{Drake1979}. It is interesting to notice anyway that the oscillatory pattern of Callisto is already quite easily separable when the elongation is sufficiently large (see Fig. ~\ref{fig:LS_incl}). Apart from this sorting put a pattern from the data is rather hard. To isolate the outermost satellite, Callisto, one could only select data being more than four diameters away from Jupiter (see Fig.~\ref{fig:hist}) where only Callisto can reach. After fitting these data with a sinusoid, the points compatible with the fitted Callisto trajectory could be then removed in the innermost part. There one would
have been left with three contributions and iterate. Yet looking at the data it is clear that this was not at all straightforward.

We now consider how the four frequencies might emerge using the Lomb-Scargle algorithm which is a classical algorithm to single out the emergence of particular frequencies in a discrete and sparsed dataset (\cite{LSP0, LSP1}). 
\subsection{Lomb-Scargle periodograms}
\label{secLS}

\begin{figure}[ht]
\centering
\includegraphics[width=\linewidth]{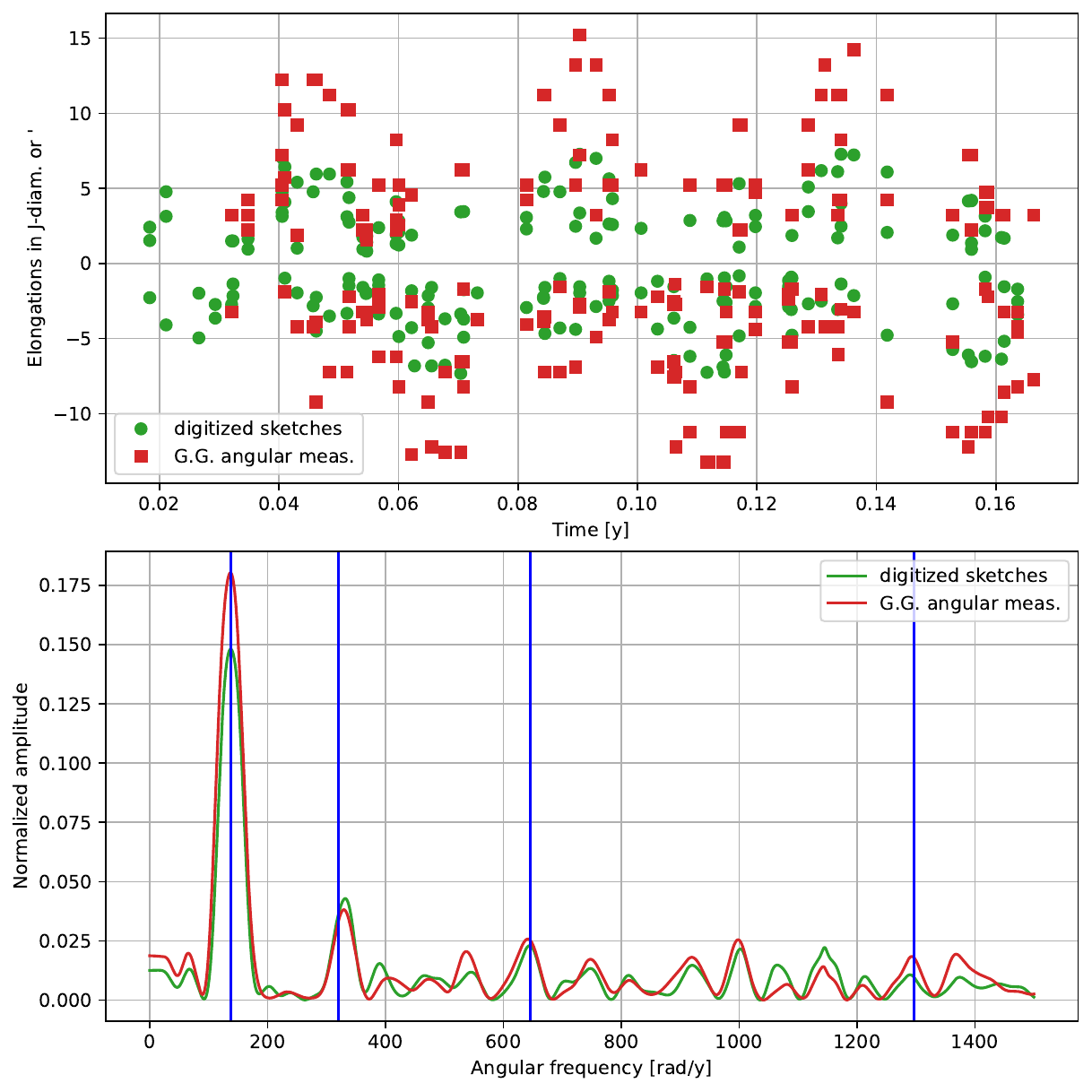}
\caption{\label{fig:LS_incl} Lomb-Scargle periodogram of all the
  data. Top: the time-series. Bottom: the periodogram. The coloured vertical lines denote the position where peaks are expected to occur for Callisto, Ganymede, Europa and Io, from left to right.}
\end{figure}

We show in Fig.~\ref{fig:LS_incl} the Lomb-Scargle (LS) periodogram of the full dataset. 
The top plot shows the
time-series ($x_i$ vs $t_i$, with $t_i$ in fraction of years since 1/1/1610) while at the bottom show the corresponding LS periodogram. The python \textsc{scipy.signal} package was used (\textsc{lombscargle}). Vertical coloured lines indicate the value of the expected
frequencies based on modern data. From left to right, blue is
Callisto, green Ganymede, light green Europa and red, Io. It can be
seen that peaking structures in the periodogram actually emerge at the correct locations. There is also a spurious one at about 1000 rad/year with a significance similar to the one of the satellites. The peak corresponding to
Callisto, is, as expected, the most prominent and with the best
signal-to-noise ratio while the other three have a weaker evidence. We will later re-apply the algorithm to the satellite-tagged four datasets that will be described in the next section.

\section{Association of the satellites using a sky simulator}
\label{association}
This association allows inspecting the expected pattern of satellites
using the \textsc{Stellarium}~\cite{stellarium} online sky simulator.

\begin{figure}
    \centering
    \includegraphics[width=\linewidth]{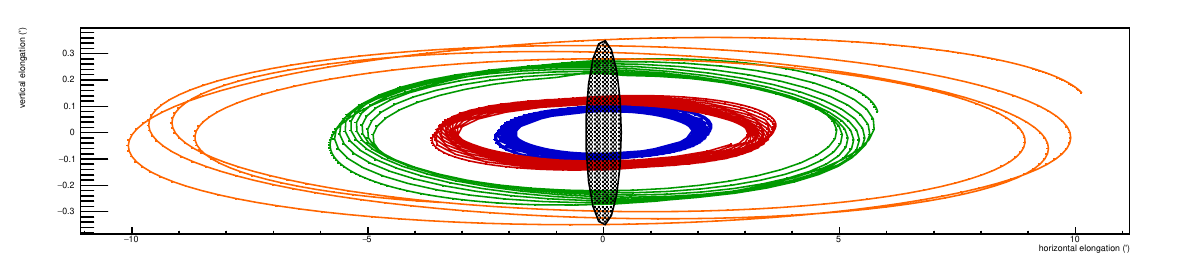}
    \caption{Foreseen orbits of the Galileian satellites in the considered period\cite{PDSRings_Ephem3Jup}.}
    \label{fig:ellipses}
\end{figure}

\begin{figure}[ht]
\centering
\includegraphics[width=100mm]{figs/g07.png}\\
{~~~~~~~~~}\includegraphics[width=100mm]{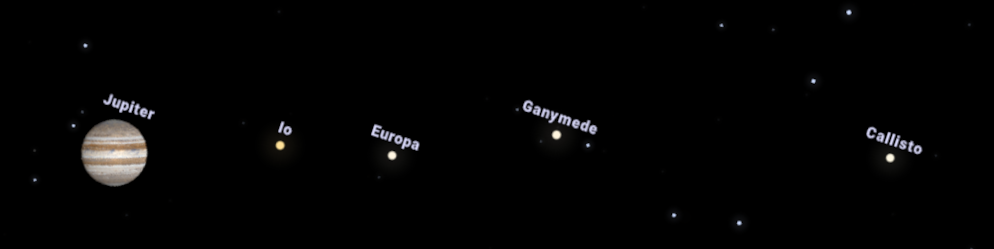}
\caption{\label{fig:s1example} Observation 7, 15 Jan. 1610, 19.55 CET: drawing vs simulation.}
\end{figure}

An example of such an exercise is shown in Fig.~\ref{fig:s1example} for the observation number 7, performed on Jan. 15 1610 at the third hour (Galilei's time) or 19.55 CET. The correspondence is remarkable and this happens almost for the entire set of observations as it will be shown later (see Appendix \ref{fullcomp}). It is worth noticing
also the level of attention of the recordings where the slight
displacement of Callisto from the ecliptic plane is reported. The orbits are almost perfectly circular and lying almost exactly at the the Jupiter equatorial plane. The rotation axis of Jupiter forms an angle of 3.1$^\circ$ with respect to the Ecliptic and the inclination of the Jupiter's orbit forms with it an angle of 1.3$^\circ$. Displacements in the ``vertical'' direction arise from 
the combinations of these effects and the satellites are describing very narrow ellipses. Figure \ref{fig:ellipses} shows these trajectories in the considered period as predicted (\cite{PDSRings_Ephem3Jup}). The vertical displacement of Callisto near Jupiter amounted to about $\pm$20$^{\prime\prime}$.

It is interesting to notice that, according to the simulator, the satellites' magnitudes were, in decreasing brightness: Ganymede (4.75), Io (5.11), Europa (5.43) and Callisto (5.69). Galilei reports instead in decreasing brightness Callisto, Ganymede, Europa, Io. It is possible that the closeness to Jupiter might have played a role in biasing the perception of the brightness\footnote{It should also be noted that the accuracy of the predictions of Stellarium for the magnitudes of the satellites have not been cross-checked with other simulators.}. 

Another evident feature is the fact that the size of Jupiter's disk is generally overestimated by Galilei. This is probably due to the glare of the planet and resolution effects. This interpretation is corroborated by the fact the often satellites were missed when too close to the planetary disk, as it will be detailed later. 

The typographic symbol employed to denote Jupiter is circular but the inner white part seems to hint at the fact the indeed the disk has some equatorial bulging. On the other hand it should be noted that in those days Jupiter was far from opposition so the left limb was in shadow (see for example Fig.~\ref{fig:s1example}) making it more circular-shaped. Furthermore it should be noticed that the equatorial bulging, is a 7\% effect, so it needs an appropriate resolution to be discernible.

Another feature that is apparent already from this single image is that the positions of the outer satellites Ganymede and Callisto seems to be underestimated by Galilei.

Each sketch was put side by side with the output of \textsc{Stellarium}
after taking care to see the values to those that were translated in modern time. All the images from \textsc{Stellarium} are shown together with the sketches of Galileo in Appendix \ref{fullcomp}. The observations 5.1, 40.1, 40.2, 42.1, 46.1, 52.1, 53.1, 61.1, 64.1 i.e. those where we only have the angular recordings but not the sketches are compared to the simulator in Appendix ~\ref{fullcomp_nosketch}.

The output of the associations is given in Tabs.~\ref{tab:ass1} and
\ref{tab:ass2}.  In several cases two satellites were too close to be
separated by Galileo. When this happens we decided to perform the
association with one of the two. In general we find an almost perfect
correspondence between the notes of Galileo and the simulator with the exception of a glitch for observation 46 of Feb 12 that will be discussed later.

\begin{table}
\centering
\begin{tabular}{ r r r r r}
\toprule
id & comment & pattern& not seen \\ \midrule
1 &  & CI~$\star$~G&\\ 
2 & C unseen & ~$\star$~IEG & C\\ 
3 & G-E unresolved. I too close to J & CE~$\star$~ & I \\ 
4 & IE too close to J & GC~$\star$~&\\ 
5 & I-C unresolved & GC~$\star$~E&\\ 
5.1 & I-C too close to J & G~$\star$~E&IC\\ 
6 &  & E~$\star$~GIC &\\ 
7 &  & ~$\star$~IEGC &\\ 
8 &  & EGC&\\ 
9 & G-E unresolved & I~$\star$~EC&\\ 
10 & G-E unresolved. I too close to J & G~$\star$~C& I\\
11 & I behind & GE~$\star$~C &\\ 
12 & I-E close in transit & G~$\star$~C&\\ 
13 & I unresolved too close  & G~$\star$~EC& I\\ 
14 &  & GI~$\star$~EC&\\ 
15 & G unresolved too close at right & E~$\star$~IC&G\\ 
16 & C/G unresolved & E~$\star$~CI&\\ 
17 &  & E~$\star$~CGI&\\ 
18 &  & EIC~$\star$~G&\\ 
19 &  & C~$\star$~IEG&\\ 
20 &  & C~$\star$~IEG&\\ 
21 & G/E unresolved & CI~$\star$~E &\\ 
22 & IEG too close to J & C~$\star$~&IEG\\ 
23 & I too close to J on the right & CEG~$\star$~&I\\ 
24 & E-G unresolved, I behind& CG~$\star$~&\\ 
25 & IE too close to J (right) & CG~$\star$~ &IE\\ 
26 & I behind J & CG~$\star$~E&\\ 
27 &  & CGI~$\star$~E&\\ 
28 & EI too close to J, G in transit & C~$\star$~&EI\\ 
29 & G-E unresolved & I~$\star$~GC &\\ 
30 & I (right) too close to J& EG~$\star$~C&I\\ 
31 & I (right) too close to J &EG~$\star$~I &I\\ 
32 & E-I unresolved & GE~$\star$~C&\\ 
\bottomrule
\end{tabular}
\caption{Summary of satellites association with comments.}
\label{tab:ass1}
\end{table}

\begin{table}
\centering
\begin{tabular}{r r r r}
\toprule
id & comment & pattern & not seen\\ 
\midrule
33 &  & G~$\star$~EC &\\ 
34 &  & GI~$\star$~EC&\\ 
35 & G too close to J (right) & E~$\star$~IC&G\\ 
36 &  & EI~$\star$~GC&\\ 
37 &  & EI~$\star$~GC&\\ 
38 & C too close to J (right), E-G unresolved & I~$\star$~G&C\\ 
39 & C-E unresolved, I too close to J & EG~$\star$~&I\\ 
40 & I-E unresolved  & CGE~$\star$~&\\
40.1 & & CGEI~$\star$~&\\
40.2 & & CGE~$\star$~& I too close to J\\
41 & I too close to J (right) & CG~$\star$~E&I\\ 
42 & I-E on transit & CG~$\star$~&\\ 
42.1 & E on transit, I, G too close to J & C~$\star$~& IG\\ 
43 & I too close to J (left) & CE~$\star$~G &I\\ 
44 &  & CEI~$\star$~G&\\ 
45 &  & CEI~$\star$~G&\\ 
46 & I seen on the other side than expected (see text) & CI~$\star$~EG&\\ 
46.1 & E too close to J& C~$\star$~IG&E\\ 
47 &  & CI~$\star$~EG&\\ 
48 & C (right) too close to J & GEI~$\star$~&C\\ 
49 & I too close to J (left), C close (right)& GE~$\star$~&IC\\ 
50 & I too close (left) & GE~$\star$~C&I\\ 
51 & I behind & G~$\star$~EC&\\ 
52 & I too close (left), E too close (right) & G~$\star$~C &IE\\ 
52.1 & I too close (right), E transit & G~$\star$~C &I\\ 
53 & I too close (left) & E~$\star$~GC &I\\ 
53.1 & I too close (left) & E~$\star$~GC &I\\ 
54 &  &  EI~$\star$~GC&\\ 
55 & I too close to J (right), E in transit & ~$\star$~GC &I\\ 
56 & G behind & E~$\star$~IC&\\ 
57 & I behind & CE~$\star$~G&\\ 
58 & E (left) and I (right) too close to J & C~$\star$~G&EI\\ 
59 & E too close to J (right) & C~$\star$~IG&E\\ 
60 & E too close to J (right) & C~$\star$~IG&E\\ 
61 &  & CI~$\star$~EG&\\
61.1 & & CI~$\star$~EG&\\ 
62 & E (left) and G (right) too close to J& 
C~$\star$~I&EG\\ 
63 & G behind & CE~$\star$~I&\\ 
64 &  & CGEI~$\star$~&\\ 
64.1 & I/E not resolved & GC~$\star$~(I/E)&\\ 
\bottomrule
\end{tabular}
\caption{Summary of satellites association with comments.}
\label{tab:ass2}
\end{table}

\subsection{Periodograms of the satellite-tagged datasets}

\begin{figure}[ht]
\centering
\includegraphics[width=0.5\linewidth]{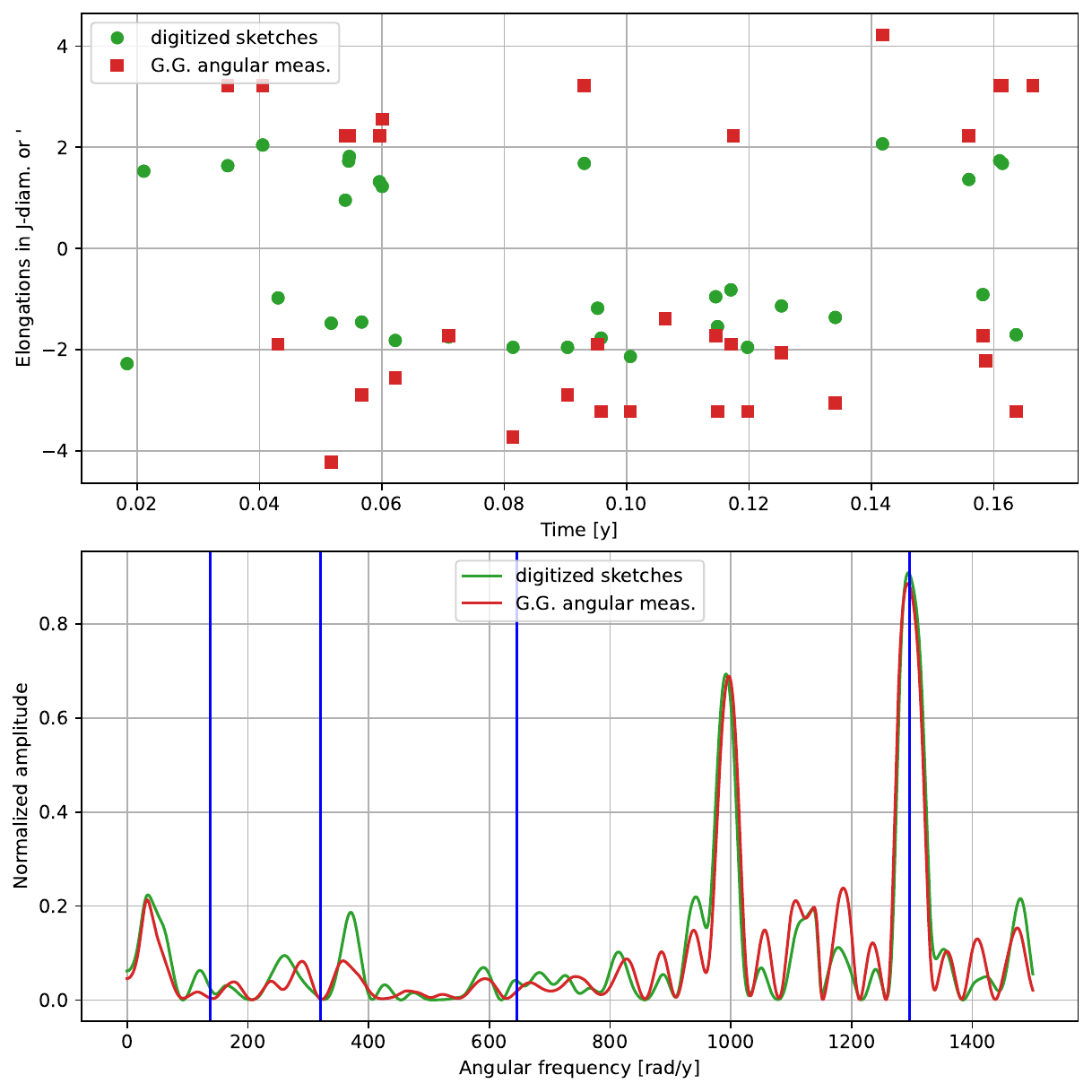}%
\includegraphics[width=0.5\linewidth]{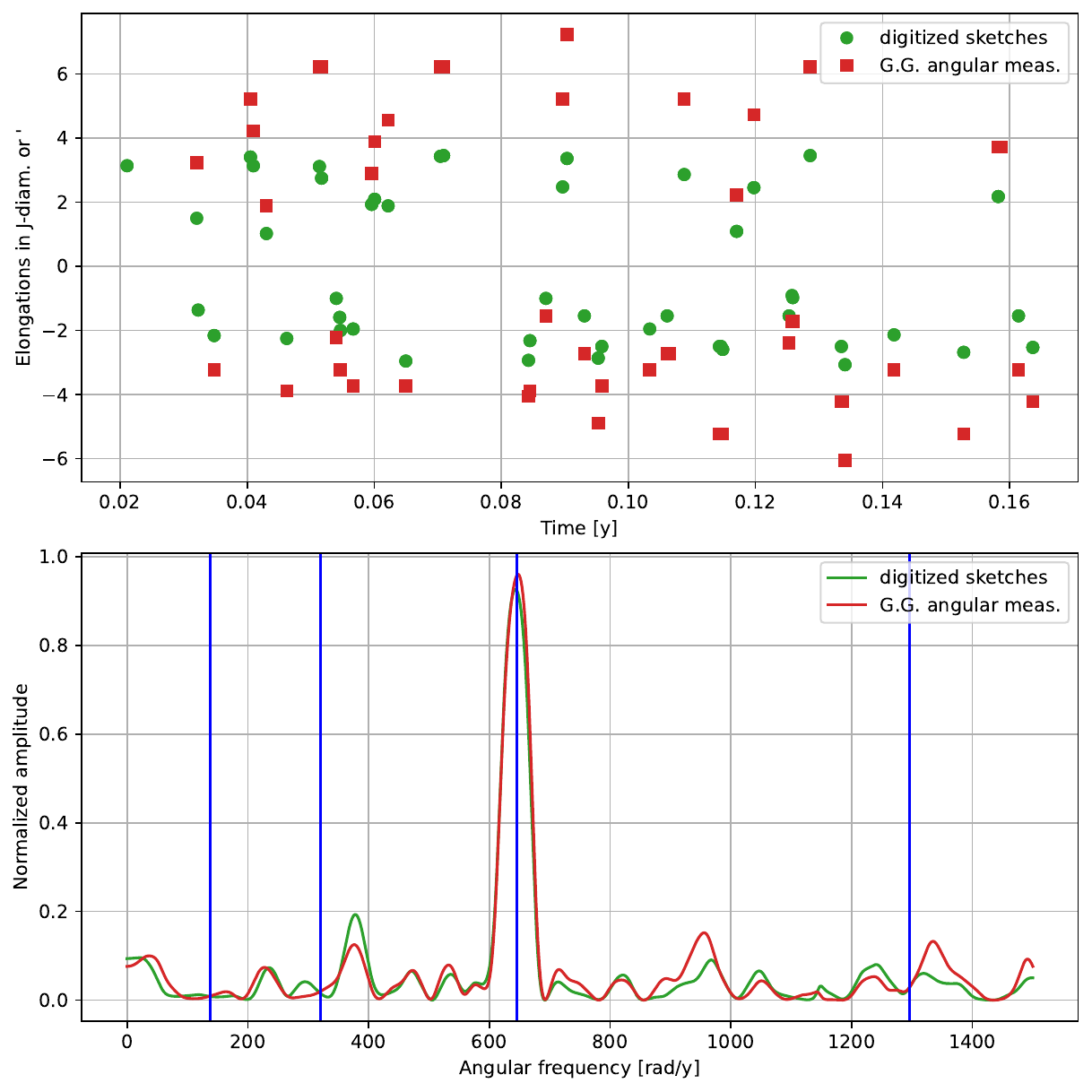}
\includegraphics[width=0.5\linewidth]{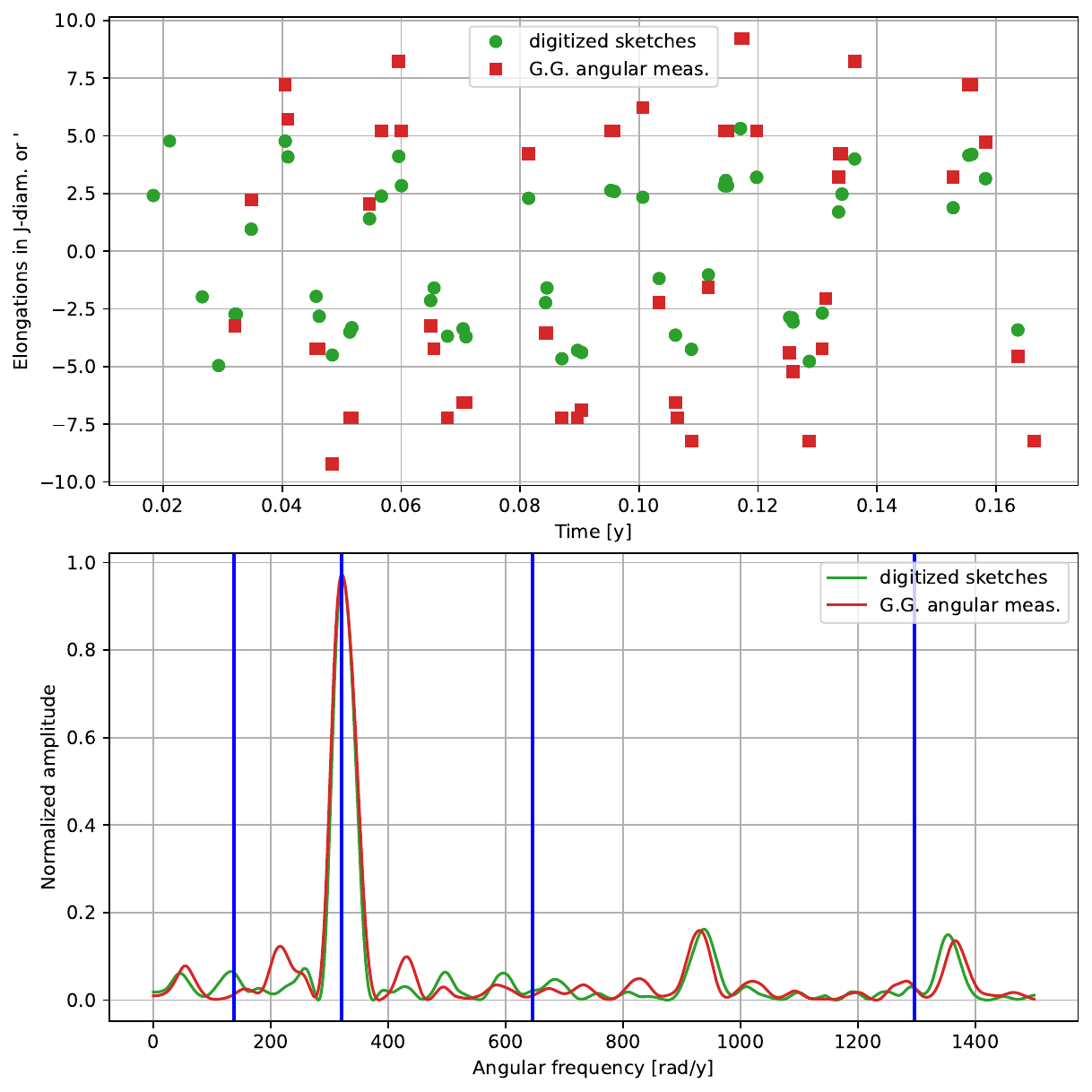}%
\includegraphics[width=0.5\linewidth]{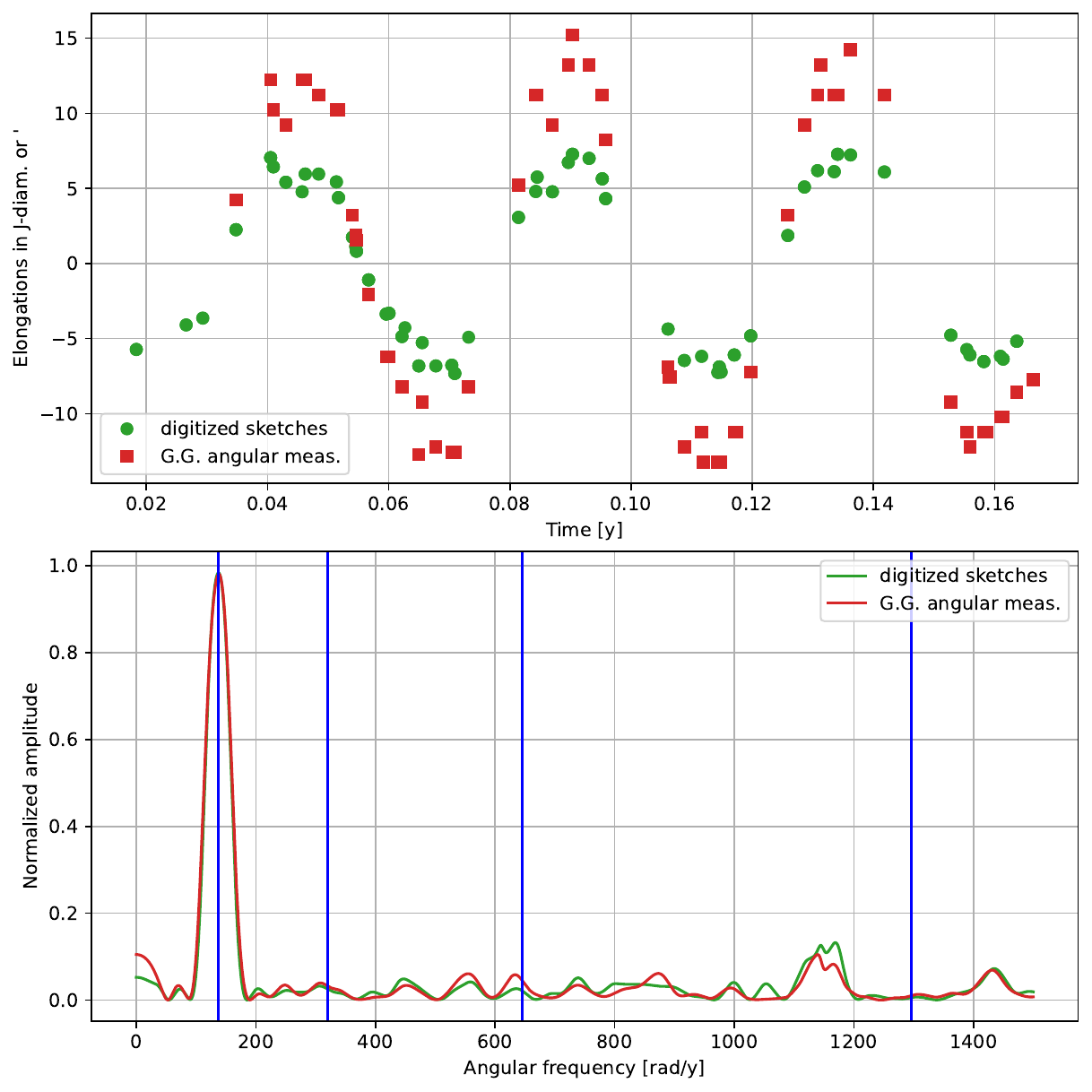}
\caption{\label{fig:LS_al} Lomb-Scargle analysis for the Io (top left), Europa (top right), Ganymede (bottom left) and Callisto (bottom right) datasets as determined from the association performed with the simulator. Both time-series and periodograms are shown. The lines indicate the position of the expected frequencies.}
\end{figure}

The periodograms of the time series separated by satellite using the sky simulator are shown in Fig.~\ref{fig:LS_al}. 
In this case the peaks of each satellite emerge without ambiguity at the expected position. Only for Io there is a less-significant, but still important, peak at about 1000 rad/y that had already been observed in the inclusive analysis of Fig.~\ref{fig:LS_incl}. In general the smaller peaks do not occur in correspondence of the expectations for other satellites supporting the fact that the rate of wrong-associations is very low.

\subsection{Sinusoidal fits and orbital parameters extraction}
\label{results}

The elongations of dataset-1, ($x_i$, $t_i$), separated by satellite have been fitted with a function of the form $x(t)=A\sin(\omega (t-t_0) + \phi_0)$ using the ROOT \cite{ROOT} fitter libraries. For dataset-2 the model also accounts for a variation of the angular size of the Jovian system due to the farthening from Earth: $\theta(t)=
\Theta(t)\sin(\omega (t-t_0) + \phi_0)$ where $\Theta(t)$ is a know function that can be obtained from the ephemerides. Indeed in the considered period (7 January-2 March) the distance of Jupiter passed from 4.28 to 5.03 AU and the phase passed from 100\% to 99\%. The angular diameter hence passed from 
45.0 to 38.3~arcsec, i.e. Jupiter disk was 15\% smaller at the end of the observation period.
This effect does need to be taken into account for the analysis where the measurements are normalized to the Jupiter disk while it is relevant for the analysis in absolute angular distances as reported by Galilei. The variation of the equatorial diameter of Jupiter (from \cite{PDSRings_Ephem3Jup}) 
was fitted by a function of the form  $\Delta(^\prime) = 46.2842-0.100186 \times t                       -0.00103484 \times t^{2}+9.86778\times10^{-6} \times t^3$ with $t$ expressing the number of days since the first observation. We will later check if this model describes better the dataset-2 with respect to the simplified model.

We have tuned the uncertainty on the points to match the r.m.s. of the residuals in practice letting the errors determined by the fit itself by imposing to get a normalized $\chi^2$ of about one. The errors hence result as follows:

\begin{table}
    \centering
    \begin{tabular}{ r r r r r}
\toprule
         & $\sigma_{I}$ & $\sigma_{E}$ & $\sigma_{G}$ & $\sigma_C$ \\
 \midrule
         dataset-1& 0.49 & 0.66 &0.59&0.71\\
         dataset-2&0.99$^\prime$&0.86$^\prime$&1.02$^\prime$&1.44$^\prime$\\
         \bottomrule
    \end{tabular}
    \caption{Errors associated to the measurements in the sinusoidal fits.}
    \label{tab:posteriori_errors}
\end{table}
This can be considered as an estimate of the pointing accuracy of the observations.

\begin{figure}
    \centering    \includegraphics[width=\linewidth]{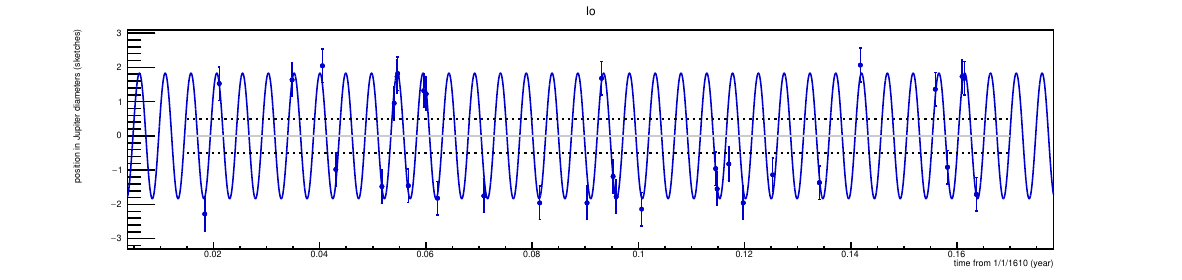}
\includegraphics[width=\linewidth]{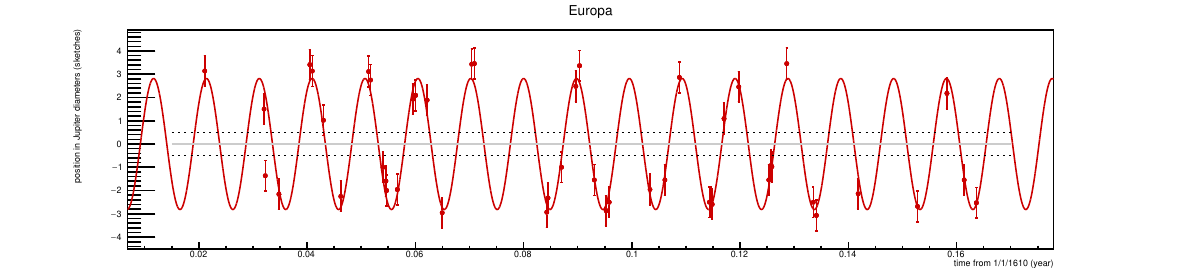}    \includegraphics[width=\linewidth]{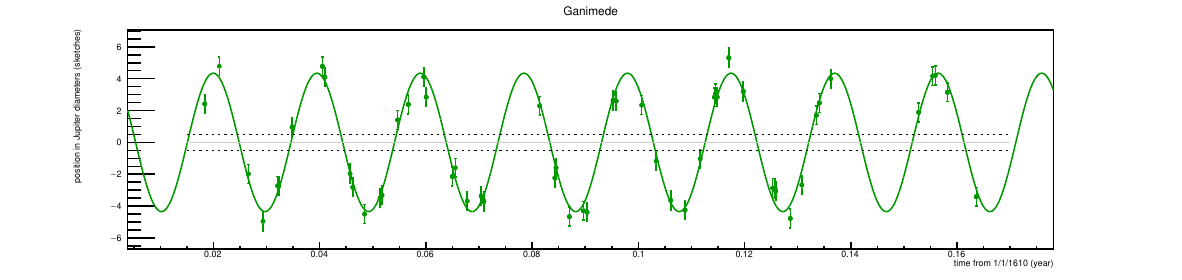}
\includegraphics[width=\linewidth]{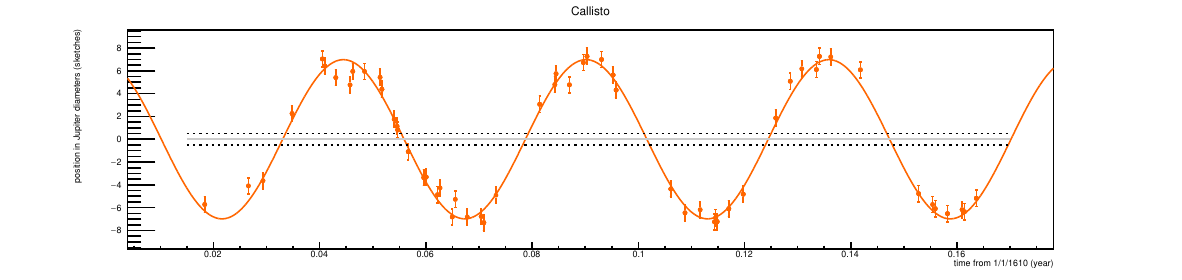}
    \caption{Fitted data points from dataset-1 shown separately for each dataset associated to individual satellites. From top to bottom Io, Europa, Ganymede and Callisto. The units on the $x$-axis are fraction of years starting from 1/1/1610 and in the $y$-axis the elongation of the satellites in GJD units.}
    \label{fig:singlefits_1}
\end{figure}

\begin{figure}
    \centering    \includegraphics[width=\linewidth]{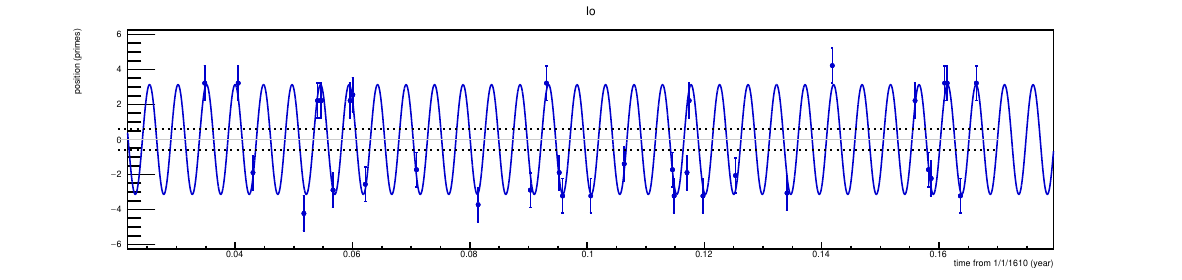}
\includegraphics[width=\linewidth]{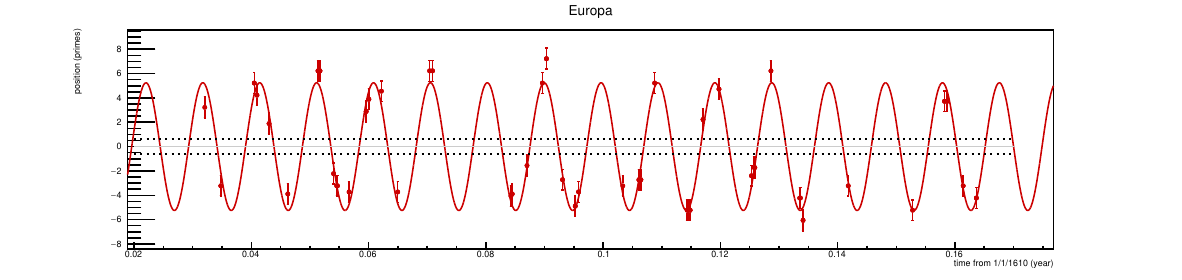}    \includegraphics[width=\linewidth]{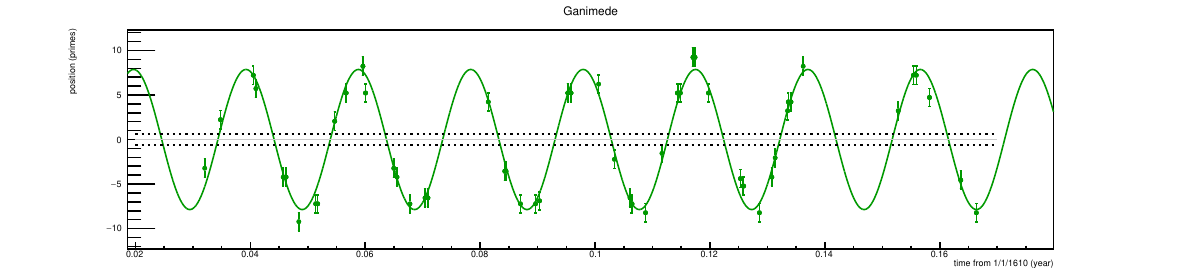}
\includegraphics[width=\linewidth]{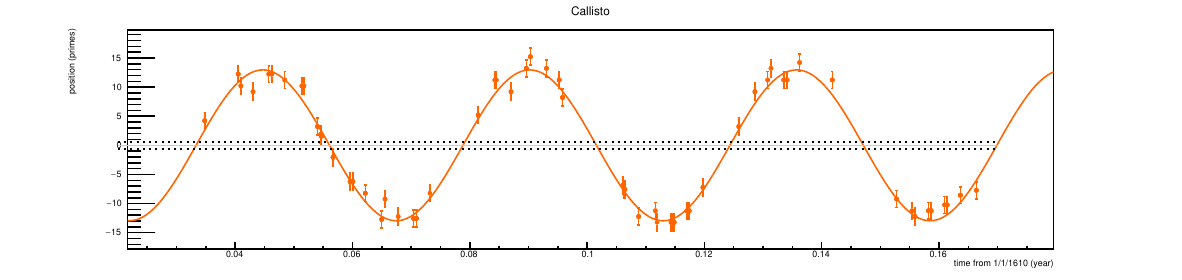}
    \caption{Fitted data points from dataset-2 shown separately for each dataset associated to individual satellites. From top to bottom Io, Europa, Ganymede and Callisto. The units on the $x$-axis are fraction of years starting from 1/1/1610 and in the $y$-axis the elongation of the satellites in GJD units.}
    \label{fig:singlefits_2}
\end{figure}

\begin{figure}
    \centering
\includegraphics[width=\linewidth, angle=0]{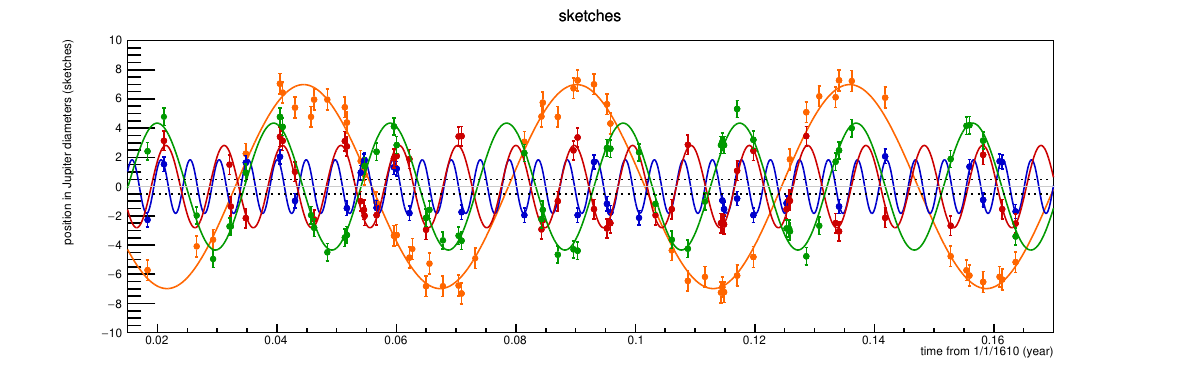}\\
\includegraphics[width=\linewidth, angle=0]{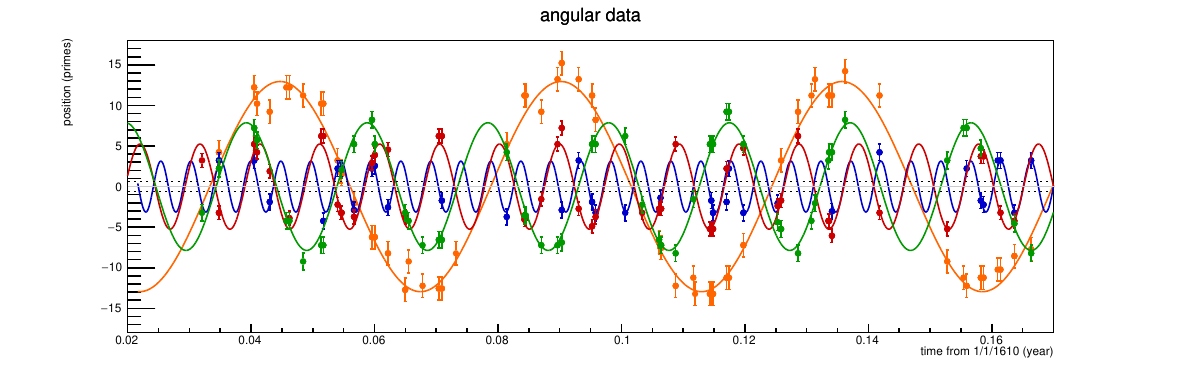}
    \caption{Fitted data points from dataset-1 (top) and dataset-2 (bottom) shown separately for each dataset associated to individual satellites: Io (red), Europa (gray), Ganymede (orange) and Callisto (magenta).  The units on the $x$-axis are fraction of years starting from 1/1/1610 and in the $y$-axis the elongation of the satellites in GJD units.}
    \label{fig:oscillo}
\end{figure}

The data with the superimposed best fit are shown in Fig.~\ref{fig:singlefits_1} (dataset-1) and \ref{fig:singlefits_2} (dataset-2).
The same results are shown superimposing all satellites together in Fig.~\ref{fig:oscillo} for dataset-1 and dataset-2.

A video showing the result of the sinusoidal fits and the satellites' associations is available in \cite{video}. There the filled points show the prediction of the fit. At the time of observation the animation is briefly pause and the measurements from 
Galilei are superimposed together with an image of the original sketch.

\begin{figure}
    \centering
\includegraphics[width=\linewidth]{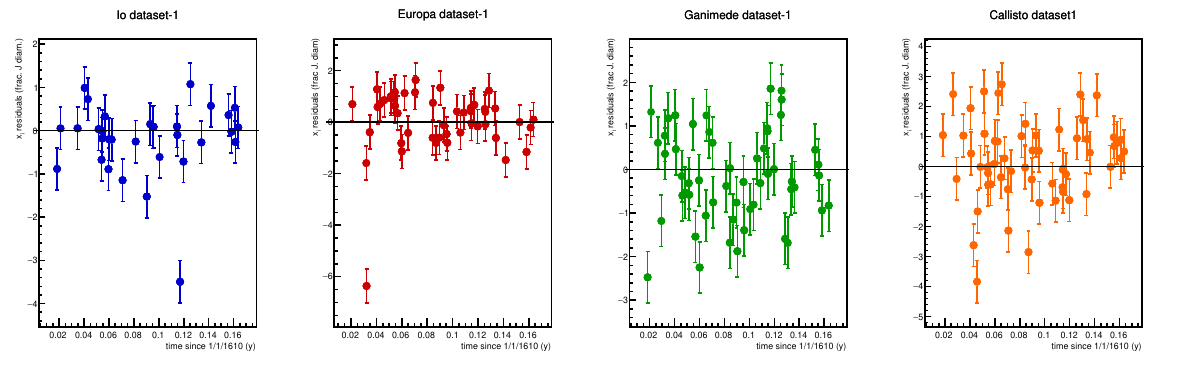}\\
\includegraphics[width=\linewidth]{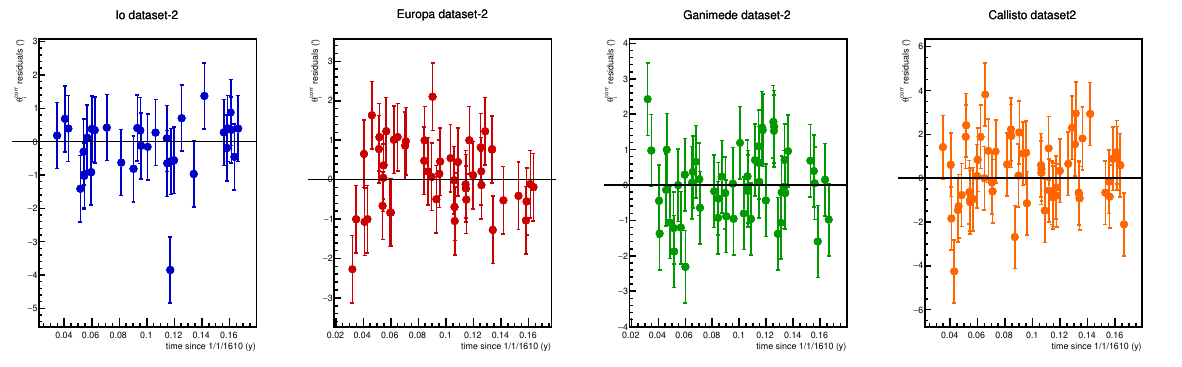}
    \caption{Residuals for sinusoidal fits for Io (red), Europa (gray), Ganymede (orange) and Callisto (magenta). In the top row the units on the $x$-axis are fraction of years starting from 1/1/1610 and in the $y$-axis the elongation of the satellites in GJD units.}
    \label{fig:resid_time}
\end{figure}

The residuals of the fits ($x_i-x(t_i)$, $\theta_i^{corr}-\theta(t_i)$) are shown for each satellite in Fig.~\ref{fig:resid_time} as a function of time for the two datasets. 

\begin{figure}
    \centering
\includegraphics[width=\linewidth]{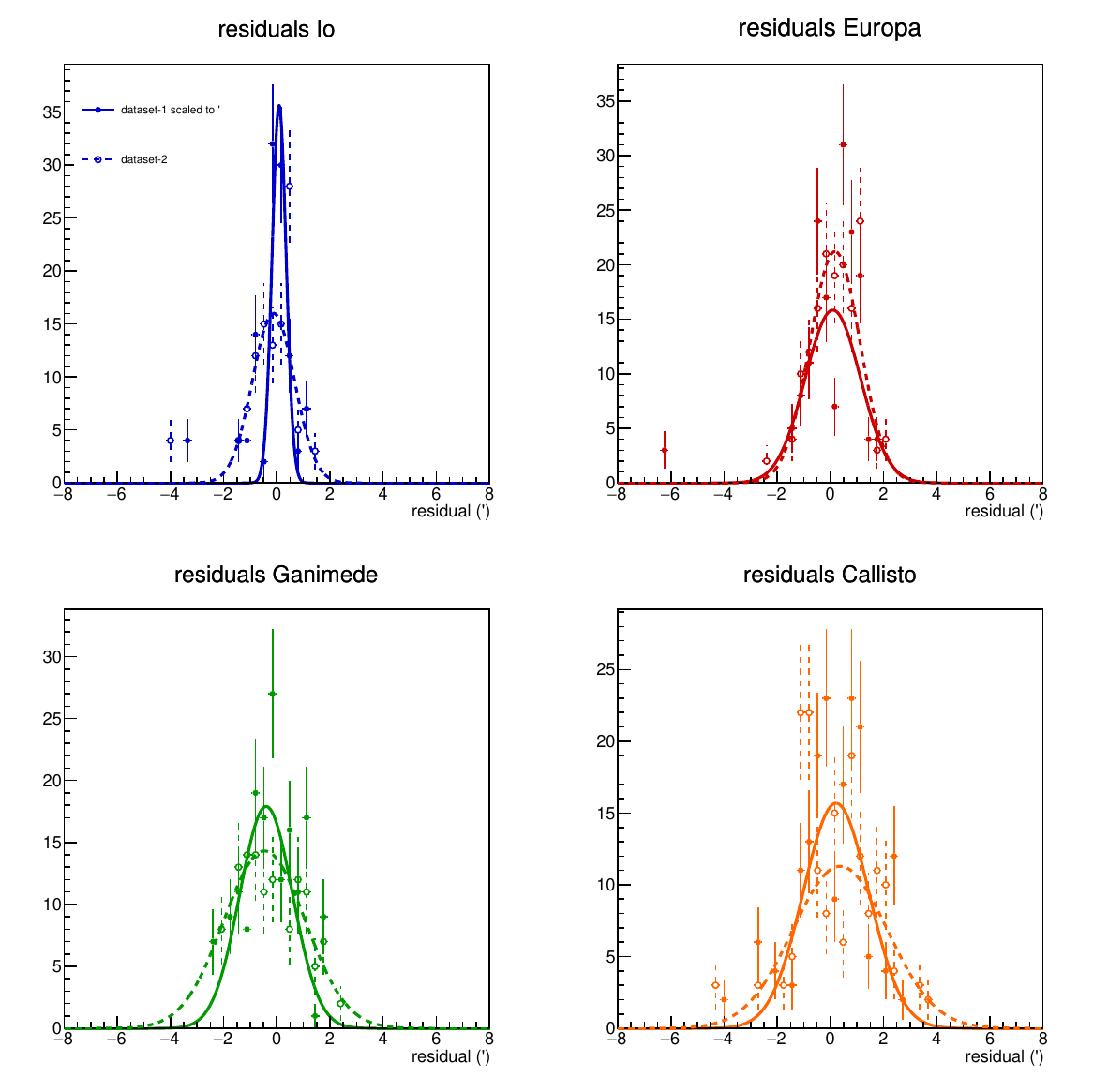}
    \caption{Distributions of the residuals for sinusoidal fits for Io (red), Europa (gray), Ganymede (orange) and Callisto (magenta) for the two datasets after scaling GJD units to primes.}
    \label{fig:resid_distr}
\end{figure}

The distribution of the residuals fitted with a Gaussian model is shown in Fig.~\ref{fig:resid_distr}. In this case we have multiplied the dataset-1 by 1.825$^\prime$ (see Fig.~\ref{fig:correlation}) in order to make them more easily comparable. 

These distributions show that for Io the dataset-1 seems significantly better fitted than dataset-2. It should be noticed anyway that the gaussian fit is not particularly good at modeling the data so this effect might appear overestimated. As we move to larger elongations the difference is moderate but the dataset-1 tends to be fitted a bit better. The reason for this is not easy to imagine and it would be important to know the relation between the notes of Galileo and the sketches in the Sidereus. We can suppose that in the printing process the drawings were drawn starting from the angular measurements (except for the first measurements). Still it is interesting to notice that dataset-1 seems significantly less discretized than the dataset-2 (see Fig.~\ref{fig:correlation}).

The models are reproducing the data very well without particular trends in the residuals or anomalies. The sigma of the gaussian fits are (0.3, 1.0, 1.0, 1.2)$^\prime$ for dataset-1 and (0.8, 0.9, 1.5 and 1.8)$^\prime$ for dataset-2. The fitted values with uncertainties are given in Tab.~\ref{tab:results}.

As we said we converted the observation times to modern hours in the analysis whenever available. It is interesting to cross check if this correction is effective. As a test we have tried to assume that all the measurements were done at 19.00 CET. Indeed as expected, we observe a very significant increase in the $\chi^2$ in the latter case, especially for Io for which this correction is generally quite important. Applying the correct observation the $\chi^2$ changes as follows: from 180.9 to 106.5 for Io, from 181.6 to 164.7 for Europa, from 155.0 to 158.4 for Ganymede and from 195.6 to 184.7 for Callisto.

%

\subsection{Compatibility of datasets}
\begin{figure}
    \centering
\includegraphics[width=\linewidth]{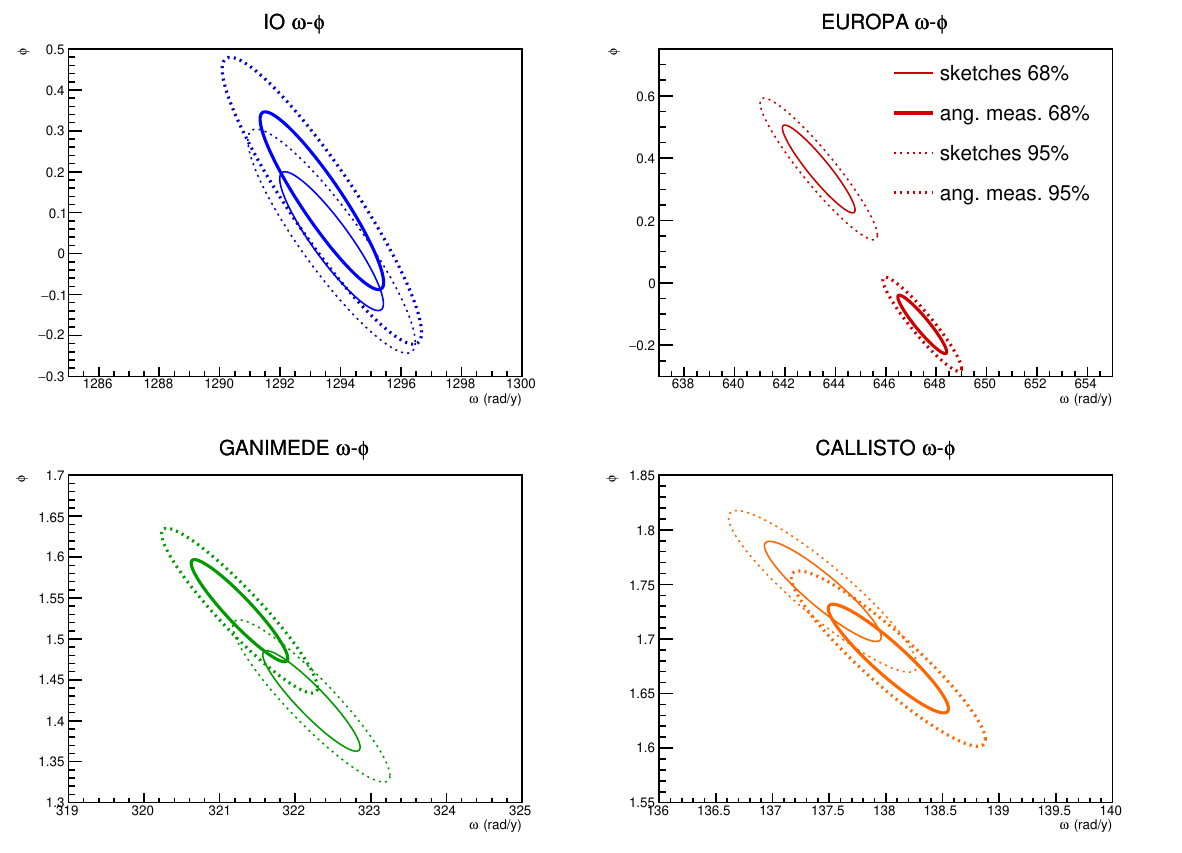}
    \caption{Confidence regions for $\omega$ and $\phi$ for dataset-1 and 2.}
    \label{fig:contours}
\end{figure}

As we have seen in Fig.~\ref{fig:super} the two datasets are very close. Here we evaluate how compatible are the periods and phases of the sinusoids extracted from the fits. The results are shown in Fig.~\ref{fig:contours}. As expected the extracted values are compatible to a very good level with the exception of Europa where there is a mild tension between the two datasets.

\subsection{Effect of varying distance}

\begin{figure}
    \centering
\includegraphics[width=0.8\linewidth]{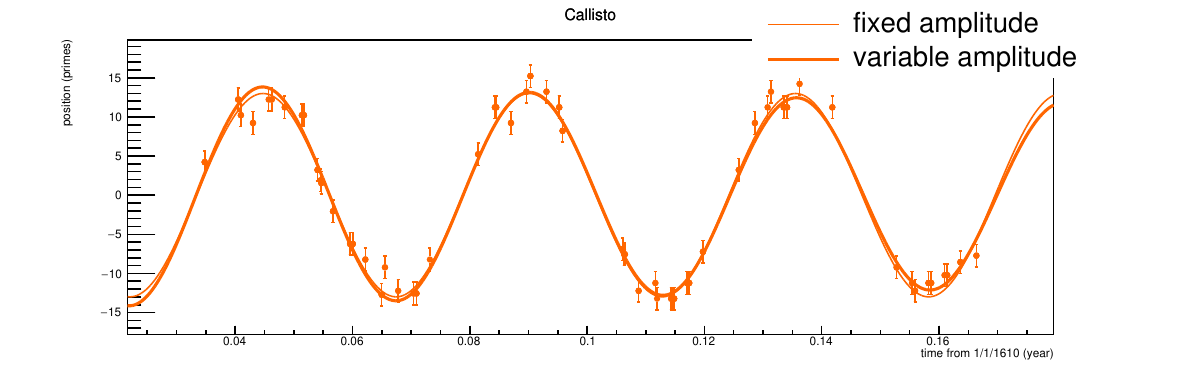}
    \caption{Difference in the fit of the Callisto points for dataset-2 with the fixed amplitude and variable amplitude fit.}
    \label{fig:varfit}
\end{figure}

The result of the fits with variable or fixed amplitude for Callisto dataset-2 is shown in Fig.~\ref{fig:varfit}. A better description is achieved with the fixed amplitude fit. The $\chi^2$ is consistent in the two fits but slightly lower for the fixed amplitude (about 4 units of $\chi^2$). This supports the fact that also the absolute measurement are in reality relative to the size of Jupiter.

\subsection{Comparison with ephemerides for the angular dataset}

\begin{figure}
    \centering
\includegraphics[width=\linewidth]{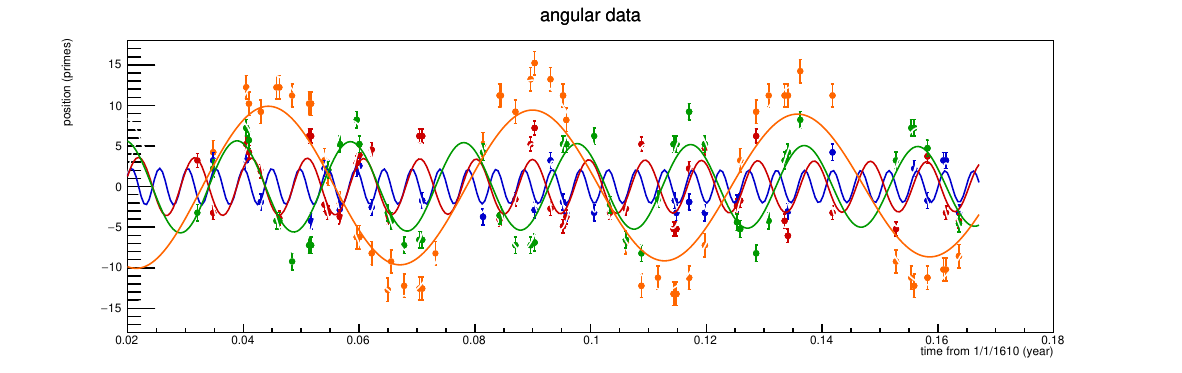}
    \caption{Comparison between dataset-2 and the ephemeris of Jupiter providing the absolute elongations.}
    \label{fig:confronto_assoluto}
\end{figure}

So far the elongations have been treated as a free parameter but in the case of dataset-2 we can check if the fitted elongations agree with the absolute prediction provided by the ephemerides\cite{PDSRings_Ephem3Jup}. The comparison of the data with the ephemerides is shown in Fig.~\ref{fig:confronto_assoluto}.

\begin{figure}
    \centering
\includegraphics[width=\linewidth]{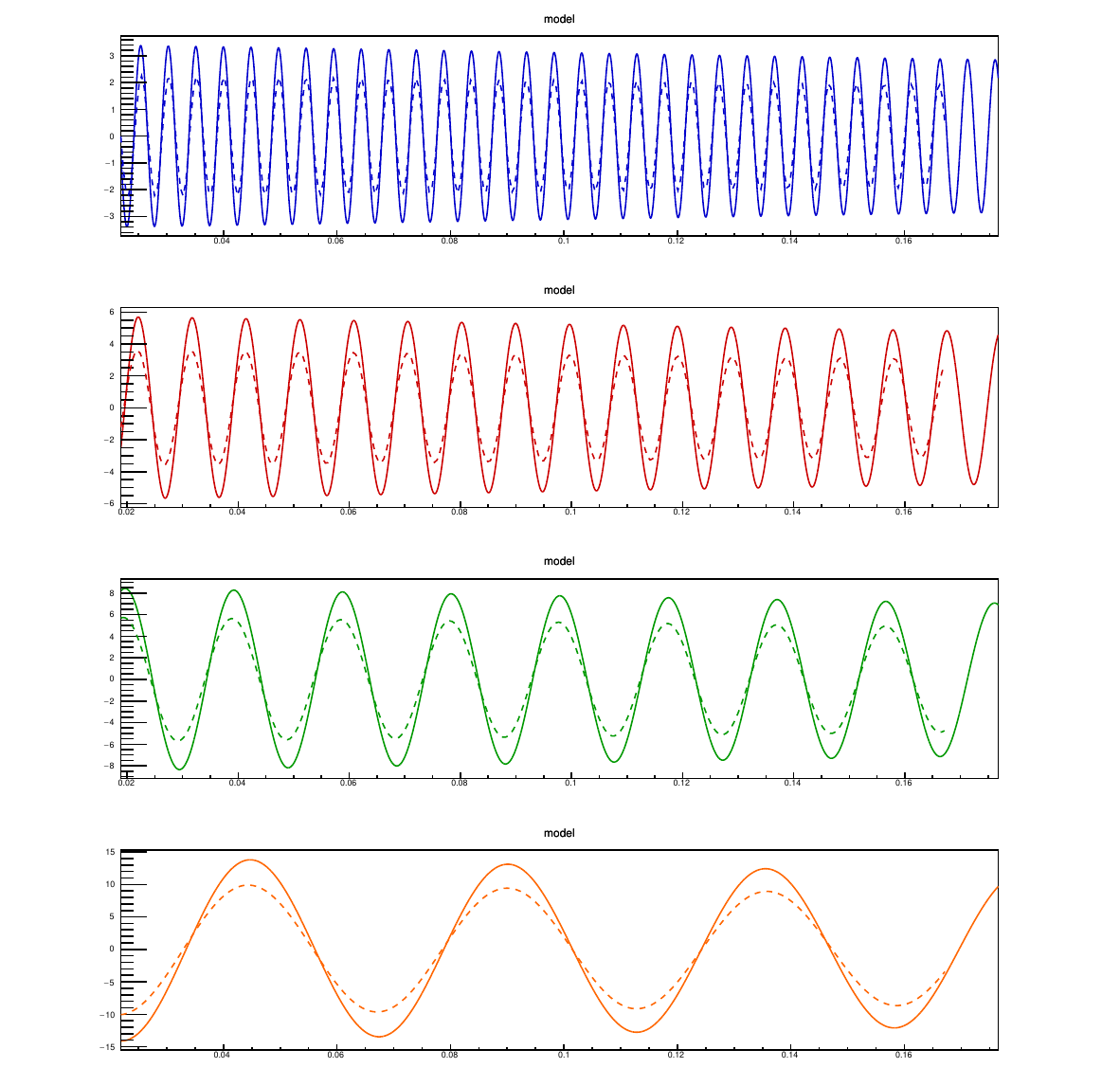}
    \caption{Comparison of the absolute predictions (dashed line) from the ephemerides and the fit (solid line) for dataset-2.}
    \label{fig:fit-eph}
\end{figure}

The data appear to be in excess by a significant fraction. To evaluate this factor we have compared these absolute predictions with the fit function with variable width. These two curves are compared in Fig.~\ref{fig:fit-eph}.
The fit and ephemerides are perfectly matching in terms of periods and initial phase. To match the amplitudes we had to scale the ephemerides down by a factor of (0.68, 0.65, 0.69, 0.73) for the four satellites and the result is shown in Fig.~\ref{fig:fitS-eph} in appendix. This result matches with the findings of Roche \cite{Roche1982}
that quotes a factor 1.4 (quoting the the inverse).

According to Giovanni Alfonso Borelli the elongations were estimated using a ``micrometer'' i.e. a scale that could slide along the tube of the telescope until the apparent size of a division, seen with the other naked eye, matched the diameter of Jupiter. In this case the overestimation might be due to the assumption that the angular size of Jupiter was 1$^\prime$ wide, hence 70\% larger than the real value of 38-45$^{\prime\prime}$. In this case these measurements would also be relative to the diameter of Jupiter and then it is not surprising that the amplitude of the orbits seems to be constant in the data. On the other hand in the Sidereus Galilei describes how the absolute measurement could be absolutely calibrated with terrestrial objects of known distance and dimensions. Probably the 30\% discrepancy might due to a systematic uncertainty in the calibration or simply a deliberate choice dictated by the limited avaliable time that led him to choosing for Jupiter a diameter of one prime.
It is interesting to notice that the estimation of the uncertainty diven by Galilei in the same section of the Sidereus is \emph{just one minute, or two}. The estimation is in perfect agreement with our analysis (see f.e. Tab.~\ref{tab:posteriori_errors}). 

Roche proposes a different explanation related to the fact that the measurement might have used the knowledge of the field of view of the telescope and to the fact that this field of view depends (\cite{Katz2007}) on the aperture of the eye pupil. The pupil adapts to smaller diameters during daytime or when trying to make a calibration on the Moon hence producing a different f.o.v. with respect to the one that is achieved when observing Jupiter. This explanation is indeed intriguing and not completely easy to verify. Taking pictures naturally does not help as the effect of the pupil is not taken into account. It is certainly an interesting and clever interpretation that would deserve more scrutiny. On the other hand the fact that the fitted amplitudes do not show hints for reducing during towards the end of the considered period supports that the measurement are indeed relative to the perceived angular size of Jupiter. This explanation is discarded in by Roche (\cite{Roche1982}, pag. 25) on the basis that according to him Galilei estimated 2$^\prime$ for the size of the disk of Jupiter. This would make the overestimation too gross and fail this interpretation. Our analysis (Fig.~\ref{fig:correlation} and Fig.~\ref{fig:chi2corr}) indeed indicates that Galileo was assuming a value of about 1$^\prime$ which is indeed about a factor 1.4 larger than the real angular extension.

\subsection{Kepler III law and 1:2:4 resonance}

We now try to imagine how these data can be used to test the validity of Kepler third law for the Jupiter system and the 1:2:4 Laplace resonance
between the periods of Io, Europa and Ganymede. Kepler published it nine years after the first observations of Galilei, in 1619 in \emph{Harmonices Mundi}. For this test we have converted the radiuses of satellites' orbits with a scaling factor chosen such that the scaled value for Ganymede corresponds to the modern measured value of 1.0704~million of km\footnote{The scaling factor is 1.0704~Gm/$(4.482 \pm 0.084)$~GJD = 2.39$ \times $10$^{8}$~m/GJD.}.

\begin{figure}
    \centering
    \includegraphics[width=\linewidth]{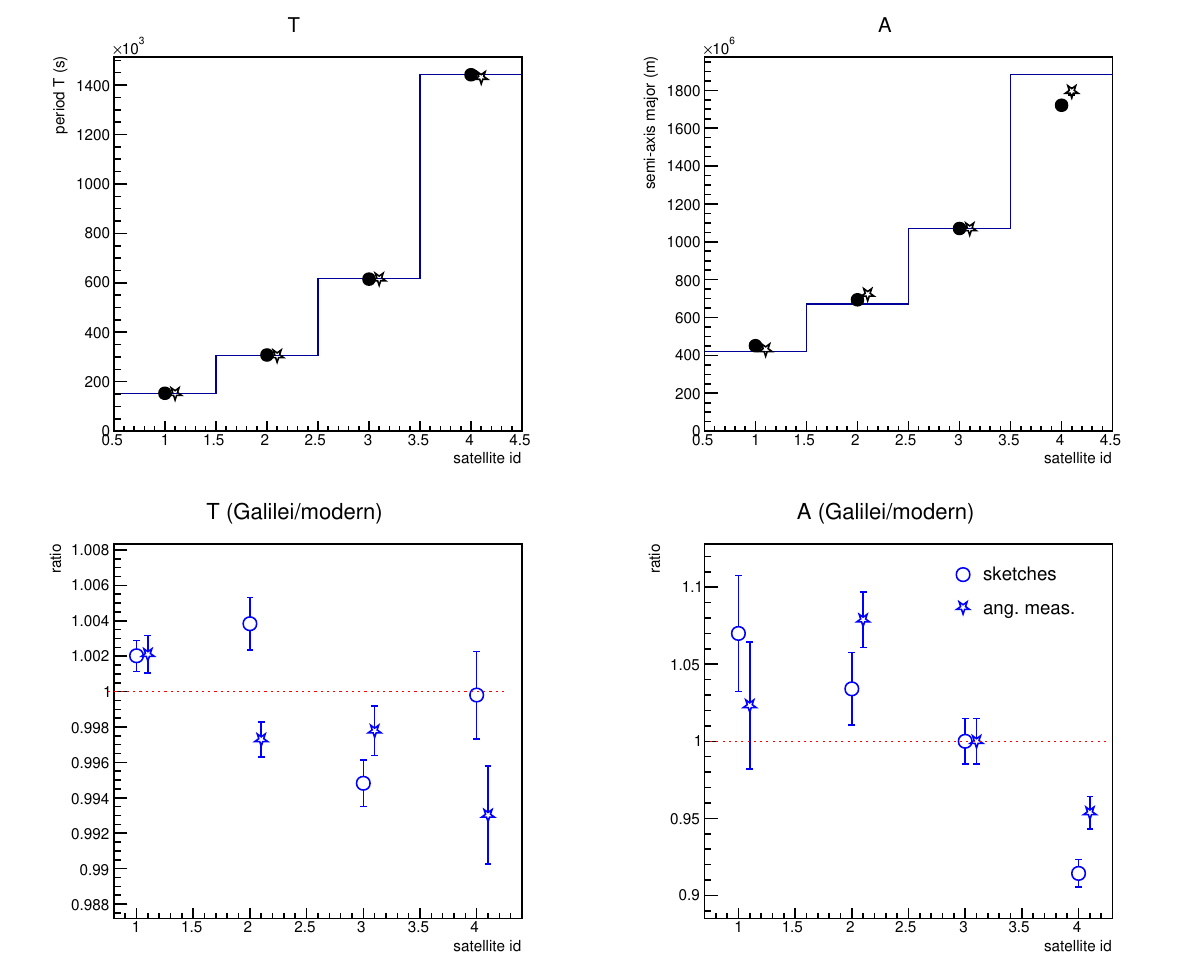}
    \caption{Top left: satellites periods as derived from the sinusoidal fit to Galilei's sketches (bullets) superimposed to the modern values (histogram). Top right: same as top left but for the major semi-axes. Bottom left: ratios between Galilei and modern for satellites' periods. Bottom right: same as bottom left but for major semi-axes.} 
    \label{fig:kepl4}
\end{figure}

Figure \ref{fig:kepl4} shows a comparison between the values of the satellites' orbits radiuses $A$ and their periods $T=2\pi/\omega$ as found by Galilei (bullets) and by modern determinations (histogram). 

\begin{table}[h]
\centering
\begin{scriptsize}
\begin{tabular}{llccc}
\toprule
 & Satellite & Value & Uncertainty & e.r. (\%) \\
\midrule
\multicolumn{5}{c}{\textbf{Dataset-1}} \\
\midrule
$A$ (/J. drawn diameter)& Io  & 1.830  & 0.064  & 3.5 \\
$A$ (/J. drawn diameter)& Europa  & 2.884  & 0.065  & 2.2 \\
$A$ (/J. drawn diameter)& Ganymede& 4.342  & 0.064  & 1.5 \\
$A$ (/J. drawn diameter)& Callisto& 6.985  & 0.069  & 0.98 \\
$T$ [days] & Io  & 1.7727  & 0.0015  & 0.087 \\
$T$ [days] & Europa  & 3.5551  & 0.0049  & 0.14 \\
$T$ [days] & Ganymede& 7.1175  & 0.0094  & 0.13 \\
$T$ [days] & Callisto& 16.6693 & 0.0395  & 0.24 \\
\midrule
\multicolumn{5}{c}{\textbf{Dataset-2}} \\
\midrule
$A$ ($^\prime$)& Io  & 3.13  & 0.12  & 3.9 \\
$A$ ($^\prime$)& Europa  & 5.242  & 0.086  & 1.6 \\
$A$ ($^\prime$)& Ganymede& 7.86  & 0.11  & 1.4 \\
$A$ ($^\prime$)& Callisto& 12.95 & 0.13  & 1.0 \\
$T$ [days] & Io  & 1.774  & 0.0018  & 0.10 \\
$T$ [days] & Europa  & 3.5415  & 0.0035  & 0.10 \\
$T$ [days] & Ganymede& 7.1403  & 0.0095  & 0.13 \\
$T$ [days] & Callisto& 16.613 & 0.043  & 0.26 \\
\bottomrule
\end{tabular}
\caption{Fitted amplitudes ($A$) and orbital periods ($T$) with uncertainties and relative errors for the four Galilean satellites.}
\label{tab:results}
\end{scriptsize}
\end{table}

\begin{figure}
    \centering
    \includegraphics[width=0.55\linewidth]{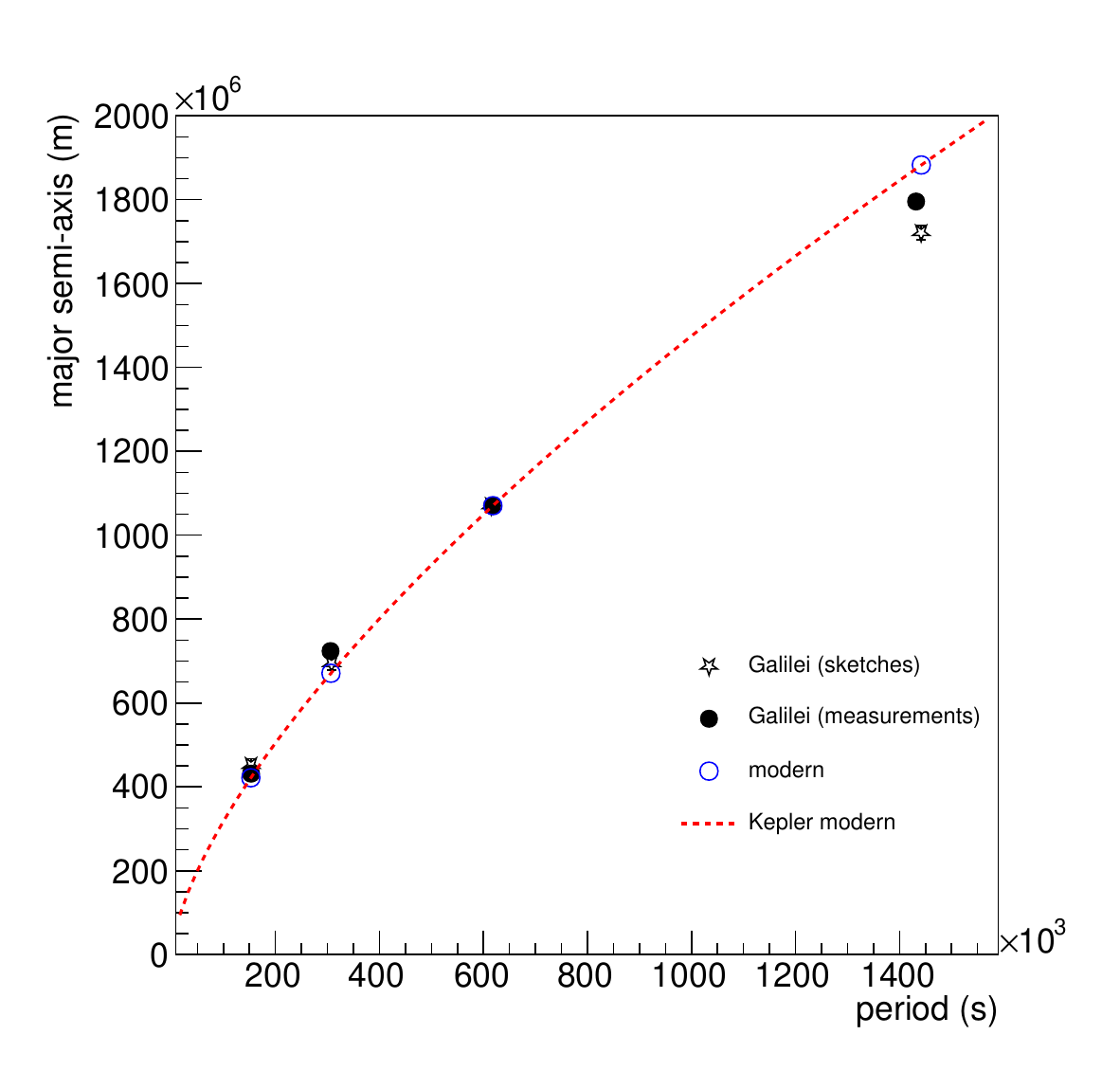}%
    \includegraphics[width=0.55\linewidth]{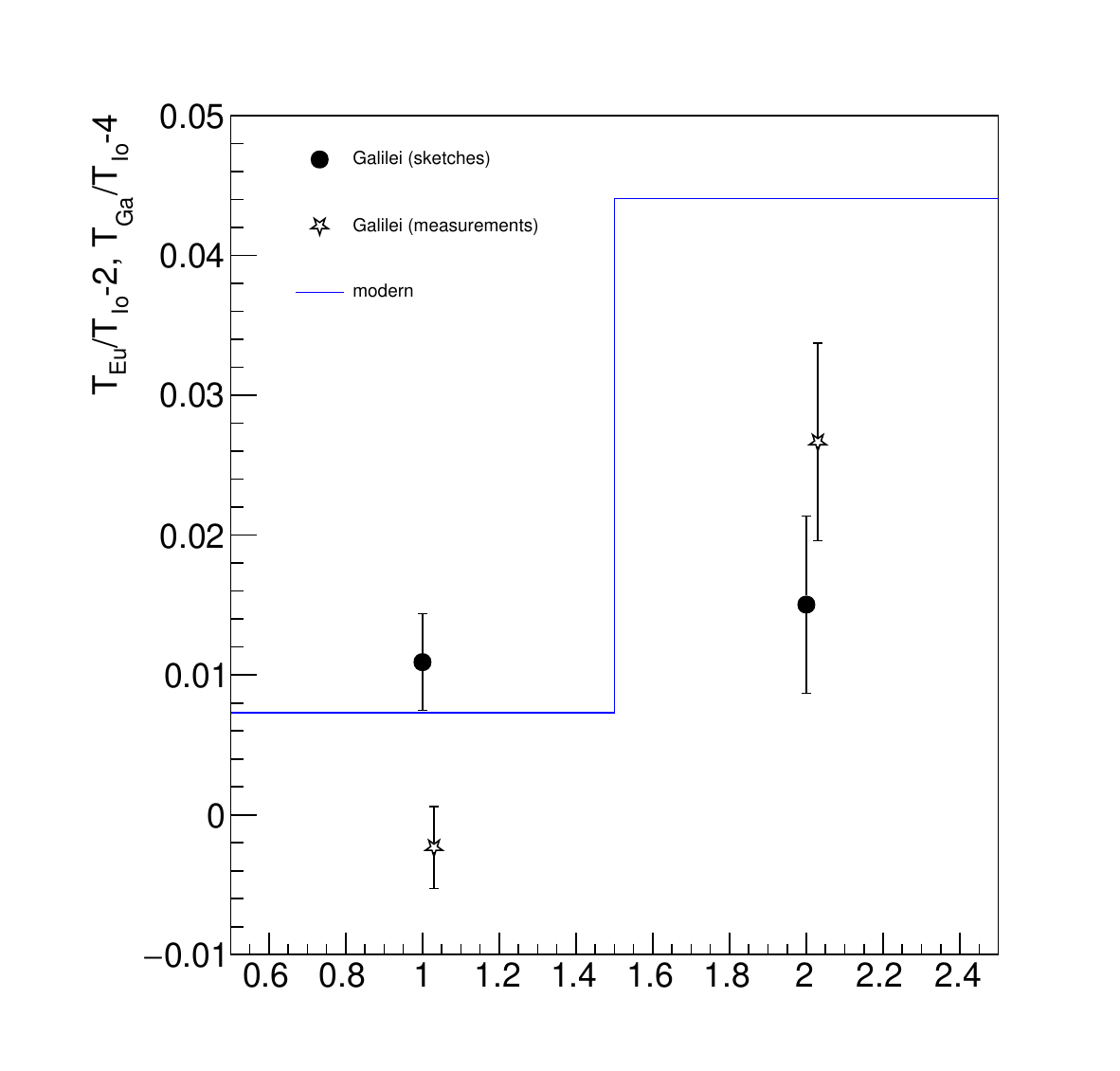}
    \caption{Left: Kepler third law for the Jupiter system. Right: Test of the 1:2:4 Laplace resonance for Io, Europa and Callisto.}
    \label{fig:kepl3}
\end{figure}

On the $x$-axis we show the satellite identifier (from left to right Io, Europa, Ganymede, Callisto). Periods are in seconds, are shown in the top left plot while radiuses (in meters) in the top right plot. The bottom plots show the ratio between the measured value by Galilei and the modern value. The periods are compatible with modern values within their errors that range from 0.1\% to 0.3\%. As far as the radiuses are concerned the ratios are compatible with unity with 1-2 $\sigma$ for Io and Europa. The one of Ganymede is one by construction while for Callisto $A$ is underestimated by about 10\% for dataset-1 and 5\% for dataset-2. The statistical errors on these parameters range from 1.9\% for the outer satellites to 4.3\% for Io. There is hence a trend for a significant underestimation of the major semi-axes for Callisto (See Fig.~\ref{fig:kepl3}, bottom right plot).

Figure \ref{fig:kepl3} (left) shows a scatter plot of
$A$ vs $T$. Superimposed a curve of the type $T^\frac{2}{3}$\footnote{The curve is the function: $A(T)=k_JT^{\frac{2}{3}}=\left(\frac{G M_J}{4\pi^2}\right)^\frac{1}{3}T^{\frac{2}{3}}$ with 
$G = 6.67430\times 10^{-11}$ Nm$^2$kg$^{-2}$ and $M_J = 1.898\times 10^{27}$~kg hence $A(T)= 5.02\times 10^5 \times T^{\frac{2}{3}}$.}.

In determining the law in physical units we have used our knowledge on the orbit of Ganymede to pass from angles to distances. It is interesting to notice that, if you use the knowledge on the distance of the Earth and Jupiter from the Sun, this measurement allows determining the ratio between the mass of Jupiter and that of the Sun just by comparing $k_J$ and $k_{Sun}$. Still you have to know that the Kepler constants depend on the mass to one third which is something that will not be clear until Newton's law.

In Fig.~\ref{fig:kepl3} (right) we compute the ratios $T_{E}/T_{I}-2$ and $T_{E}/T_{I}-4$ to verify the level of accuracy in determining the orbital resonance effects. Black filled bullets indicate the measurements of Galilei and the blue hollow ones the modern values. Good agreement is found. Deviations from an exact 1:2:4 ratios, also present in modern data, are due to the slow precession of Io's orbit's closest point to Jupiter (perijove).

\section{Efficiency near the disk and separating power}
\label{cannocchiale}
\begin{figure}
    \centering
\includegraphics[width=0.65\linewidth]{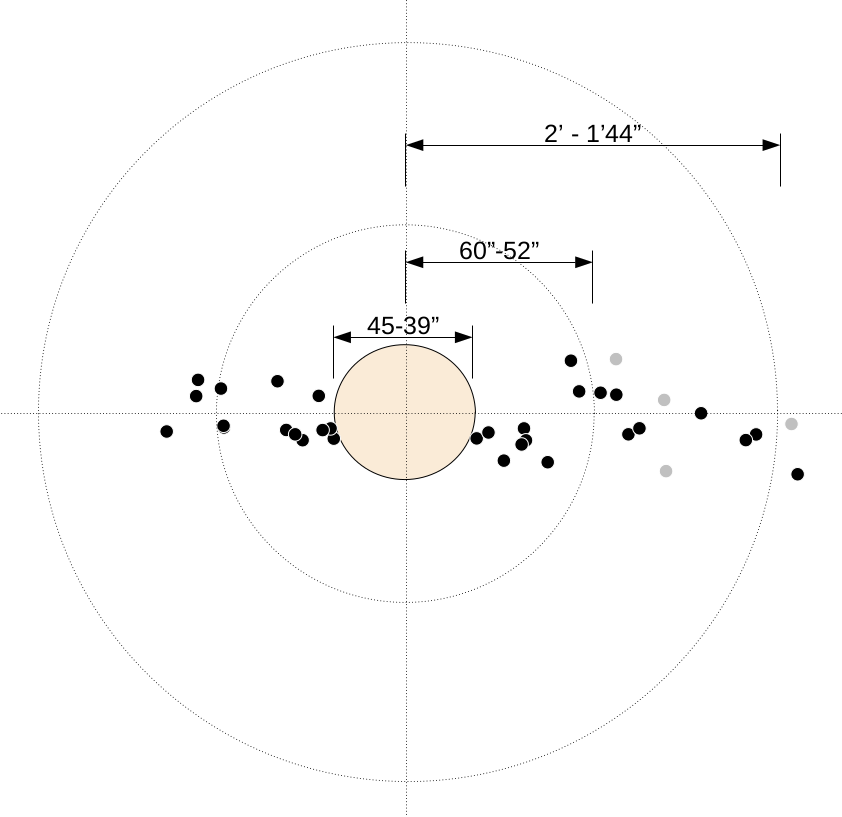}
    \caption{Map of satellites that went undetected probably due to being too close to the disk of Jupiter. The disks in gray represent that are less reliable (obs. 23, 38, 39, 48) as the hour is not specified hence the satellite might have been closer to Jupiter than what is assumed in the simulator (observation at one hour after sunset).}
    \label{fig:undetected}
\end{figure}

In Fig.~\ref{fig:undetected} we show the distribution of the positions of satellites that went undetected probably due to being too close to the disk of Jupiter. There is a sort of indication of a lower efficiency for the western side. This is the side that sits in the field of view of the telescope for less time due to Earth's rotation. It might still be a statistical fluctuation.

In Fig.~\ref{fig:undetected} we show the distribution of the positions of pairs of satellites that were not resolved into a single object.  The largest unresolved case has an angular separation of 31-37$^{\prime\prime}$
but other typical cases are in the order of 15-17$^{\prime\prime}$. This estimate is somewhat affected by the uncertainty in the exact time of the observation, especially when Io was involved or in the proximity of Jupiter where the velocity is higher. It is interesting to notice that the pairs in cyan color and  dark blue, were observed as a single object even if they fall in the region where efficiency was low (see Fig.~\ref{fig:undetected}).
\begin{figure}
    \centering
\includegraphics[width=1.\linewidth]{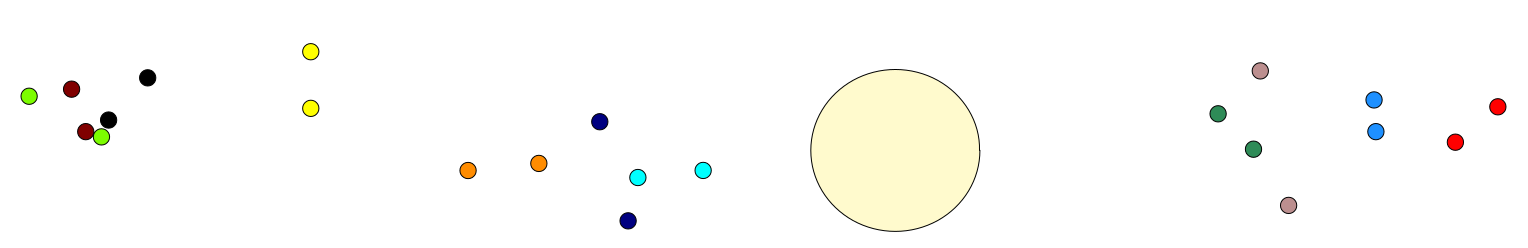}
    \caption{Map of pairs of satellites that were perceived as a single object.}
    \label{fig:unresolved}
\end{figure}
There are then the cases where the satellites were quite close to the diameter but were anyway detected. This happened up to angular distance distances of about one equatorial diameter or less from the limb (see observation 5 and 18, both with a double pair unresolved, and 32, 45 with a single satellite quite close but seen). 

For observation 32 Galileo mentions a distance from the limb of 20 arcsec which corresponds to about half disk. The dimension of the disk reported in the drawing in that case is similar to the real one. This makes us thing that the problem for the loss of efficiency was likely more related to the glare of the planet rather than an enlarged disk due to aberrations.
It should be noted that in a publication of 1992 Greco et al. \cite{Greco1992GalileoLenses} showed that Galileo’s lenses, though simple (a positive objective and a negative eyepiece), were made with remarkable skill for the early 17th century. The objective lenses showed near-diffraction-limited quality, with errors only about 1/4 wavelength. Eyepieces were of poorer quality, but their small aperture angles minimized the effect on image quality. Aberrations were limited by applying collimators to the objective also using elliptical profiles (\emph{…that the opening left uncovered should be oval in shape, for in this way the objects will be seen much more distinctly.”}) \cite{GalileiOpere,VanHelden1977GalileoTelescope} (pag. 278).

In summary the efficiency is practically zero when the satellite is less than one diameter from the limb, it rises then above 1 diameter but there is still a significant inefficiency until more than two diameters from the limb, especially on the west side.

\begin{figure}
    \centering
    \includegraphics[width=\linewidth]{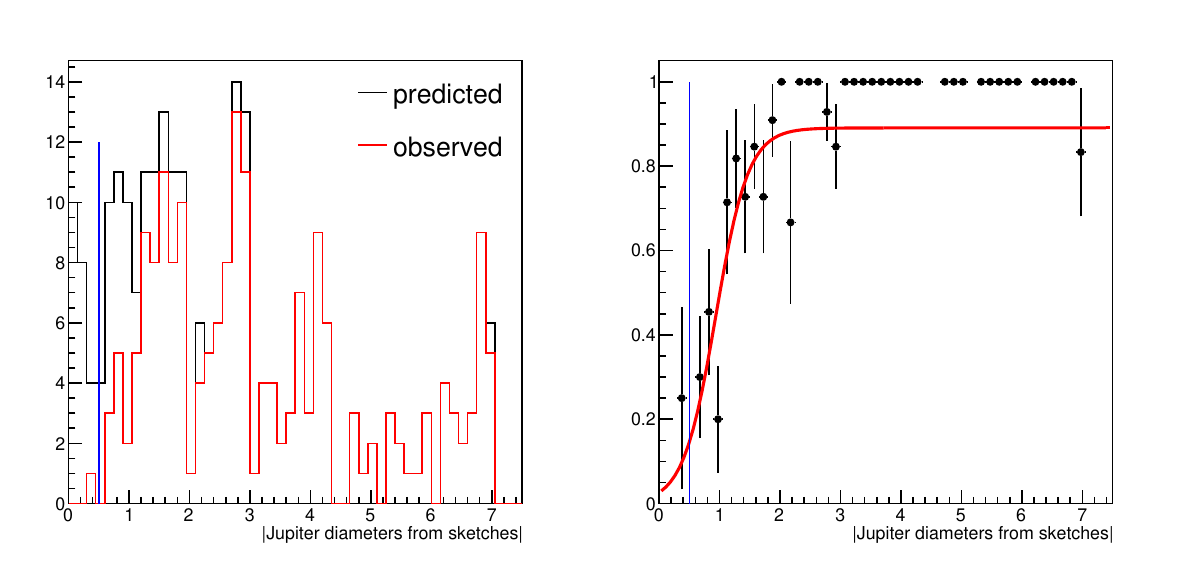}\\
    \includegraphics[width=\linewidth]{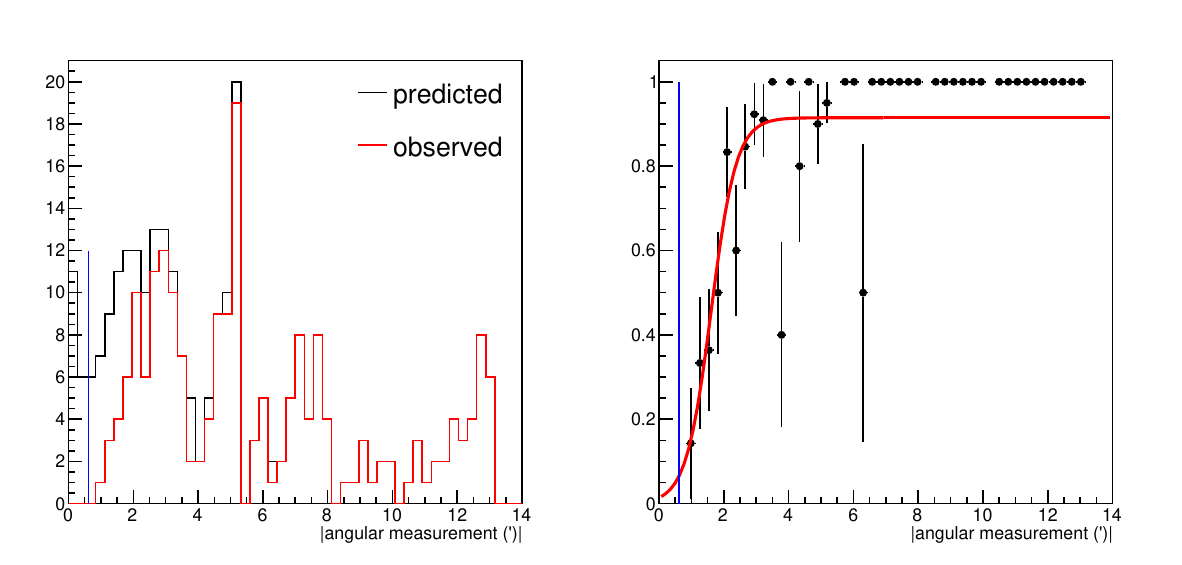}
    \caption{Top left: distribution of all the predicted elongations (black histogram) and of those that correspond to an observation for dataset-1. Top right: detection efficiency obtained as the ration between the red and black histogram. Bottom, same as top but for dataset-2.}
    \label{fig:efficiency}
\end{figure}

We have finally performed a more quantitative evaluation of the dependency of efficiency on the distance from Jupiter.
We have taken the sinusoidal fits of the four satellites and evaluated the elongations at the time of each observation
for all four satellites. The distribution of the absolute value of the elongations is shown in the black histogram of Fig.~\ref{fig:efficiency}. The red histogram shows the same predictions only taken when the satellite was actually observed according to the pattern described in Tab.~\ref{tab:ass1} and \ref{tab:ass2}. The efficiency is obtained by taking the ratio of the two distributions and is shown in the top right plot of the same Figure as a function of the predicted elongation of the satellite. The drop of efficiency due to the proximity to the disk is very evident and shows a progressive trend as the distance is lower. The inefficiencies at higher distances are basically due to the fact that sometimes satellites are too close so the efficiency goes to 50\% when they are not resolved.

There are cases where two very close satellites were resolved. For example Io and Europa are at about one Jupiter diameter from each other for observation 37 or in observation 48 they are at about 3/4 of the equatorial diameter or Ganymede and Europa that in observation 64 are again at about 3/4 of the diameter (i.e. $\sim$. 30$^{\prime\prime}$ being this observation done a the beginning of March when the diameter was about 30 arcsec). 

In total the satellites that were potentially visible but are not
recorded are 33 (Io 17 times, Europa 8 times, Ganymede 4 times and
Callisto 4 times).  We have excluded the cases in which the satellites
were transiting the disk of Jupiter or were behind it. The reason for
the inefficiency of the recordings is almost always due to the
proximity of the satellite to the disk of the planet that must have
been blurry. We can estimate that Galileo could see the satellites is
they were away from the limb of the planet by at least approximately
2.5 diameters or about 1'40".
It should be noted that on January 7 1610 Jupiter was 4.28 AU away corresponding to about 45" for the disk angular diameter while on March 1 it was 5.02 AU away corresponding to 38".
This change in angular diameter of the system by 18\% is not taken into account in the following analysis as we always report the elongations of satellites normalized to the angular diameter of the disk. Indeed the result support the validity of this assumption.

Observation 46 is probably one of the only that is difficult to reconcile with the prediction of the simulator.  Galilei observes right after sunset and sees two satellites on the left and two on the right. It seems that either Io or Europa are on opposite sides of the disk while the simulator predicts both being on the right. We have checked if this could be related to a mismatch in the recorded hour but in the night of Feb 11 the pattern was the same as observation 45 (both Io and Europa left). Even assuming an observation in the morning of 12/2 does not explain the anomaly. The transit of both happens during the daytime of Feb 12 and during the early night of 12 they are both on the right side. It is interesting to notice that this anomaly, independently observed in this work, it had been already noticed in the literature (Drake, Levi/Levi-Donati, Bernieri). Galileo was traveling back from Venice to Padua that day, the implication being that he might have been hasty due to the constraints of the travel, or maybe drew the sketch some time later, from memory. This explanation seems very likely. Indeed we think that a close-by moon was there (Europa) and it was just recorded at about the correct elongation but on the wrong side. What remains strange is that 3h~20' later (at the fifth hour) the satellite on the East had disappeared while in principle Europa had become more visible to the West.

For observation 49 it is interesting to note that Callisto is not
recorded even if it is less close than Europa that was seen instead.

\section{Other observations in the Sidereus Nuncius}
\label{otherobservations}

\subsection{The Pleiades}
\begin{figure}[ht]
\centering
\includegraphics[width=0.8\linewidth]{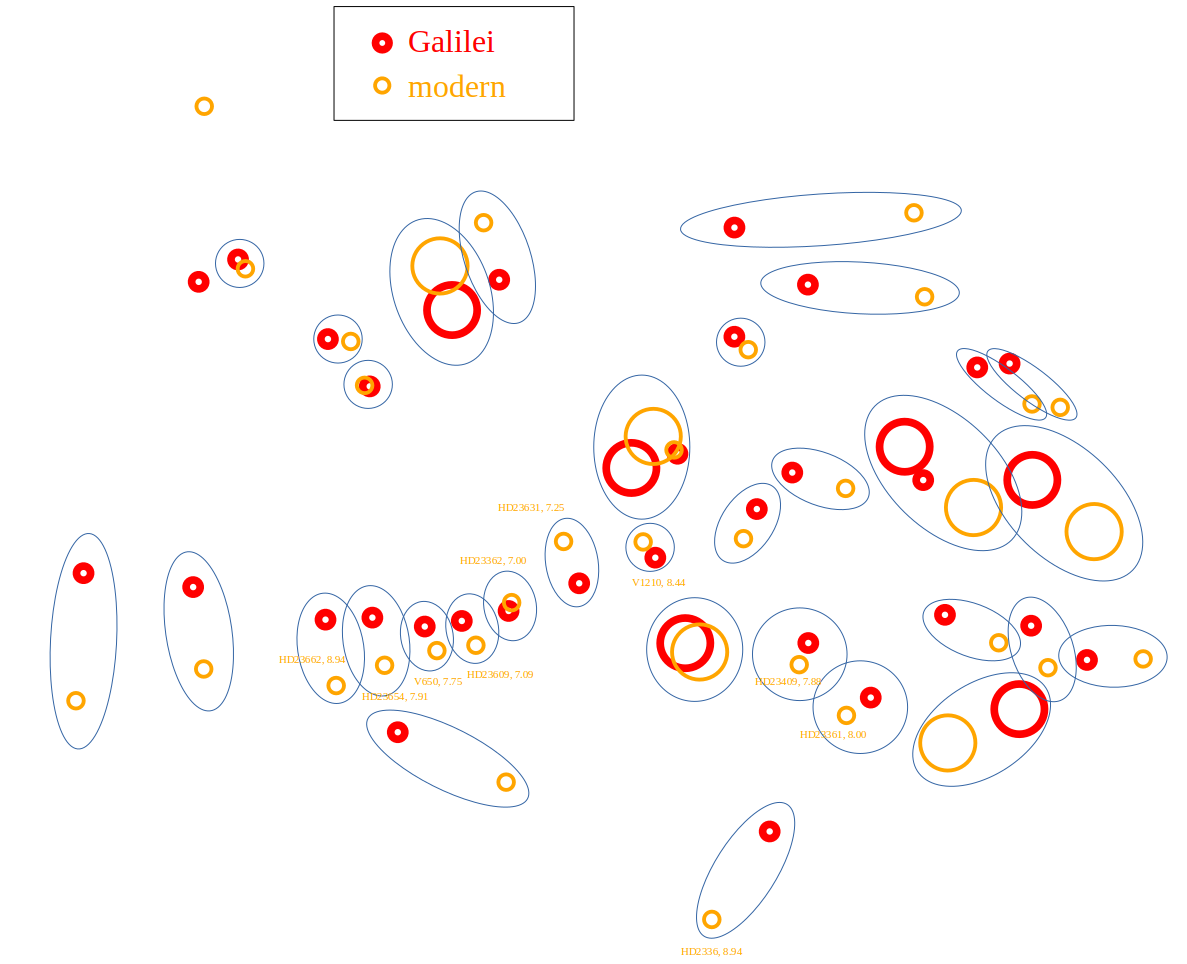}
\caption{\label{fig:pleiades} Pleiades: comparison with modern positions. The magnitudes are marked besides the stars. The faintest shines at 8.94.}
\end{figure}

In Fig.~\ref{fig:pleiades}, left, we show Sidereus Nuncius image of the iconic Pleiades star open cluster in the constellation of the Bull, the same where Jupiter was in those days. It was observed on Jan.~31. On the right a modern image. The two patterns has been superimposed in the bottom part where we have attempted to maximize the overlap by tuning the relative scaling and rotation. The cluster extends for about 1.5$^\circ$ while the field of view of the telescope was likely about one fifth of that. This can explain why
the two patterns show several systematic distortions as it could not
be visible all together at once. Nevertheless it is easy to establish
an unambiguous correspondence with the stars in the modern picture in almost all cases. The names and magnitudes of some of these stars are shown in the figure.  

Galileo could see stars up to magnitude almost 9 (i.e. 8.94 for HD2336). This is quite remarkable considering the small aperture of the telescope and it was helped by the total absence of light pollution in Padua, unlike nowadays. 

It is interesting to wonder if the positions of some of the stars might really have changed in 400 years such that some differences might be real. Considering that the proper motions can arrive to 60~m$^{\prime\prime}$/year this would translate into an angular displacement of 24 arcsec or about half a prime corresponding to about half the size of the Jupiter disk. The contribution of this effect is hence sub-leading (about half prime on a field of view of about 90 primes).

\subsection{The Orion belt and head}
\begin{figure}[ht]
\centering
\includegraphics[width=\linewidth]{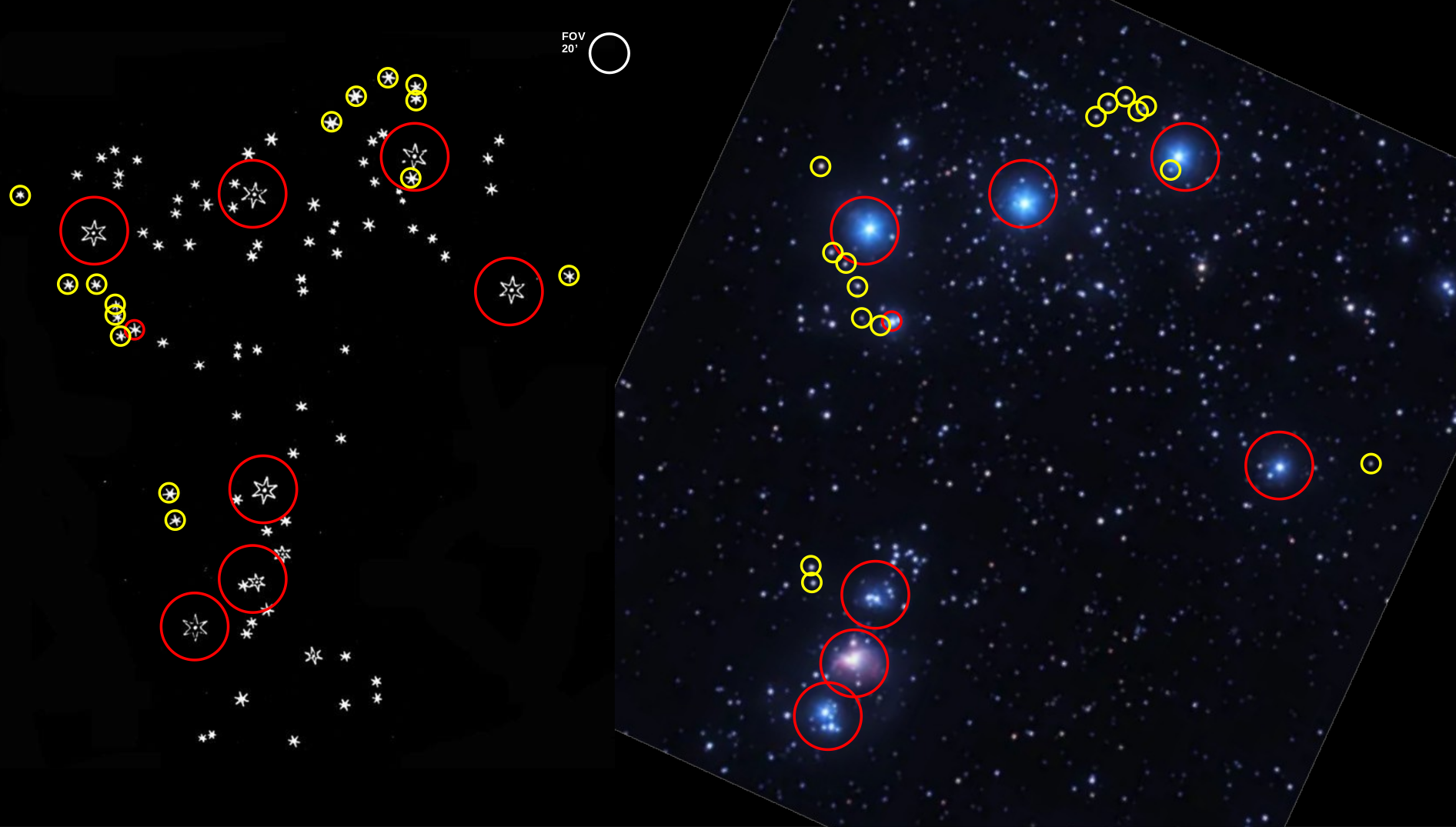}\\
\caption{\label{fig:orionbelt1} The Orion belt region. Red and yellow circles mark tentative associations between the sketch and the picture.}
\end{figure}

\begin{figure}
    \centering
\includegraphics[width=0.9\linewidth]{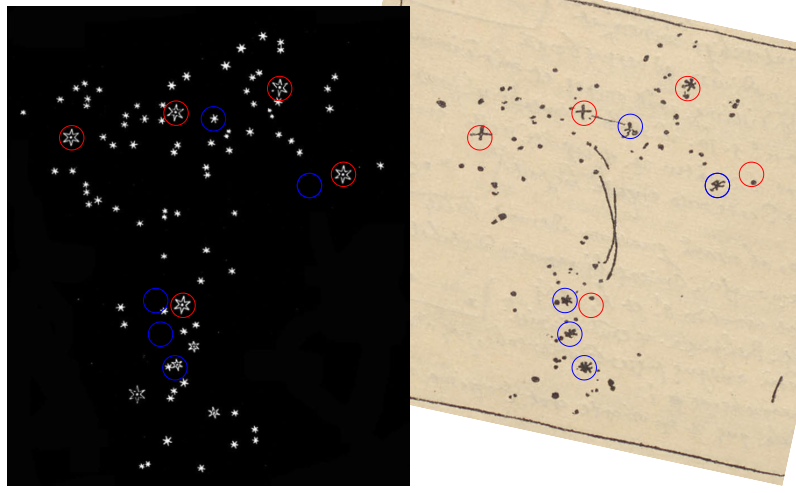}
    \caption{Comparison between the Orion belt in the Sidereus Nuncius (left) and on the original notes.}
    \label{fig:SN-sketch}
\end{figure}

Galilei also reports observations of the Orion constellation. He writes that originally he wanted to accurately map the entire constellation but the task was overwhelming due to the large amount of visible stars. He then reports the region of the Orion belt where, as for the Pleiades, he marks the ``old'' (visible by naked eye) stars with large markers. A comparison of his map and the one from \textsc{Stellarium} is in Fig.~\ref{fig:orionbelt1}.
These maps resemble a bit old topographic maps of the World where basic features are correct but a lot of deformations are visible. We can state that with respect to the Pleiades the correspondence is less accurate. It should be noted that the field of view of the cannocchiale (about 1/3 of a degree) is tiny with respect to the extension of the map (represented as the white circle in the mid-top of the figure). It is interesting to notice that the rightmost big star does not form a right angle with the belt. Indeed if we inspect the original notes~\cite{GalileiOpereVol3} of Feb. 7 1610, by which the Sidereus images are derived, it is clear that this is an error done in the printing phase as in the note the star position is more accurate (Fig.~\ref{fig:SN-sketch}).
Also the ``sword'' region is much more more realistic in the notes. It is interesting to notice the presence of two arcs connecting the belt and sword region that might indicate the fact that the sword is drawn closer than it should have been.
\subsubsection{Absolute angular separations around Alnitak}

In the same page of the note Galilei also reports angular separations between Alnitak (the leftmost star of the belt) and the surrounding stars. 
These data are rather clean as there are no uncertainties related to ephemerides and the precise knowledge of the observation time. Furthermore in this case there is no disk of Jupiter hence these data are not relative.

\begin{figure}
\includegraphics[width=0.5\linewidth]{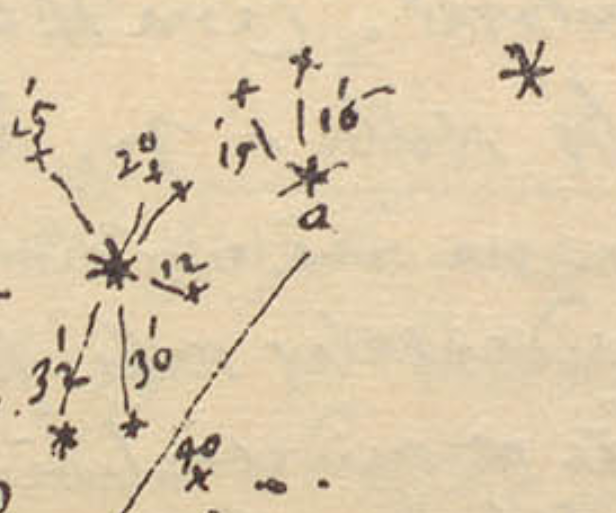}%
    \includegraphics[width=0.45\linewidth]{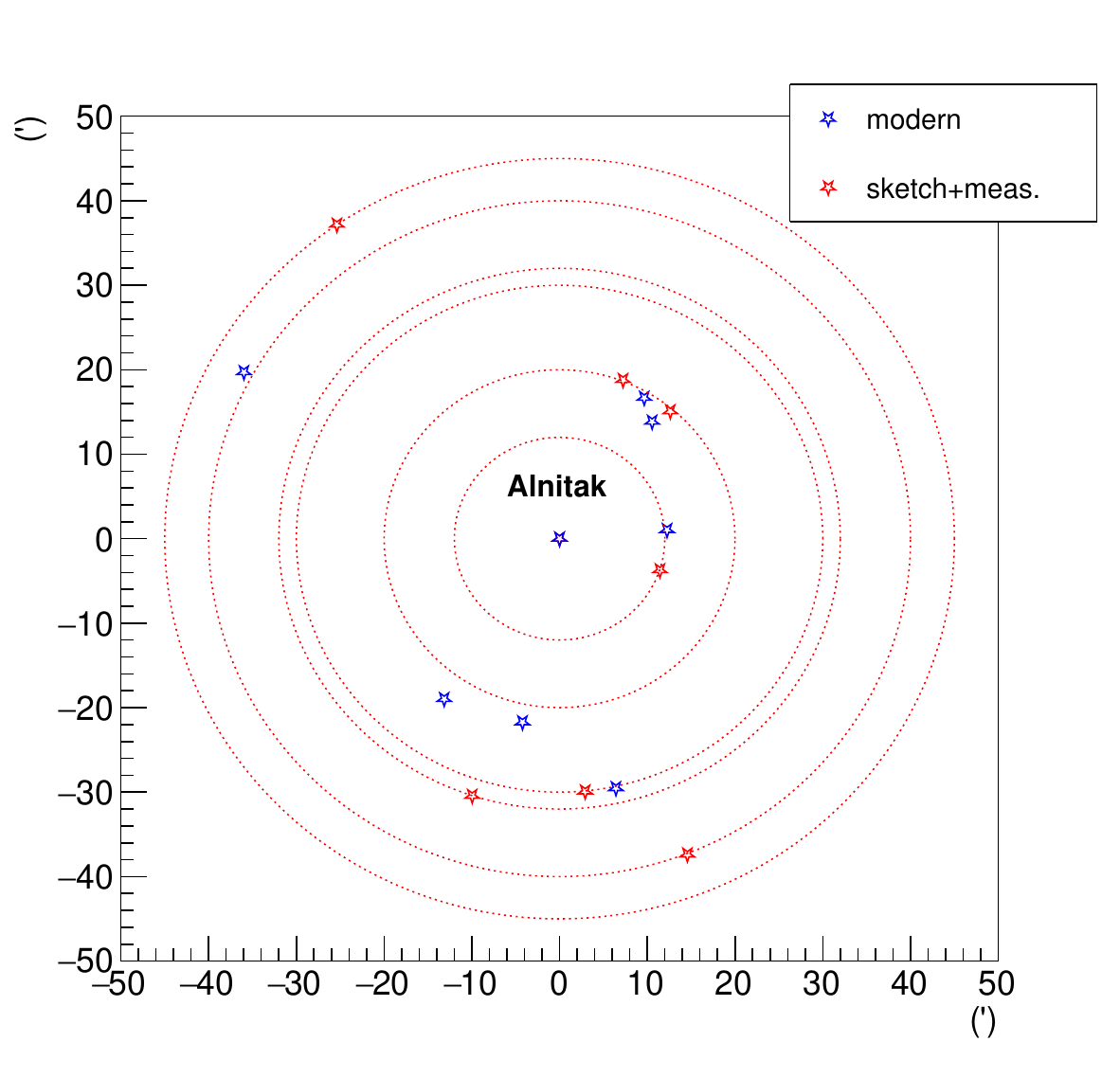}
    \caption{Alnitak and the surrounding stars with annotations on the angular separation (left). The right plot compares the modern angular distances with the ones on the left. The stars, shown by blue markers, are Alnitak (center), HD 37641 (mag.~7.53) the rightmost around $y=0$. Then, in clockwise order we find: HD37662 (mag.~8.5), HD37661 (mag.~8.8), V901-Ori (mag.~6.94), HD37903 (mag.~7.81), HD37805 (mag.~7.56) and HD37699 (mag. 7.56).}
    \label{fig:alnitak}
\end{figure}

We have hence tried to check the accuracy of these angular measurements as shown in Fig.~\ref{fig:alnitak}. In the left picture the original sketch is shown (Alnitak is the big star on the left). We have then taken the sketch and drawn the surrounding star following the written separations of 12, 30, 32, 45, 20 and 20$^\prime$ (red markers in the right plot) keeping the orientations of the original drawing. The blue markers represent modern values. A clear match emerges even though the azimuthal position correspondence is rather approximate. The distance of the closest star, (12$^\prime$, close to the maximal elongation of Callisto) is almost perfectly measured. The distances of the couple of stars at 20$^\prime$ are overestimated by few \%. At larger distances (30-32$^\prime$) the overestimation becomes at the level of 20-30\%. The separation of the farthest star at 45$^\prime$ (top-right) is also overestimated but less than before. 

This analysis supports a general overestimation of the elongations even if these occur on a scale larger than the typical elongations of the satellites of Jupiter. The closest star is quite well matched, and the agreement is way better than the 30\% discrepancy seen for the Jovian system. In this case, the overestimation at larger scale, likely comes from the need to work out of a single field of view. This might be the dominant effect.

It is worth noticing that Galileo reports in the same sheet a similar interesting sketch with distances of eight stars labelling it \emph{Canis}. We tried to match it with some obvious objects: Sirius (Canis maior), Procyon (Canis minor) or the M41 open cluster (in Canis maior) or the near-by M47 open cluster but, unfortunately, we did not obtain a satisfactory match.

\subsection{The Praesepe}
An observation of two ``Nebulae'' is also reported: one that Galilei calls the ``Orion head nebula'' and the Praesepe (nativity scene). This is also known as the Beehive cluster (Messier 44) in the Cancer constellation. 

\begin{figure}[ht]
\centering
\includegraphics[width=0.9\linewidth]{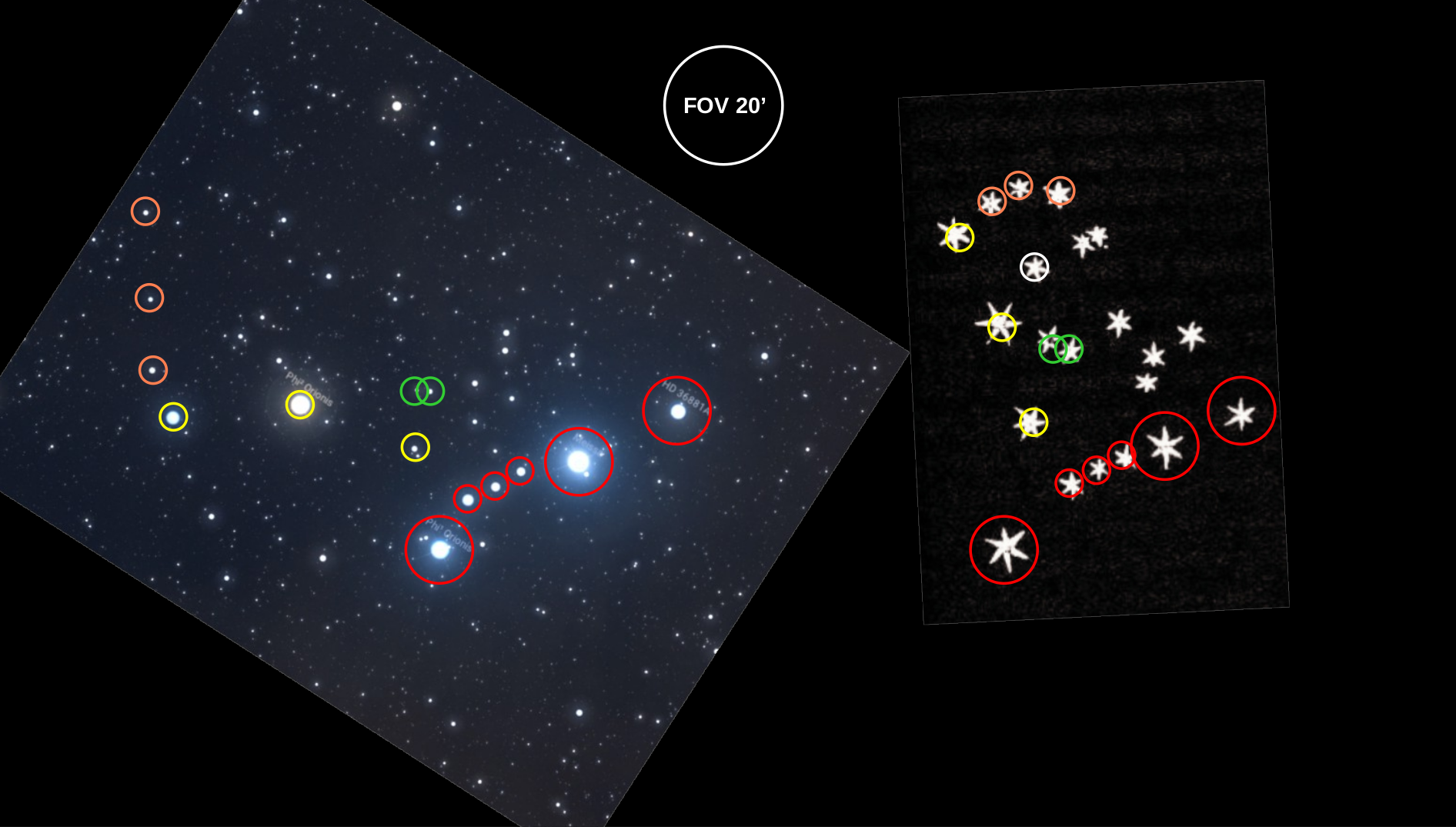}
\caption{\label{fig:orionhead} The head of Orion ``nebula''.}
\end{figure}

We have compared the sketch of Galilei with the group of stars located where the head of Orion sits. 
Despite being called a nebula this part of the constellation has nothing to do with ``the sword'' of Orion where the well-known Orion Nebula M42 lies. M42 was observed only later by Peiresc in 1610 and independently one year later by Cysat while Galilei later made a sketch of the stars of the trapezium asterism lying inside the M42 nebula (\cite{Palla2009StelleGalileo}). In Fig.~\ref{fig:orionhead} we have attempted a match with the simulator starfield. The bottom asterism (red circles) exhibits a quite convincing match while the rest of stars pattern has some resemblance but a massive distorsion effect must be assumed. Also in this case, the area is much larger than the field of view but more similar to the case of Pleiades than to the one of Orion. It is hence hard to understand why such a big distorsion effect is needed. For an independent evaluation it is interesting to mention an interesting work by Francesco Palla \cite{Palla2009StelleGalileo} done in 2009 where he also reaches similar conclusions using, in addition, real pictures taken with a replica of Galilei's telescope and installed at the Arceteri observatory.

\begin{figure}[ht]
\centering
\includegraphics[width=0.9\linewidth]{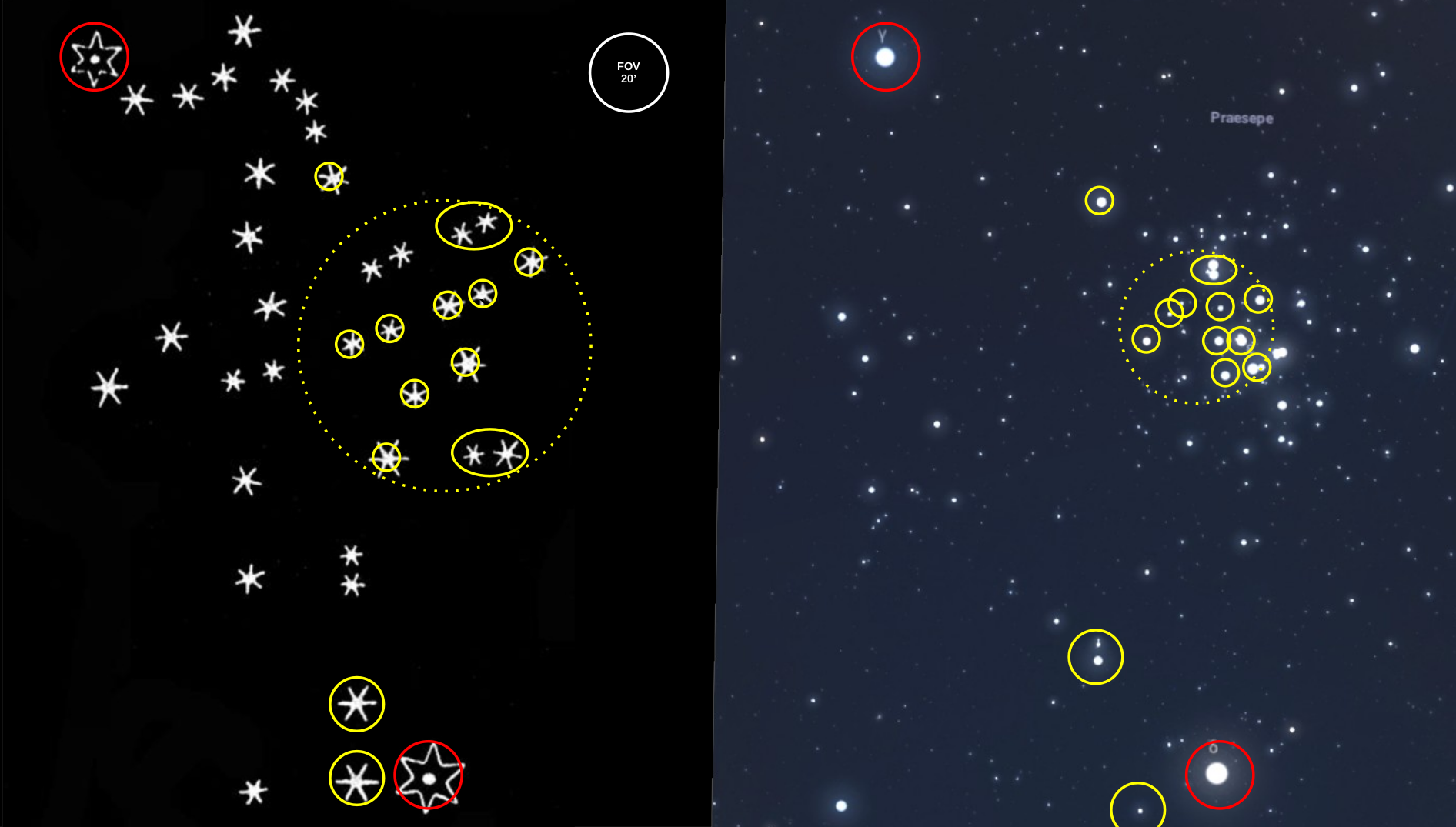}\\
\includegraphics[width=0.9\linewidth]{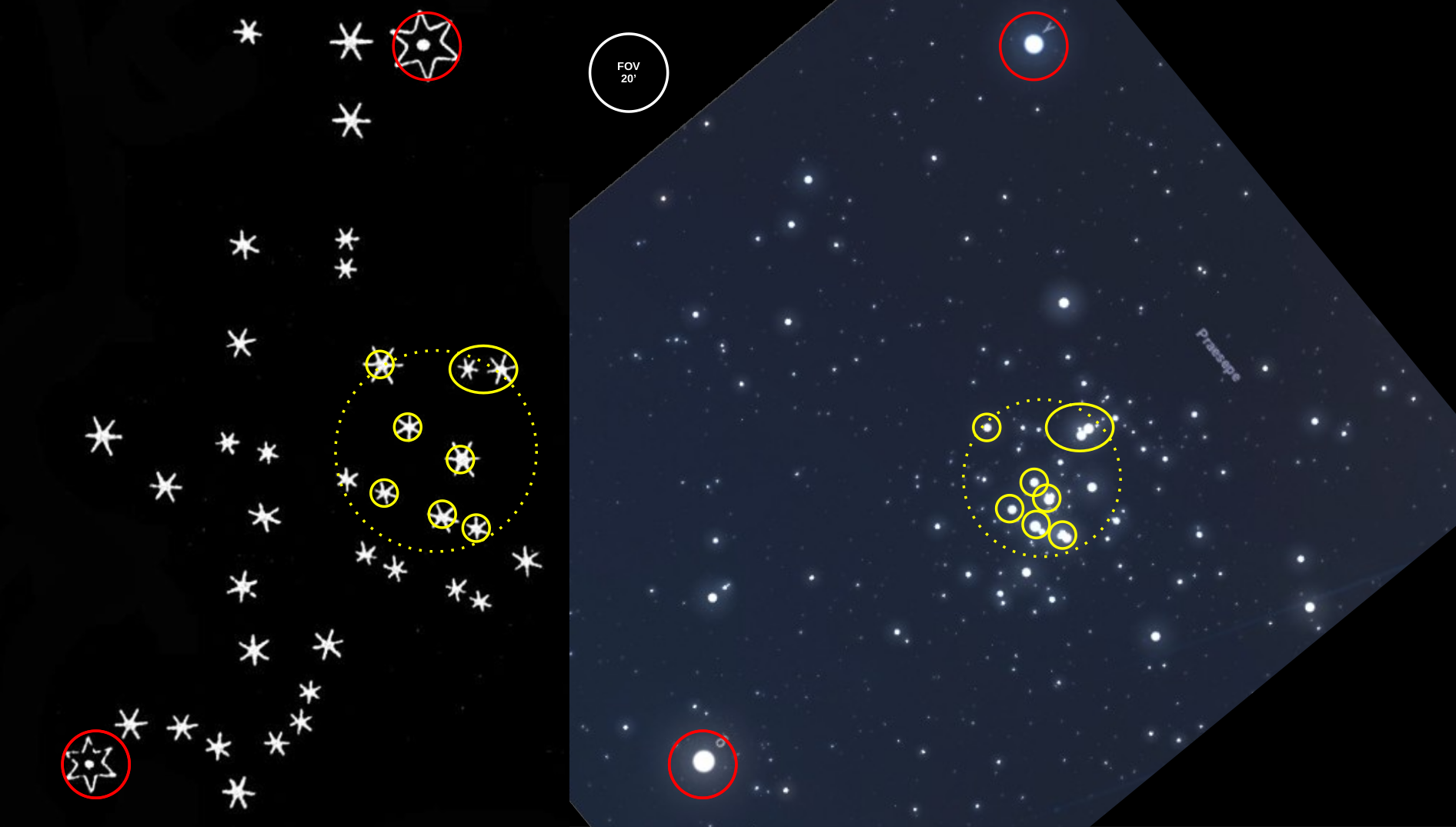}
\caption{\label{fig:beehive} Beehive cluster (M44) comparison. The top figure shows the drawing in its original form while the bottom one is flipped vertically.}
\end{figure}

The comparison for the Beehive cluster or M44 is shown in Fig.~\ref{fig:beehive}. In the following pictures the white circle highlights a field of view of 20$^\prime$.
Galilei write that the pair of ``large'' stars are the two ``little donkeys'' (\textit{Asellos}, $\gamma$ and $\delta$ Cancri), well visible by naked eye. What is striking is the absence of the little cluster of stars. The stars look more diffuse in the drawing. 
It is rather hard to establish a one-to-one correspondence unlikely to the case of the Pleiades. We have highlighted some tentative associations but there are several stars marked as bright that are hardly fitted in. The top image is the one that should correspond to the correct orientation where north is up. We have also considered, in the second image, flipping the drawing vertically (N-S). Other combinations, assuming other potential mistakes in the printing process, do not seem to give convincing matches. Also considering that the two brighter stars might not be the Asellos but two bright stars closer to the cluster does not bring much further.

\subsection{Inkwashes of the Moon}
\label{sec:moon}
A vast literature has discussed the interpretation and dating of the images of the moon \cite{GingerichVanHelden2003,Bredekamp2001,Bernieri2012, Bredekamp2019,Righini1975Lunar,Whitaker1978DatingSN}. The engravings in the Sidereus Nuncius are based on seven inkwashes that are the most faithful representation of what Galilei saw. In the engravings a large crater was added to strengthen the message that the moon was an object similar to Earth. A discussion of the debates on this subject are given by Gingerich and Van Helden in 2003 in \cite{GingerichVanHelden2003}. The accepted dates of the inkwashes are between 30 November 1609 and 17 December 1609 with the addition of an image on January 19 1610. It is remarkable that for this inkwash we can know the moment with resolution of minutes thanks to the vicinity of a star ($\theta$-Librae) that had just been reemerged after having been occulted by the Moon \cite{Molaro2013}. This happened at 6.36 AM CEST. Galileo, as we saw, had observed Jupiter also the previous evening (18/1) and he also did the following evening (19/1). Despite the urgency of the observations of Jupiter he had found the time to continue the observations of the Moon that he had started the month before.

Figures \ref{fig:moonA}, \ref{fig:moonB}, \ref{fig:moonC} (appendix) show a comparison of the seven inkwashes with the expectations of \textsc{Stellarium}. The hour and date were tuned to maximise the correspondence which, overall, is quite good. It is interesting to notice that in the last picture of Fig.~\ref{fig:moonB}, Galilei draws the Grimaldi crater a a grey small spot near the eastern limb of the waxing moon. Another interesting fact to observe is that the position of Mare Crisium is not matching so well, it seems much more extreme to the limb due to maximal libration in the simulation while in the inkwashes (especially n.3 and 4) it appears to be more central.

\section{Tests with a modern replica of the telescope}
\label{sec:replica}
\begin{figure}
    \centering
    \includegraphics[width=\linewidth]{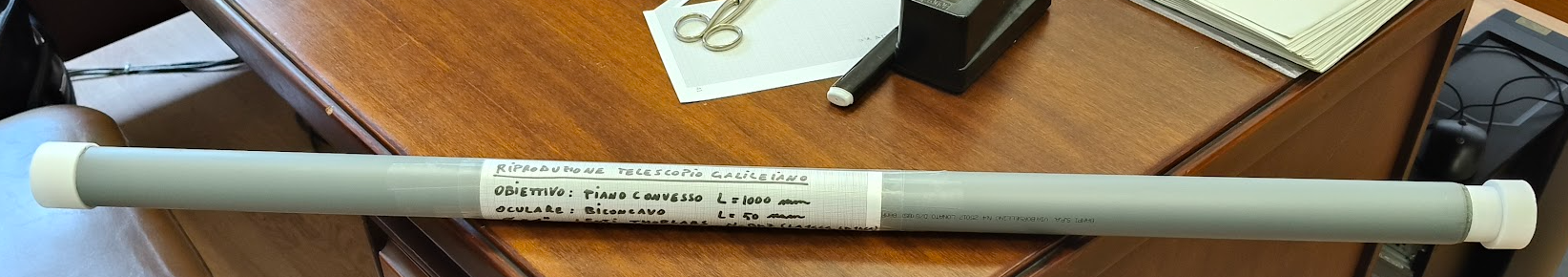}
    \caption{Replica of the telescope.}
    \label{fig:replica}
\end{figure}

In order to better understand the observations of Galilei, we have reproduced (Fig.~\ref{fig:replica}) one of the two telescopes that are currently preserved at the Museo Galileo in Florence\cite{museogalilei,GalTele,Greco1992GalileoLenses}. Among the two telescopes we have chosen to reproduce the one with the focal length of about one meter due to an easier availability of the lenses. The objective is a plane-convex lens with an aperture of 25.4~mm and a focal length of 1000~mm, the eyepiece made by a bi-concave lens with a focal length of 50 mm and a diameter 2.54~mm (from Thorlabs). It thus yields a 20$\times$ magnification. In addition to the obvious difference related to the precision of the optics, in our replica we did not put optical collimators as in the original. The plastic tube created stray light due to internal reflection so it was covered internally with a cylinder of sandpaper. 

\begin{figure}
    \centering
\includegraphics[width=\linewidth]{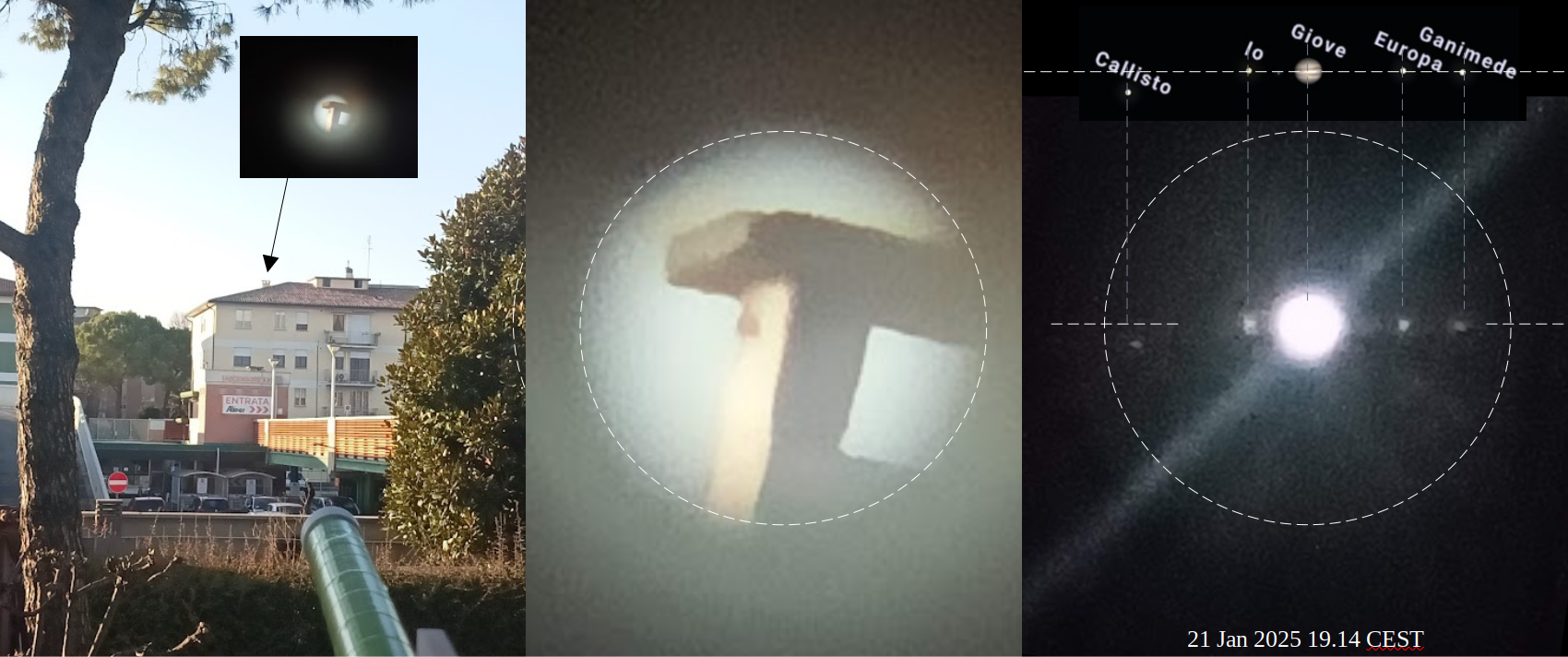} 
\caption{View of the galileian telescope view during daylight on a terrestrial target. Notice the narrowness of the field of view. Observation of the Galileian satellites with the modern reproduction of the Galileian telescope. Taken 21 Jan 2025 at 19.14 CEST.}
    \label{fig:telgal1}
\end{figure}

Figure \ref{fig:telgal1} is meant to convey the extremely narrow field of view by showing an example of the view provided for a terrestrial target in full daylight (left and central pictures). The central picture shows that there are no easily identifiable distortions near the edge of the field. The right picture shows the Jupiter system pictured on 21 Jan. 2025 and compared to \textsc{Stellarium}. The quality of the picture is rather poor as it was simply taken with a smartphone in a lucky shot. Still the presence of stray light is quite evident as well as the glare of the Jupiter disk. It is not difficult to imagine the difficulty of spotting satellites near the disk. But what is the real lesson coming from a direct experience with this object is the need for a really good support. On one hand even touching the object with the eye creates a risk for losing the target. In addition the target drifts very fast\footnote{It is easy to calculate that the target moves from the center to the edge in about 45~s with a f.o.v. of 20$^\prime$.} in the field of view and thus the possibility to apply fine corrections is needed. Pointing Jupiter might require some minutes. Callisto can easily fall out of the field of view. Initially it was tested on a support for cameras but later the equatorial mount of a modern telescope was used. It must also be stressed that, as we have seen, Galilei had to operate it with altitude angles up to about 70$^\circ$. The idea of using the micrometer techniques inputs additional challenges.

These challenges are not mentioned by Galilei in the Sidereus but he mentions them in a letter of 7 January 1610 to Antonio De Medici (or Enea Piccolomini) when he says (\cite{BucciantiniCamerotaGiudice}, pag.~62): \emph{Che lo strumento si tenga fermo, et perci\`o \`e bene, per fuggire la titubatione della mano che dal moto delle arterie et della respiratione stessa essa procede, fermare il cannocchiale in qualche luogo stabile.} i.e. ``That the instrument be held steady, and therefore it is advisable, in order to avoid the unsteadiness of the hand — which arises from the motion of the arteries and from breathing itself — to fix the telescope upon some stable support''~\cite{BucciantiniCamerotaGiudice}.
Bonifacio Vannozzi mentions the activity of Sergio Venturi, that in Rome was trying to build telescopes in that period between 1609 and 1610 and says: 
(\cite{BucciantiniCamerotaGiudice}, pag. 59) \emph{But in the sea, where these instruments could be very beneficial, they can never be used well, for the continuous motion of that setting}. Another indication of the frustration in the use of these instruments is found in a letter of Giovanni Bartoli to Belisario Vinta (\cite{BucciantiniCamerotaGiudice}, pag. 33) 
\emph{As for me, having seen several of them — and in particular one that was sold for three zecchini to the master of the post in Prague — I must confess I am not fully satisfied with it, because, being longer than an arm’s length, one has to struggle for quite a while to catch with the eye the object one wishes to see, and once found, one must hold the instrument so steady that the slightest movement makes it vanish again}.

The difficulty of the observation must have been particularly frustrating when Galilei went to Bologna to convince his rival Magini and a quite numerous group of ``well-educated people'' about his observations. Magini writes to Kepler\cite{BucciantiniCamerotaGiudice}, pag. 91: 
\emph{On the 24th and 25th of April he stayed at my house with his telescope, wishing to show the new satellites of Jupiter. He did not succeed. In fact, more than twenty most learned men were present, yet not one of them was able to clearly see the new planets}.

\begin{figure}
    \centering
\includegraphics[width=0.9\linewidth]{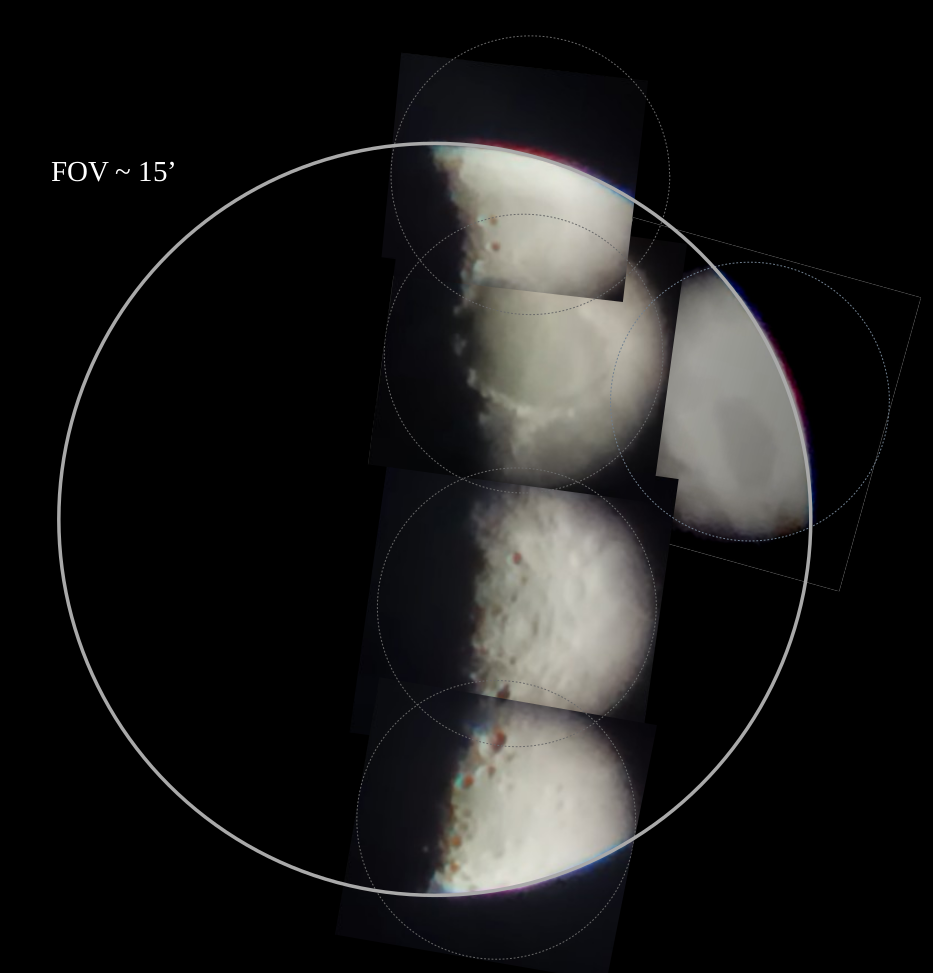}
\caption{Collage of pictures of the Moon taken with the reproduction of the Galileian telescope. Pictures of the moon taken on February 4 2025 at about 20 CEST from Padua. In the rightmost one Mare Crisium is visible. In the topo one the craters Exodus and Aristoteles (from bottom to top). In the second from the top Mare Serenitatis with Appennines and Caucasus at the terminator. In the southern region the two biggest visible craters are Maurolycus and Gemma Frisius.}
    \label{fig:telgal2}
\end{figure}

Figure ~\ref{fig:telgal2} shows a collage of pictures of the Moon taken with the replica on Feb. 4, 2025. A significant chromatic aberration can be noticed in the southern part were the craters are visible. Near the border some vignetting is visible (also in the central picture of Fig.~\ref{fig:telgal1}).The angular dimension of the Moon was 32.5$^\prime$ from which we can infer a field of view of about 15$^\prime$. Overall this picture testifies that the performance of the original was probably not much worse than the replica. The inkwashes presented in Sec.~\ref{sec:moon}
are definitely a good job considering the drift of the field of view, its limited extension, the need to realign, the vibrations and so on.

We plan to use this replica to involve students in a study attempting to reproduce the Sidereus dataset and compare their performances with those achieved by Galilei.

\section{Summary and conclusions}

In this work we have extend significantly the understanding of the Sidereus Nuncius dataset by applying an in depth analysis of the full achievable data: 
\begin{itemize}
    \item we have presented a detailed comparison of two datasets (Sec.~\ref{dataset},~\ref{digit}): the one coming from an analysis of the drawings in the published Sidereus and the angular data given by Galilei in the text. We show how digitized data suffer less from discretization and clarify how the angular measurements must be referred to the limb in order to make sense. To reconcile positive and negative elongations it is necessary to assume a Jupiter disk of 67-79$^{\prime\prime}$, as obtained from a comparison of the two sets, using a $\chi^2$ minimization. The angular data do not evidence a convincing variation of the apparent angle of the Jovian system that is expected at the 15\% level. This hints to the fact that these are not really absolute angular measurements but still relative to Jupiter, where its angular size is assumed equal to 1$^\prime$. Despite discussing a way to do an absolute calibration of the measurement Galilei seems to get the angles wrong, in excess with respect to the real ones. The normalization to the angular size of Jupiter, that is overestimated with respect to the real dimension, could also explain the overestimation of the elongations with respect to the ephemerides (about 70\% lower than the data). We performed a check of some angular measurements taken in the original notes on stars around Alnitak. In this case a separation of 12$^\prime$ is estimated quite correctly, without a 30\% discrepancy.
    \item We have presented sinusoidal fits of the elongations for both datasets as a function of the modern time of the 64 (dataset-1) or 64-5+9 (dataset-2) observations. The fitted orbital parameters are already extremely close to modern expectations. The errors are at sub-percent level for periods and at few \% level for the orbit's radiuses. By normalizing to the orbit of Ganymede we find that the elongations are accurate at better than 5\% for Io and Europa while Callisto is underestimated by about 5-10\%, depending on the considered dataset. 
    \item The residuals of the fit are used to estimate the resolution of the measurements. The errors range from 0.5-0.7 Jupiter diameters in the sketches or (0.86-1.44)$^\prime$.
    \item For the first time we have shown how a frequency analysis of sparse data (Lomb-Scargle periodograms) allows getting a fairly solid evaluation of the orbital periods of Callisto and Ganymede without having to disentangle the individual satellites;
    \item We have extend the study proposed by Bettini to this dataset showing how the data provide a stringent test of the third Kepler law for the Jovian system and a clear indication of the 1:2:4 resonance of the periods of the innermost satellites (for Io this is possible only by using the simulator to isolate the data of each satellite).
    \item We have evaluate the correctness of all the configurations and discussed the only known anomaly of Feb. 12. Each configuration drawn by Galilei was presented beside to the prediction from the software allowing to evaluate in a direct and intuitive way the reasons for missing observations.
    \item We have reevaluated the resolving power of pairs of satellites and the inefficiency introduced by the proximity to Jupiter's disk by analyzing the distribution of the positions of the missed satellites or missed pairs at ``truth-level'' using the simulator. We have shown rigorously that the onset of the efficiency drop starts already at a distance of about three diameters with respect to the center of Jupiter and falls rapidly to zero. The closest observations are typically pairs of satellites that were not resolved and hence resulted as a single satellite with double luminosity thus compensating the difficulty of detection close to the planet. We have created a 2D map of inefficient observations and of the positions of unresolved satellite pairs whose separation is typically around 15-17$^{\prime\prime}$.
    \item In Sec.~\ref{otherobservations} we have found an incredible level of accuracy for the observation of the Pleiades while for the Beehive cluster, the Orion head and belt, the association is more problematic, very likely due to the smallness of the field of view and/or time limits.
    \item The seven inkwashes of the Moon as predicted for the widely accepted dates show a relatively good match with the simulator in terms of phases and recognizable details; here we have not embarked on considerations on the pattern of Maria to infer information about the libration of the Moon.
    \item The experience accumulated with a replica of the telescope (at 20$\times$) shows that observations require a great deal of patience and expertise especially in relation to the limited field of view, the fast drift of the target, the need of a stable mounting. We have tried to recover in the literature mentions about this critical issue. The above-mentioned factors appear to be the largest obstacles in a successful data-acquisition. The optical quality of the instrument appears indeed to have been roughly comparable to modern optics, especially after the application of collimators.
\end{itemize}

This analysis allows quantifying and appreciating the level of accuracy that Galilei managed to attain with the instrumentation he had at his disposal. Even more after trying a replica of the telescope, the achieved accuracy and the diligence and completeness of these (early!) measurements is impressing and it can only add to the value of Galilei as a giant of (experimental) physics.

\vspace{0.5cm}
\noindent
{\bf{Acknowledgements}}: I would like to express my gratitude to Alessandro Bettini, Carlo Broggini, Giulio Peruzzi, Enrico Maria Corsini and Mauro Mezzetto of the Department of Physics and Astronomy ``Galileo Galilei'' and INFN Padova for their encouraging comments on an early version of this work
and guidance to the relevant literature.



\begin{appendices}
\section{Comparison of sketches and simulator}
\label{fullcomp}
\begin{center}
\begin{longtable}{N L R}
\caption{Side-by-side images with progressive numbering}\label{tab:allpairs}\\
\toprule
\textbf{No.} & \textbf{Sidereus Nuncius} & \textbf{Simulation} \\
\midrule
\endfirsthead

\toprule
\textbf{No.} & \textbf{Sidereus Nuncius} & \textbf{Simulation} \\
\midrule
\endhead

\midrule
\multicolumn{3}{r}{\small Continued on next page}\\
\bottomrule
\endfoot

\bottomrule
\endlastfoot


1  & \includegraphics[width=\linewidth]{figs/g00.png} & \includegraphics[width=\linewidth]{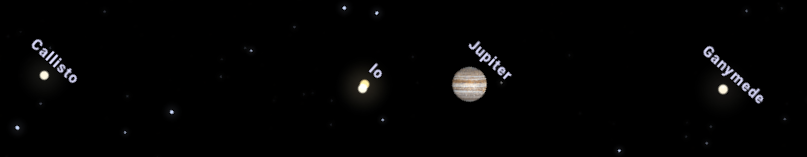} \\
2*  & \includegraphics[width=\linewidth]{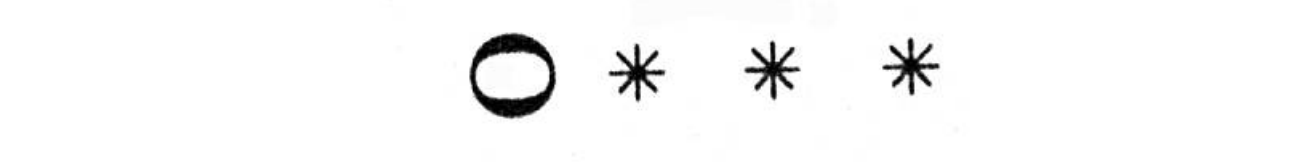} & \includegraphics[width=\linewidth]{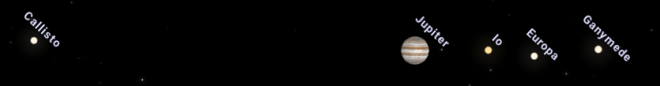} \\
3*  & \includegraphics[width=\linewidth]{figs/g03.png} & \includegraphics[width=\linewidth]{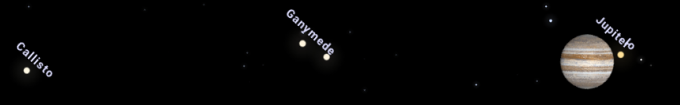} \\
4*  & \includegraphics[width=\linewidth]{figs/g04.png} & \includegraphics[width=\linewidth]{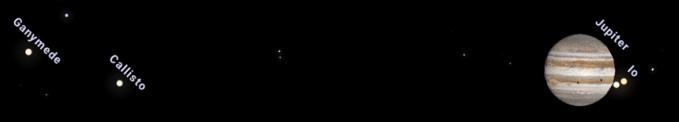} \\
5  & \includegraphics[width=\linewidth]{figs/g05.png} & \includegraphics[width=\linewidth]{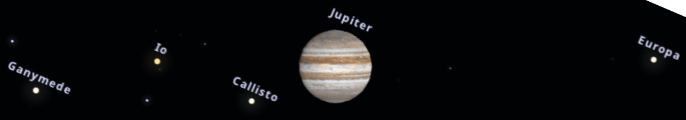} \\
6*  & \includegraphics[width=\linewidth]{figs/g06.png} & \includegraphics[width=\linewidth]{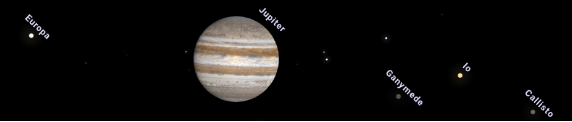} \\
7  & \includegraphics[width=\linewidth]{figs/g07.png} & \includegraphics[width=\linewidth]{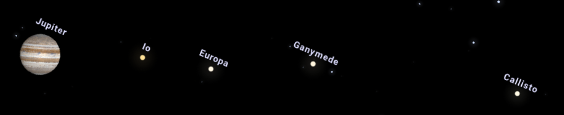} \\
8  & \includegraphics[width=\linewidth]{figs/g08.png} & \includegraphics[width=\linewidth]{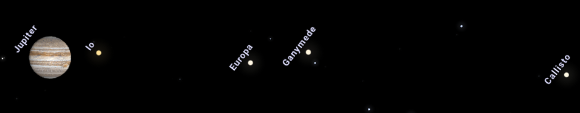} \\
9  & \includegraphics[width=\linewidth]{figs/g09.png} & \includegraphics[width=\linewidth]{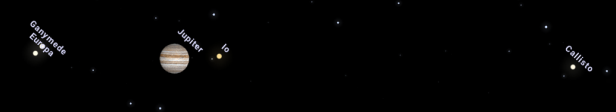} \\
10 & \includegraphics[width=\linewidth]{figs/g10.png} & \includegraphics[width=\linewidth]{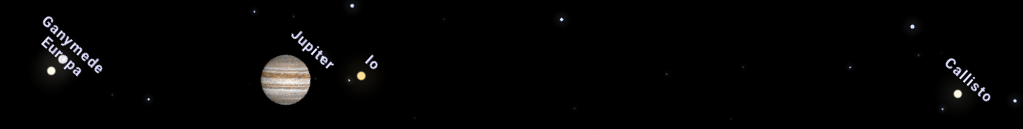} \\
11 & \includegraphics[width=\linewidth]{figs/g11.png} & \includegraphics[width=\linewidth]{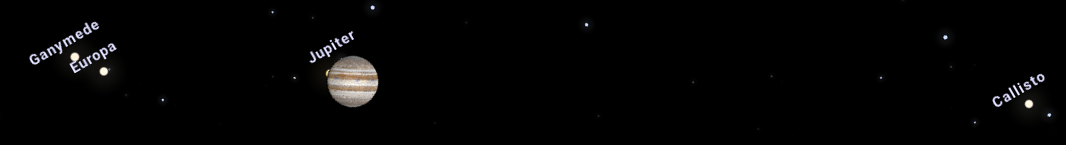} \\
12 & \includegraphics[width=\linewidth]{figs/g12.png} & \includegraphics[width=\linewidth]{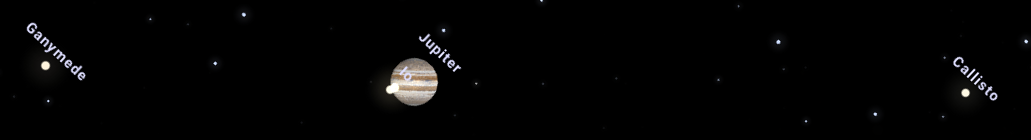} \\
13 & \includegraphics[width=\linewidth]{figs/g13.png} & \includegraphics[width=\linewidth]{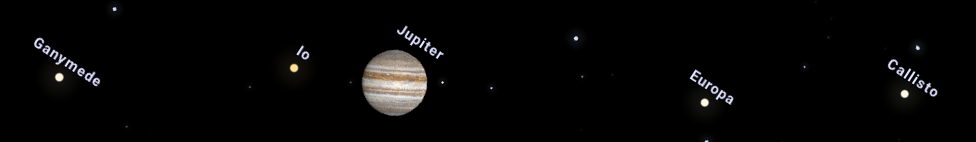} \\
14 & \includegraphics[width=\linewidth]{figs/g14.png} & \includegraphics[width=\linewidth]{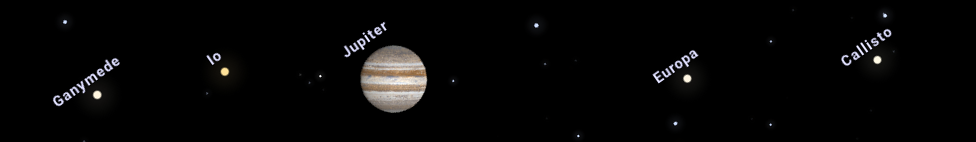} \\
15 & \includegraphics[width=\linewidth]{figs/g15.png} & \includegraphics[width=\linewidth]{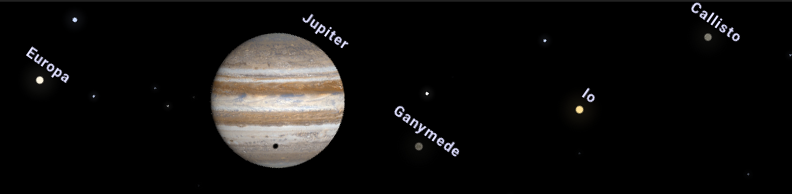} \\
16 & \includegraphics[width=\linewidth]{figs/g16.png} & \includegraphics[width=\linewidth]{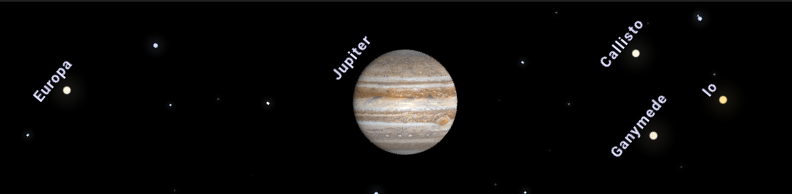} \\
17 & \includegraphics[width=\linewidth]{figs/g17.png} & \includegraphics[width=\linewidth]{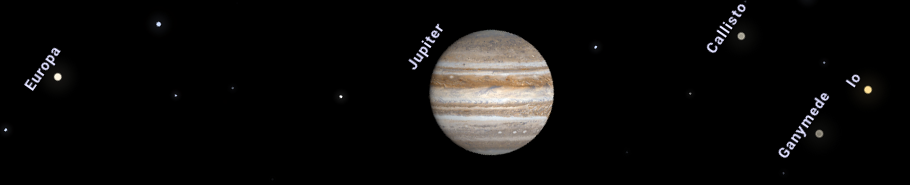} \\
18 & \includegraphics[width=\linewidth]{figs/g18.png} & \includegraphics[width=\linewidth]{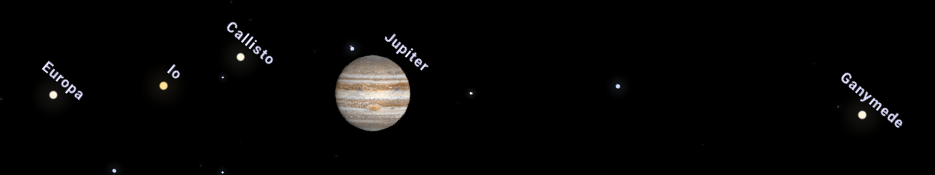} \\
19 & \includegraphics[width=\linewidth]{figs/g19.png} & \includegraphics[width=\linewidth]{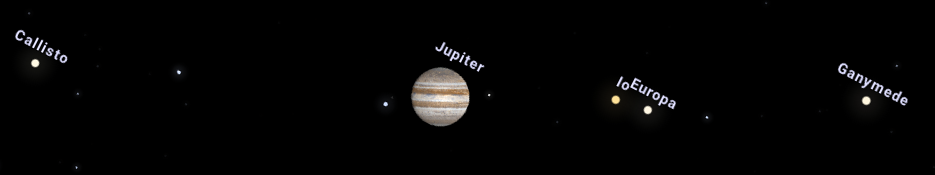} \\
20 & \includegraphics[width=\linewidth]{figs/g20.png} & \includegraphics[width=\linewidth]{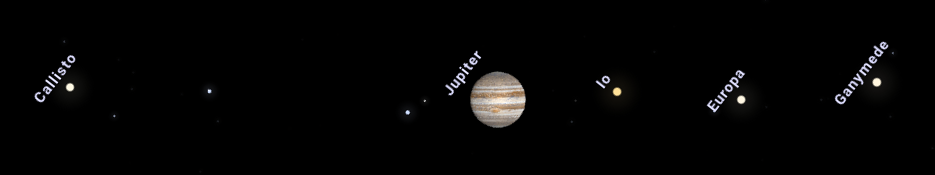} \\
21 & \includegraphics[width=\linewidth]{figs/g21.png} & \includegraphics[width=\linewidth]{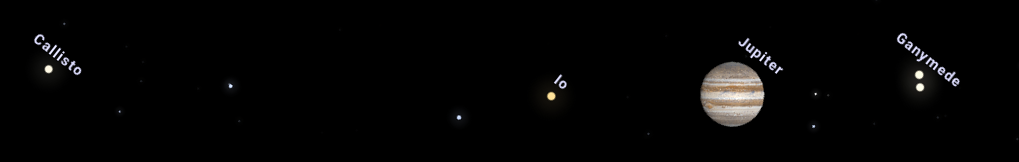} \\
22 & \includegraphics[width=\linewidth]{figs/g22.png} & \includegraphics[width=\linewidth]{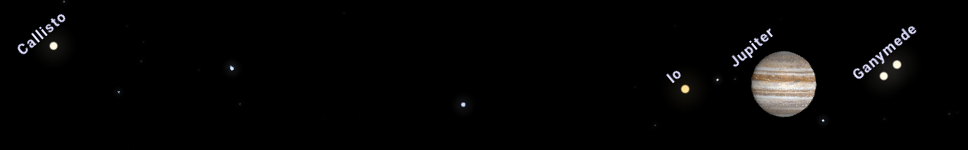} \\
23* & \includegraphics[width=\linewidth]{figs/g23.png} & \includegraphics[width=\linewidth]{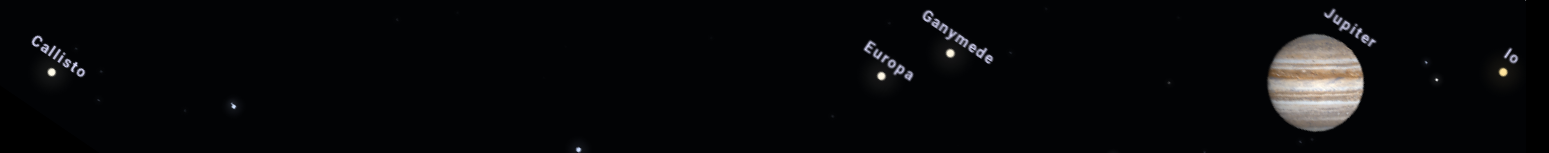} \\
24 & \includegraphics[width=\linewidth]{figs/g24.png} & \includegraphics[width=\linewidth]{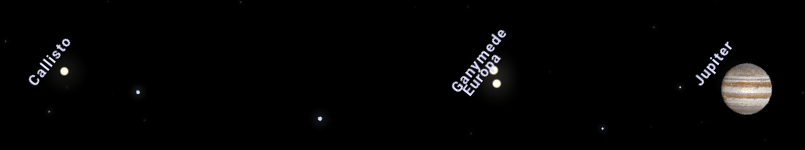} \\
25 & \includegraphics[width=\linewidth]{figs/g25.png} & \includegraphics[width=\linewidth]{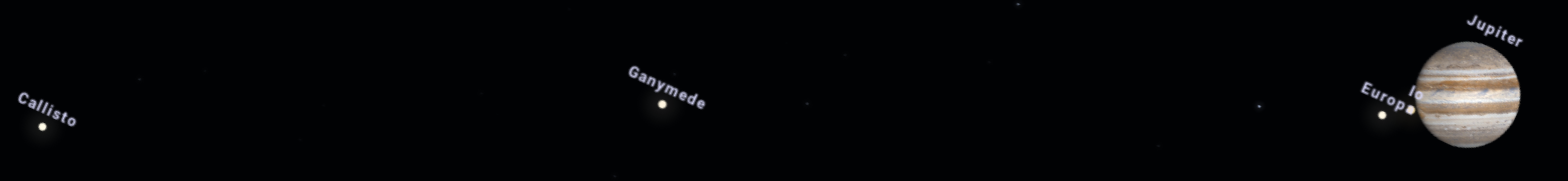} \\
26 & \includegraphics[width=\linewidth]{figs/g26.png} & \includegraphics[width=\linewidth]{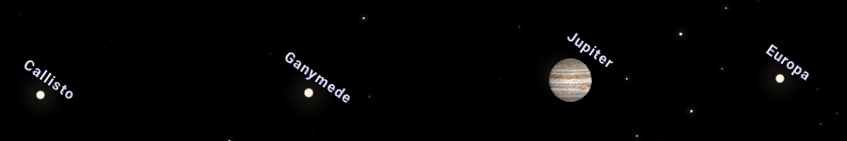} \\
27 & \includegraphics[width=\linewidth]{figs/g27.png} & \includegraphics[width=\linewidth]{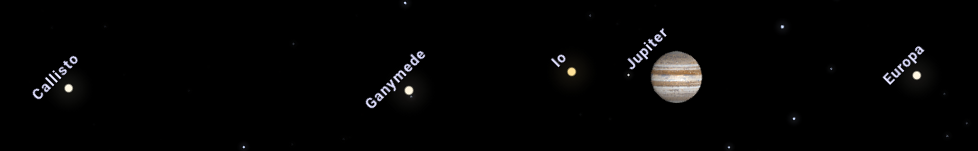} \\
28 & \includegraphics[width=\linewidth]{figs/g28.png} & \includegraphics[width=\linewidth]{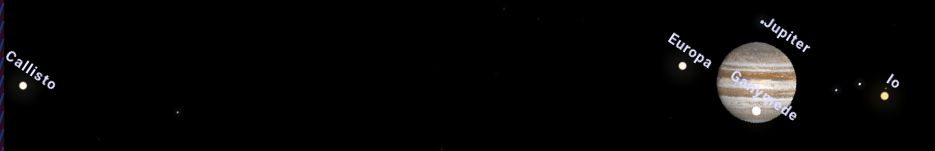} \\
29 & \includegraphics[width=\linewidth]{figs/g29.png} & \includegraphics[width=\linewidth]{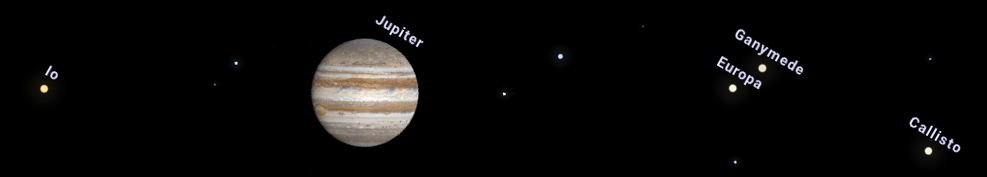} \\
30 & \includegraphics[width=\linewidth]{figs/g30.png} & \includegraphics[width=\linewidth]{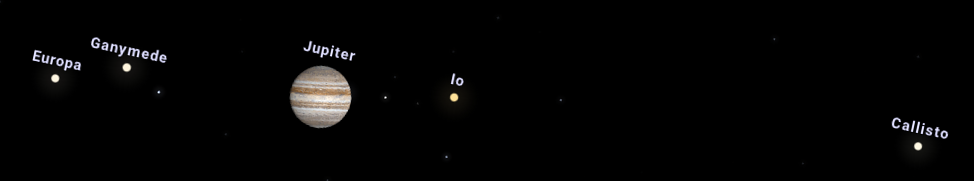} \\
31 & \includegraphics[width=\linewidth]{figs/g31.png} & \includegraphics[width=\linewidth]{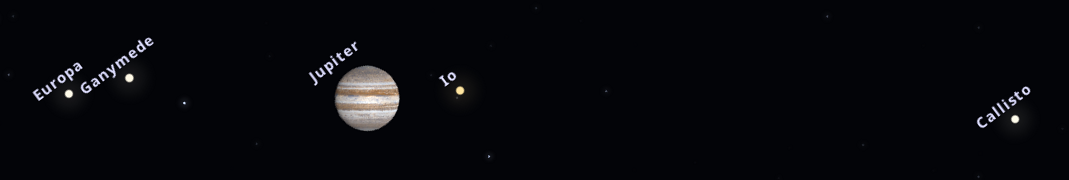} \\
32 & \includegraphics[width=\linewidth]{figs/g32.png} & \includegraphics[width=\linewidth]{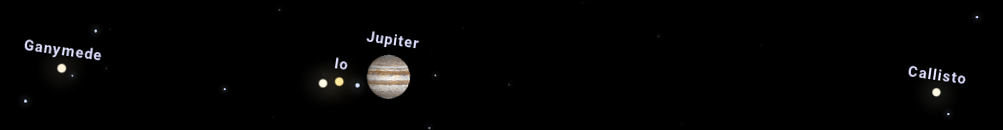} \\
33* & \includegraphics[width=\linewidth]{figs/g33.png} & \includegraphics[width=\linewidth]{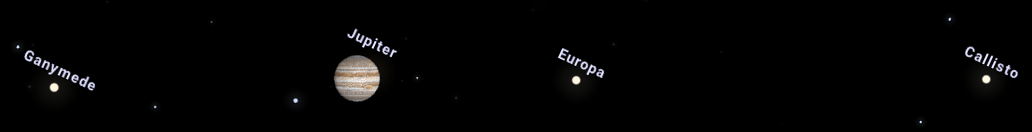} \\
34 & \includegraphics[width=\linewidth]{figs/g34.png} & \includegraphics[width=\linewidth]{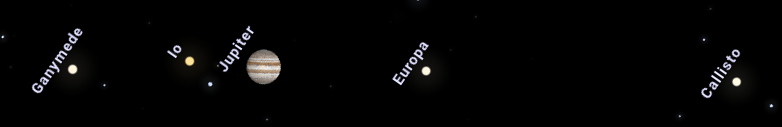} \\
35 & \includegraphics[width=\linewidth]{figs/g35.png} & \includegraphics[width=\linewidth]{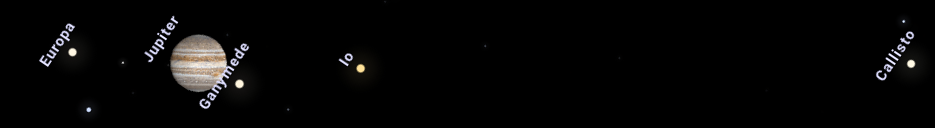} \\
36 & \includegraphics[width=\linewidth]{figs/g36.png} & \includegraphics[width=\linewidth]{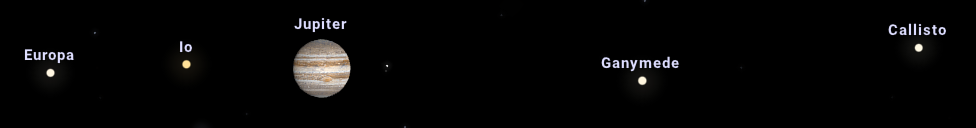} \\
37 & \includegraphics[width=\linewidth]{figs/g37.png} & \includegraphics[width=\linewidth]{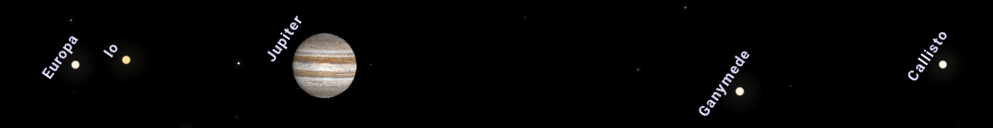} \\
38* & \includegraphics[width=\linewidth]{figs/g38.png} & \includegraphics[width=\linewidth]{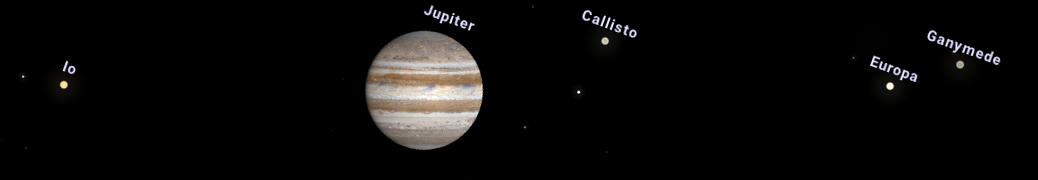} \\
39* & \includegraphics[width=\linewidth]{figs/g39.png} & \includegraphics[width=\linewidth]{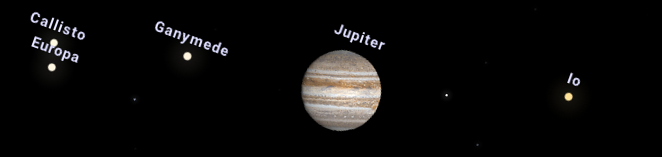} \\
40 & \includegraphics[width=\linewidth]{figs/g40.png} & \includegraphics[width=\linewidth]{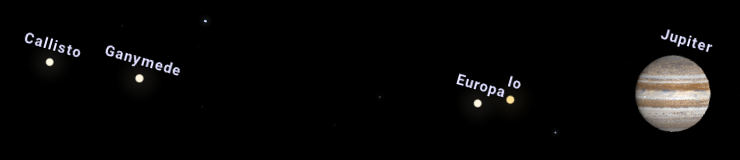} \\
41 & \includegraphics[width=\linewidth]{figs/g41.png} & \includegraphics[width=\linewidth]{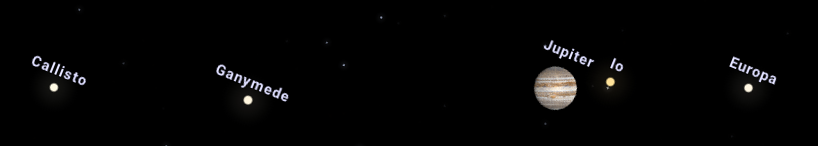} \\
42 & \includegraphics[width=\linewidth]{figs/g42.png} & \includegraphics[width=\linewidth]{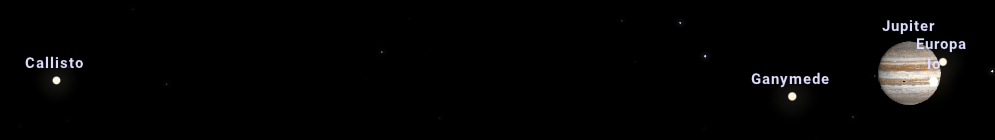} \\
43 & \includegraphics[width=\linewidth]{figs/g43.png} & \includegraphics[width=\linewidth]{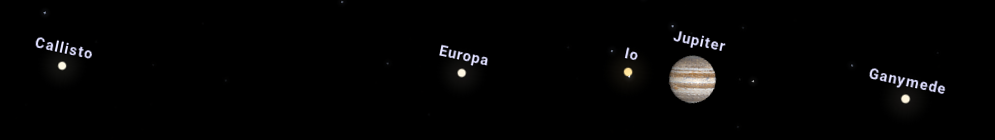} \\
44 & \includegraphics[width=\linewidth]{figs/g44.png} & \includegraphics[width=\linewidth]{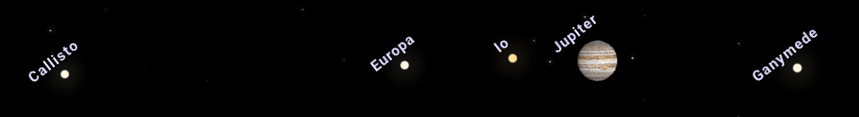} \\
45 & \includegraphics[width=\linewidth]{figs/g45.png} & \includegraphics[width=\linewidth]{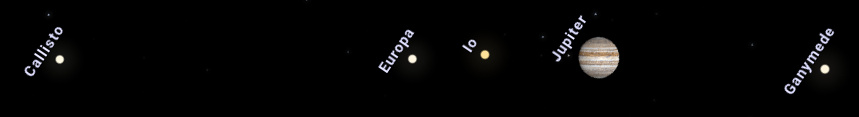} \\
46 & \includegraphics[width=\linewidth]{figs/g46.png} & \includegraphics[width=\linewidth]{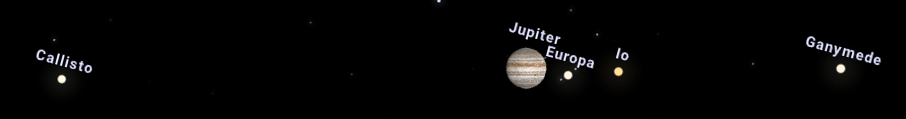} \\
47 & \includegraphics[width=\linewidth]{figs/g47.png} & \includegraphics[width=\linewidth]{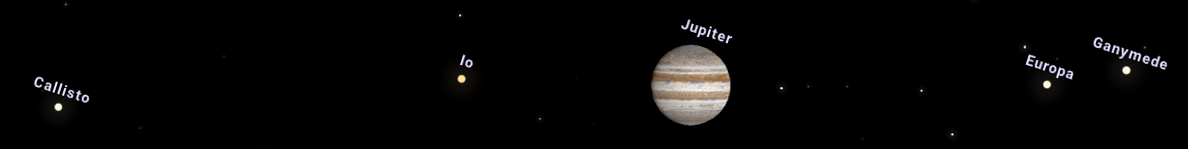} \\
48* & \includegraphics[width=\linewidth]{figs/g48.png} & \includegraphics[width=\linewidth]{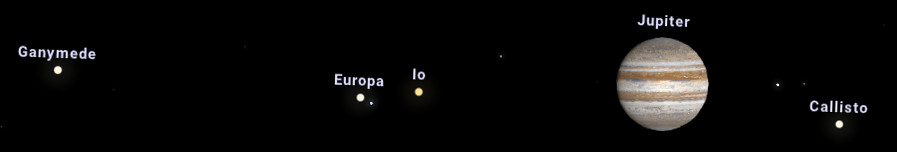} \\
49 & \includegraphics[width=\linewidth]{figs/g49.png} & \includegraphics[width=\linewidth]{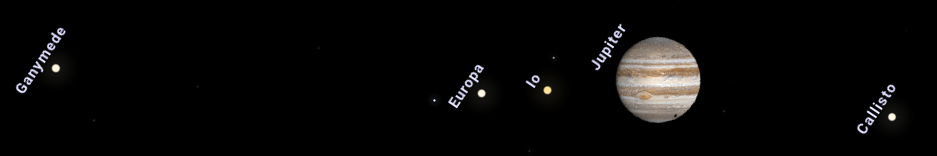} \\
50 & \includegraphics[width=\linewidth]{figs/g50.png} & \includegraphics[width=\linewidth]{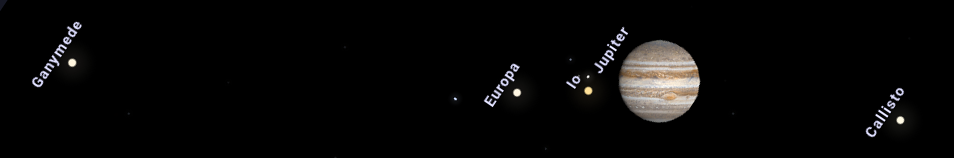} \\
51 & \includegraphics[width=\linewidth]{figs/g51.png} & \includegraphics[width=\linewidth]{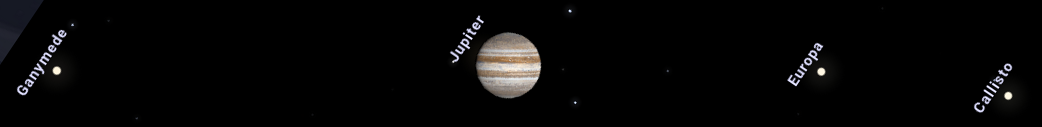} \\
52 & \includegraphics[width=\linewidth]{figs/g52.png} & \includegraphics[width=\linewidth]{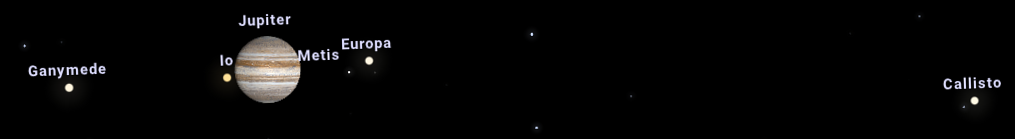} \\
53* & \includegraphics[width=\linewidth]{figs/g53.png} & \includegraphics[width=\linewidth]{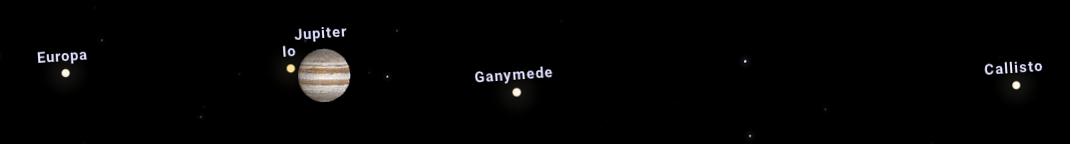} \\
54 & \includegraphics[width=\linewidth]{figs/g54.png} & \includegraphics[width=\linewidth]{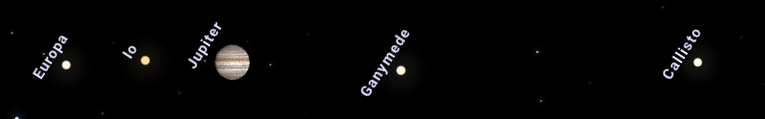} \\
55 & \includegraphics[width=\linewidth]{figs/g55.png} & \includegraphics[width=\linewidth]{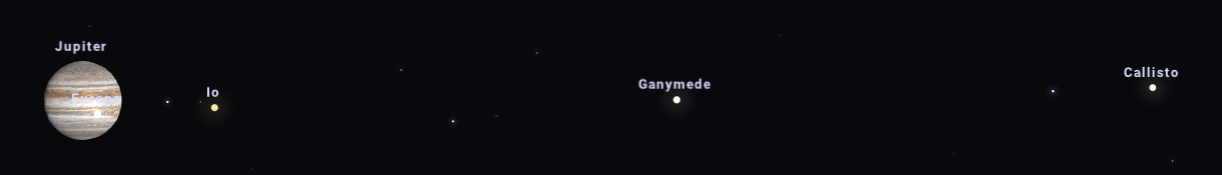} \\
56 & \includegraphics[width=\linewidth]{figs/g56.png} & \includegraphics[width=\linewidth]{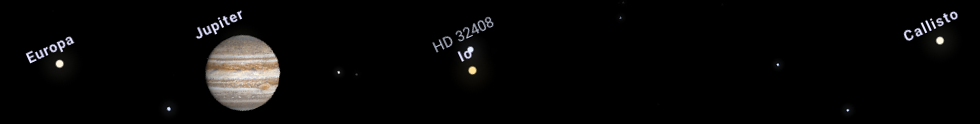} \\
57 & \includegraphics[width=\linewidth]{figs/g57.png} & \includegraphics[width=\linewidth]{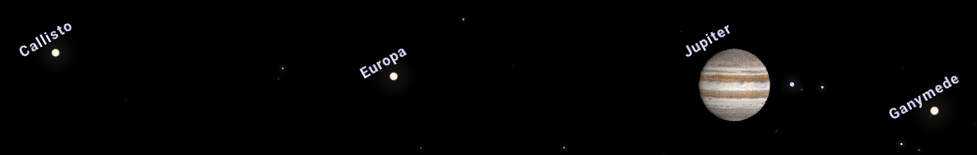} \\
58 & \includegraphics[width=\linewidth]{figs/g58.png} & \includegraphics[width=\linewidth]{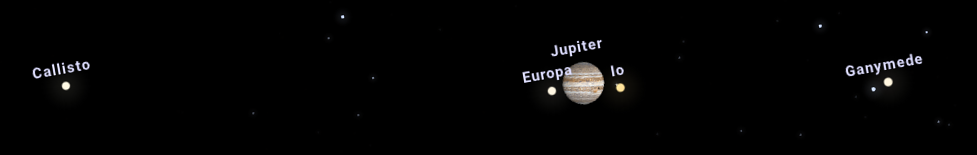} \\
59 & \includegraphics[width=\linewidth]{figs/g59.png} & \includegraphics[width=\linewidth]{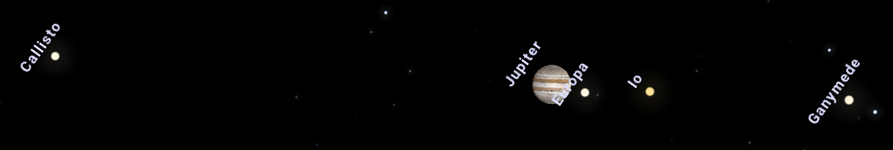} \\
60 & \includegraphics[width=\linewidth]{figs/g60.png} & \includegraphics[width=\linewidth]{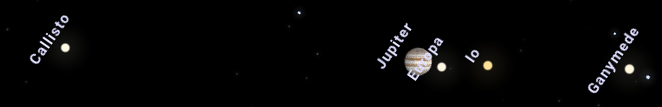} \\
61 & \includegraphics[width=\linewidth]{figs/g61.png} & \includegraphics[width=\linewidth]{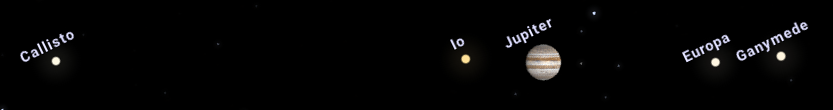} \\
62 & \includegraphics[width=\linewidth]{figs/g62.png} & \includegraphics[width=\linewidth]{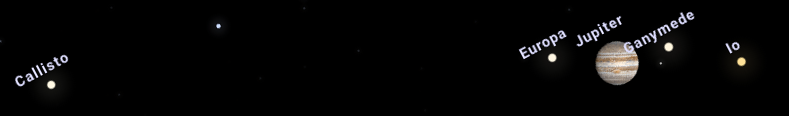} \\
63 & \includegraphics[width=\linewidth]{figs/g63.png} & \includegraphics[width=\linewidth]{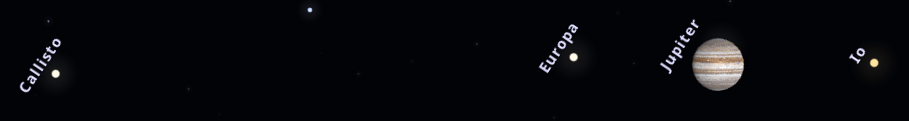} \\
64* & \includegraphics[width=\linewidth]{figs/g64.png} & \includegraphics[width=\linewidth]{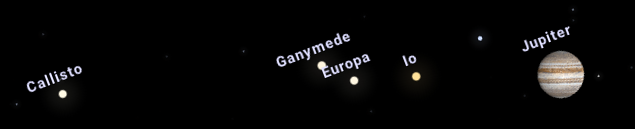} \\
\caption{The observations marked with a * are those for which the hour is not specified (assumed to be one hour after sunset).}
\label{tab:stellarium}
\end{longtable}
\end{center}

\section{Measurements without sketches}
\label{fullcomp_nosketch}
\begin{figure}[h!]
\centering
\includegraphics[width=0.4\linewidth]{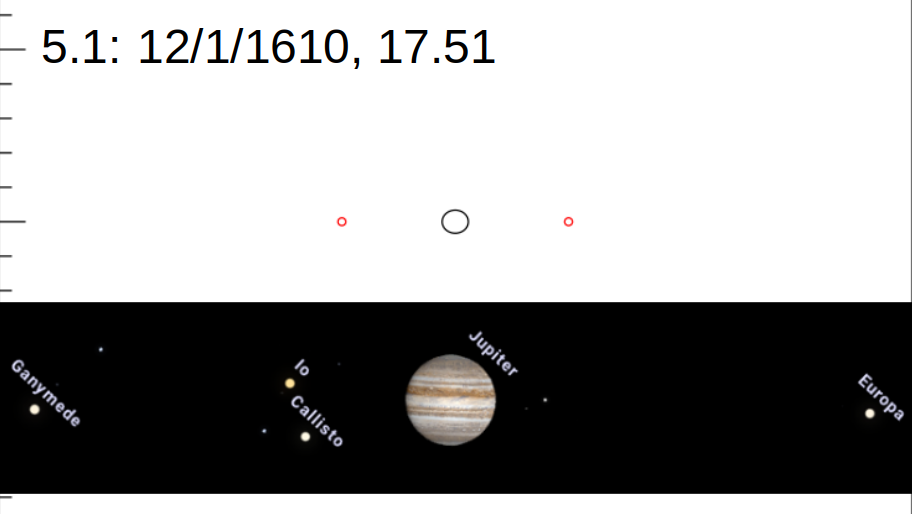}{~~~}%
\includegraphics[width=0.4\linewidth]{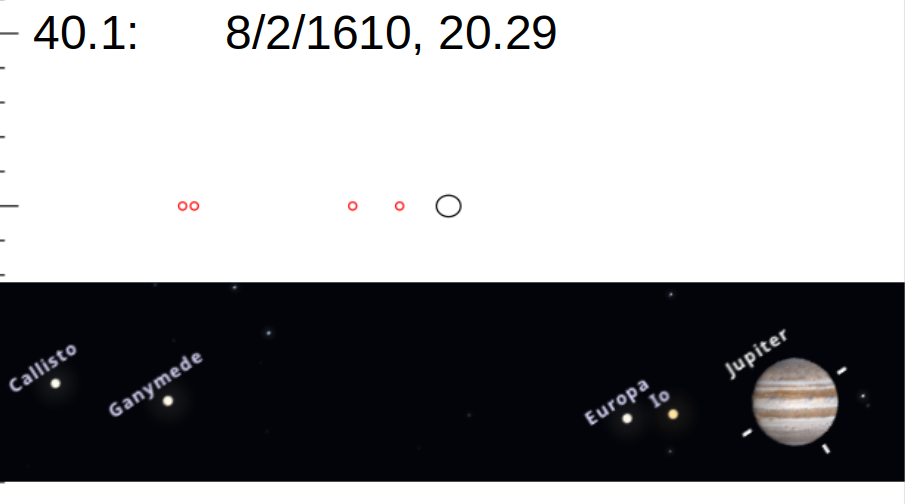}\\
\includegraphics[width=0.4\linewidth]{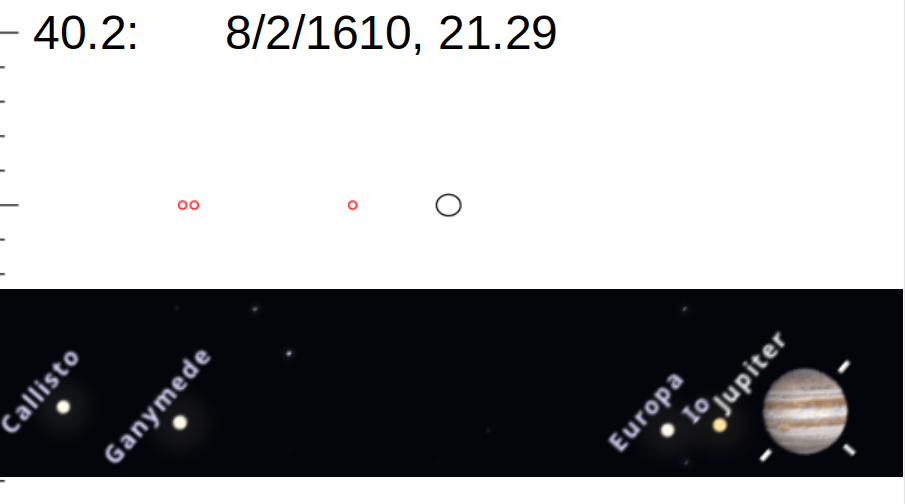}{~~~~}%
\includegraphics[width=0.4\linewidth]{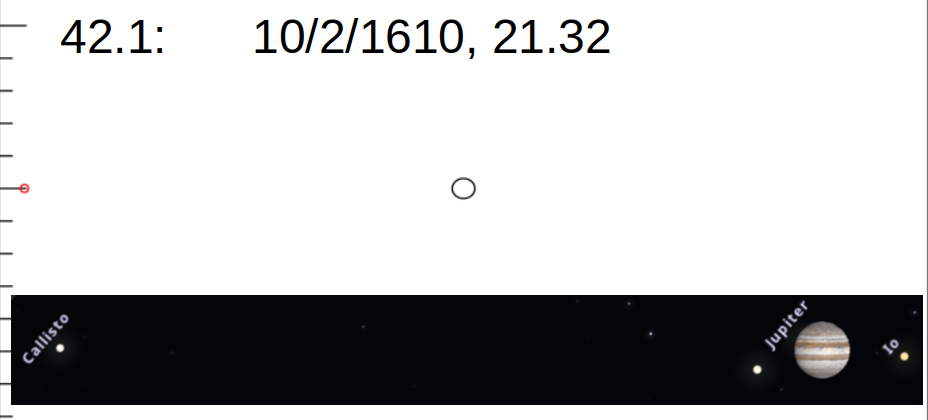}\\
\includegraphics[width=0.4\linewidth]{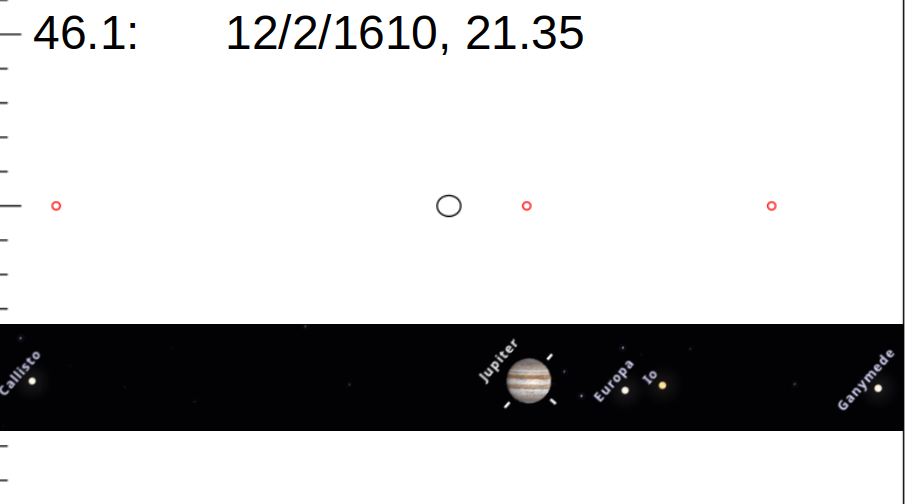}{~~~}%
\includegraphics[width=0.4\linewidth]{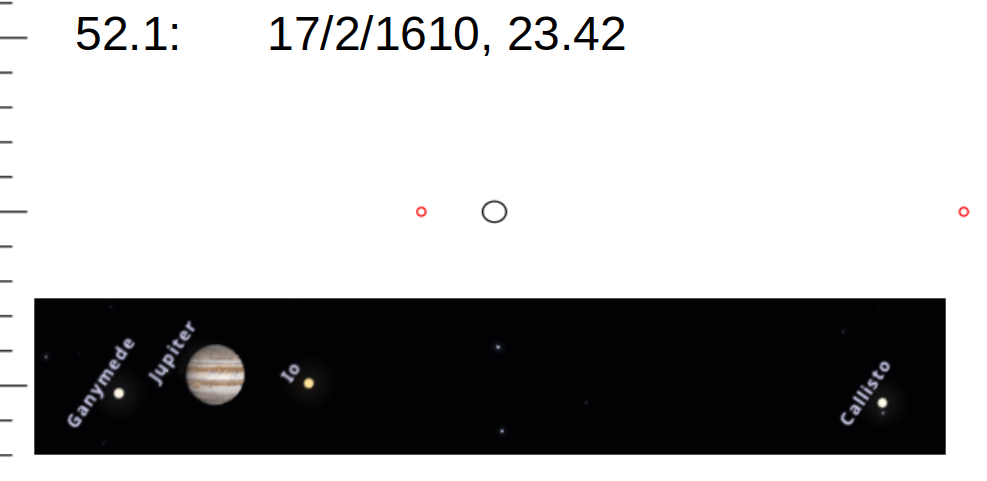}\\
\includegraphics[width=0.4\linewidth]{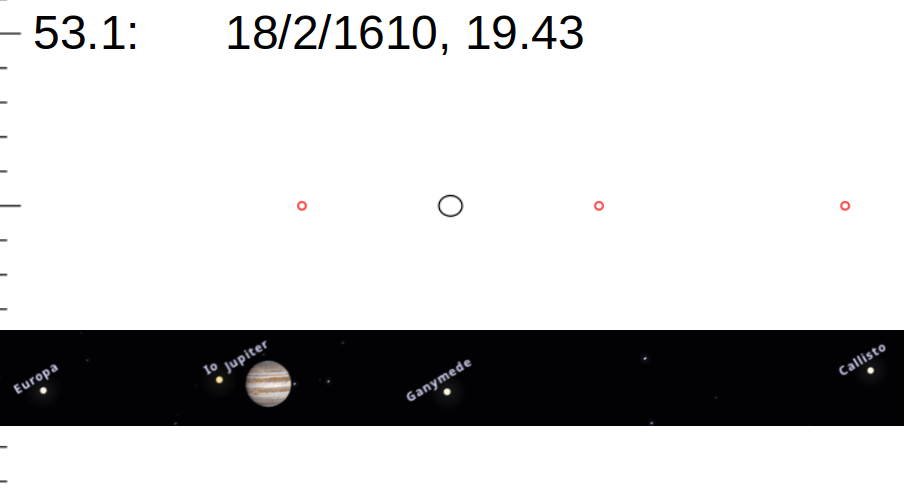}{~~~}%
\includegraphics[width=0.4\linewidth]{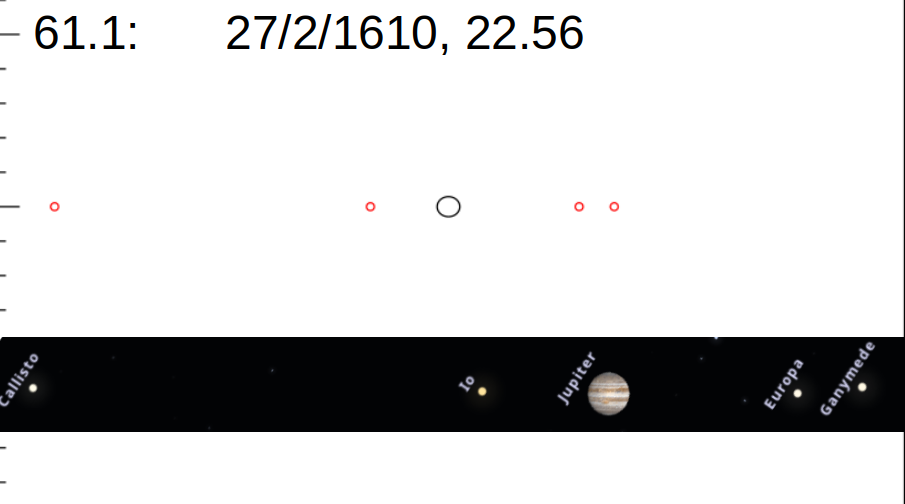}\\
\includegraphics[width=0.4\linewidth]{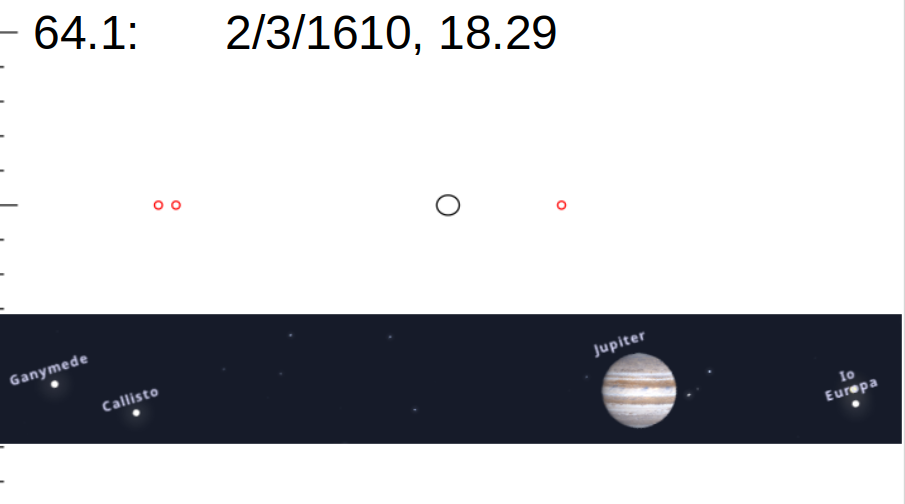}
\caption{\label{fig:nosketches} Comparison between the simulator and the observations with only angular data but no associated sketches. The top sketch is built from the angular separations by Galilei. Bottom image the prediction.}
\end{figure}
\section{Moon inkwashes}
\label{inkwashes}
\begin{figure}[h!]
\centering
\includegraphics[width=0.8\linewidth]{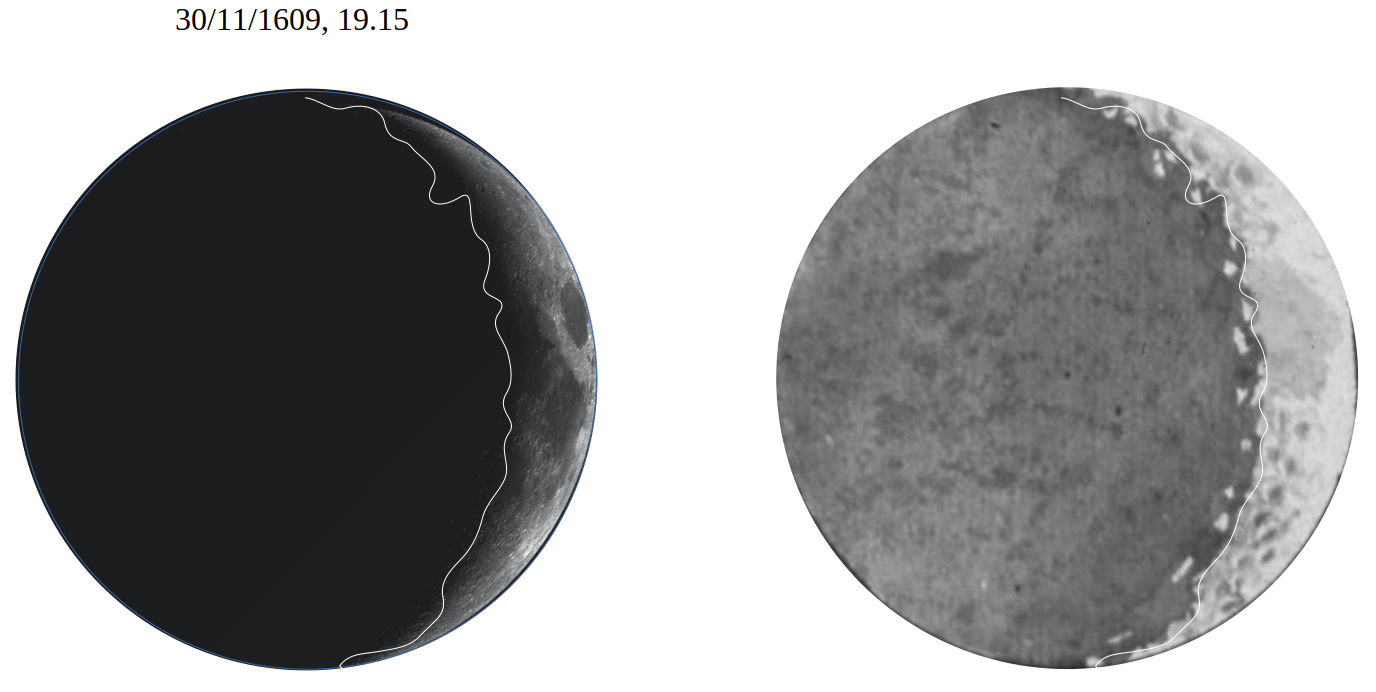}{~~~~}\\
\includegraphics[width=0.8\linewidth]{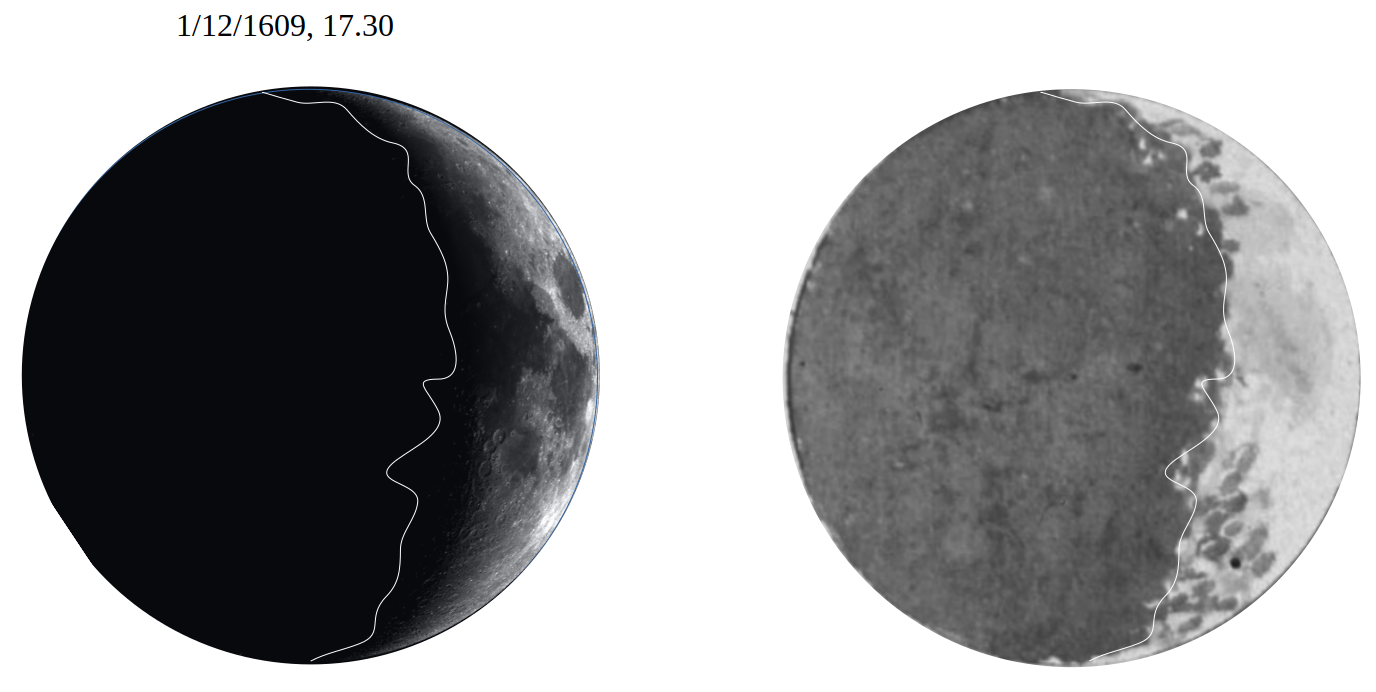}{~~~~}\\
\includegraphics[width=0.8\linewidth]{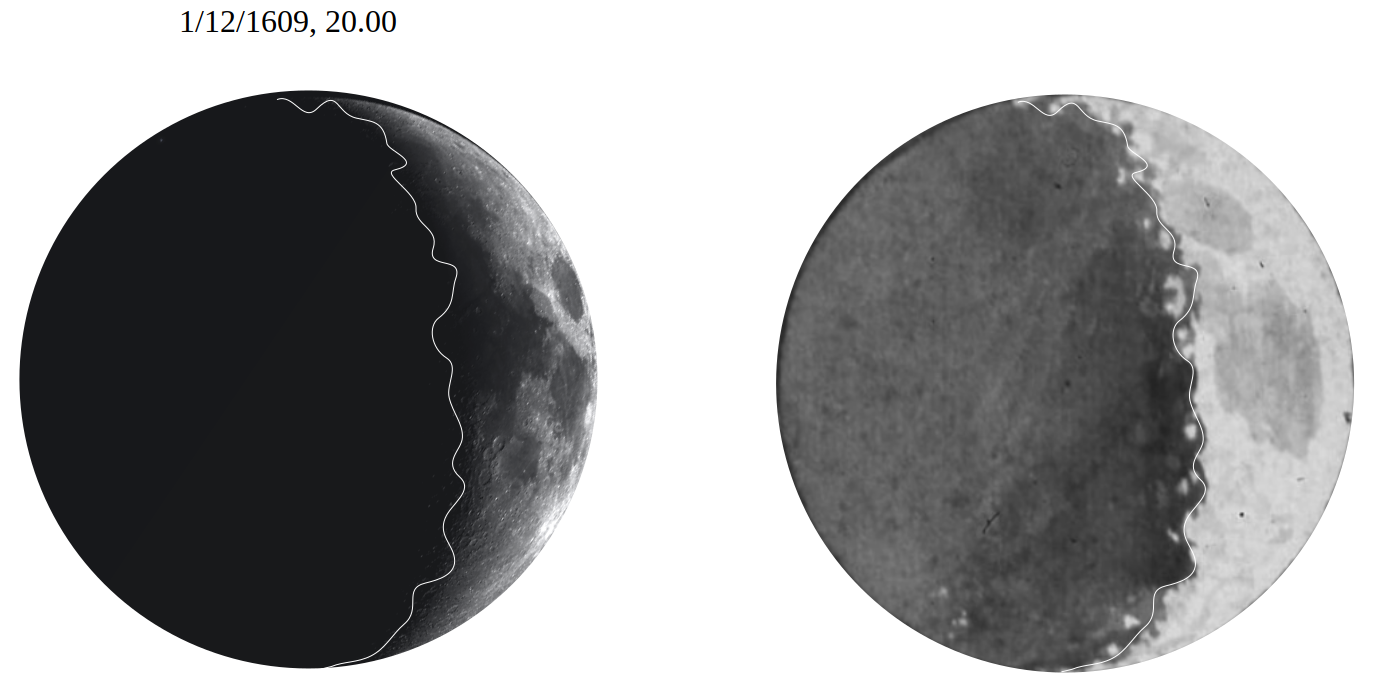}
\caption{\label{fig:moonA} Comparison of the Moon inkwashes with the prediction from the \textsc{Stellarium} simulator. The date is shown in the upper left side of the picture. The white line marks the terminator in the inkwash and it is superimposed to image from the simulation in the left to guide the eye in the comparison.}
\end{figure}

\begin{figure}[h!]
\centering
\includegraphics[width=0.8\linewidth]{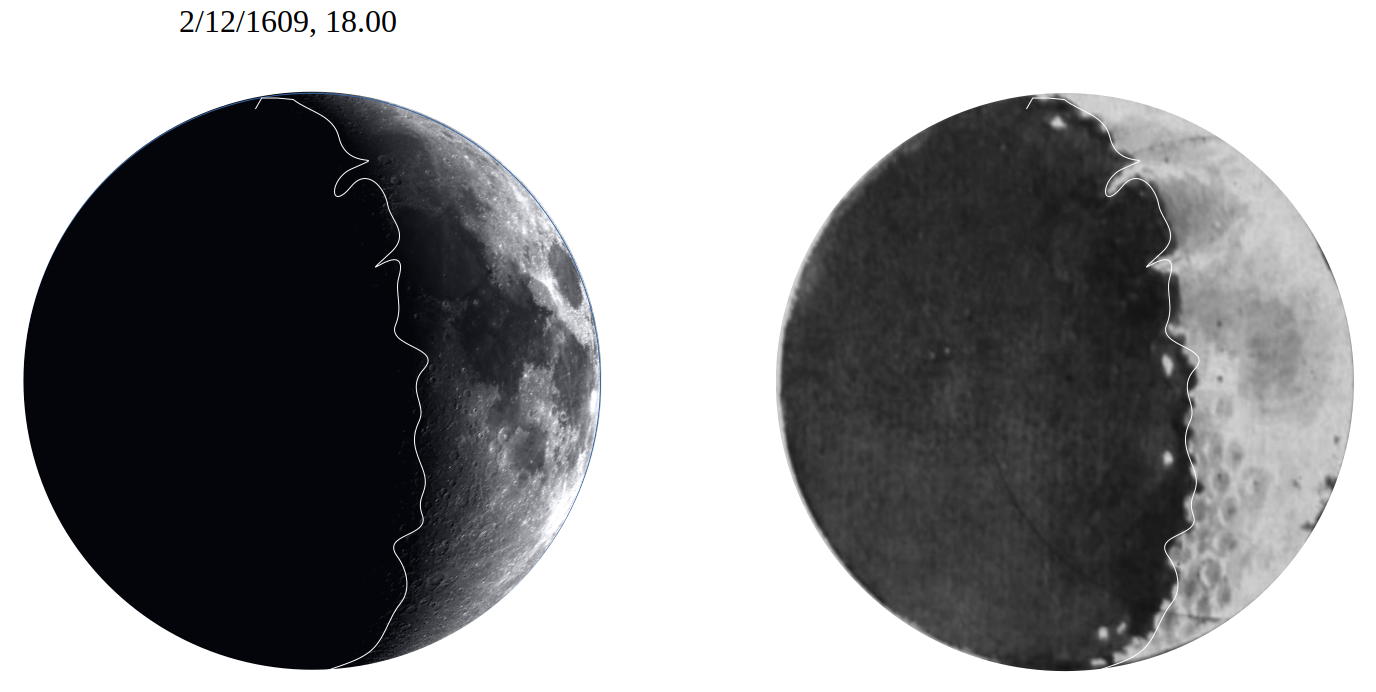}
{~~~~}\\
\includegraphics[width=0.8\linewidth]{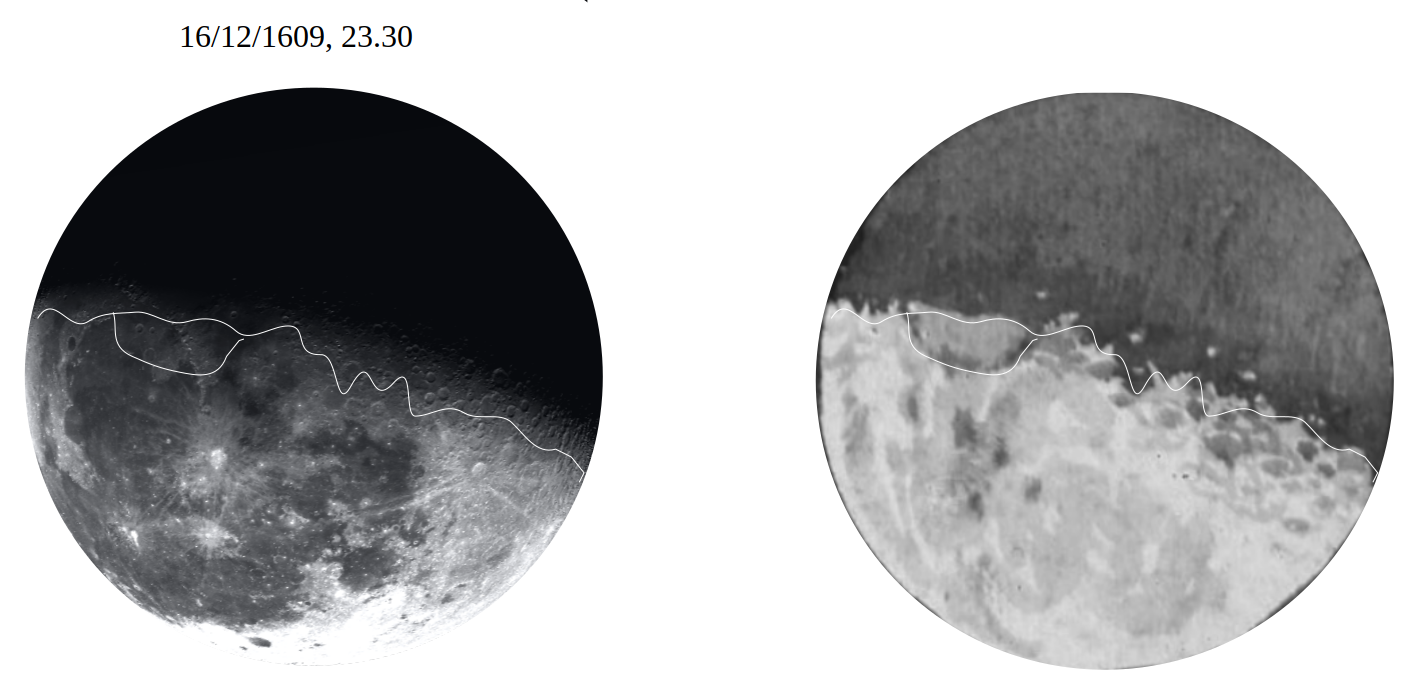}
{~~~~}\\
\includegraphics[width=0.8\linewidth]{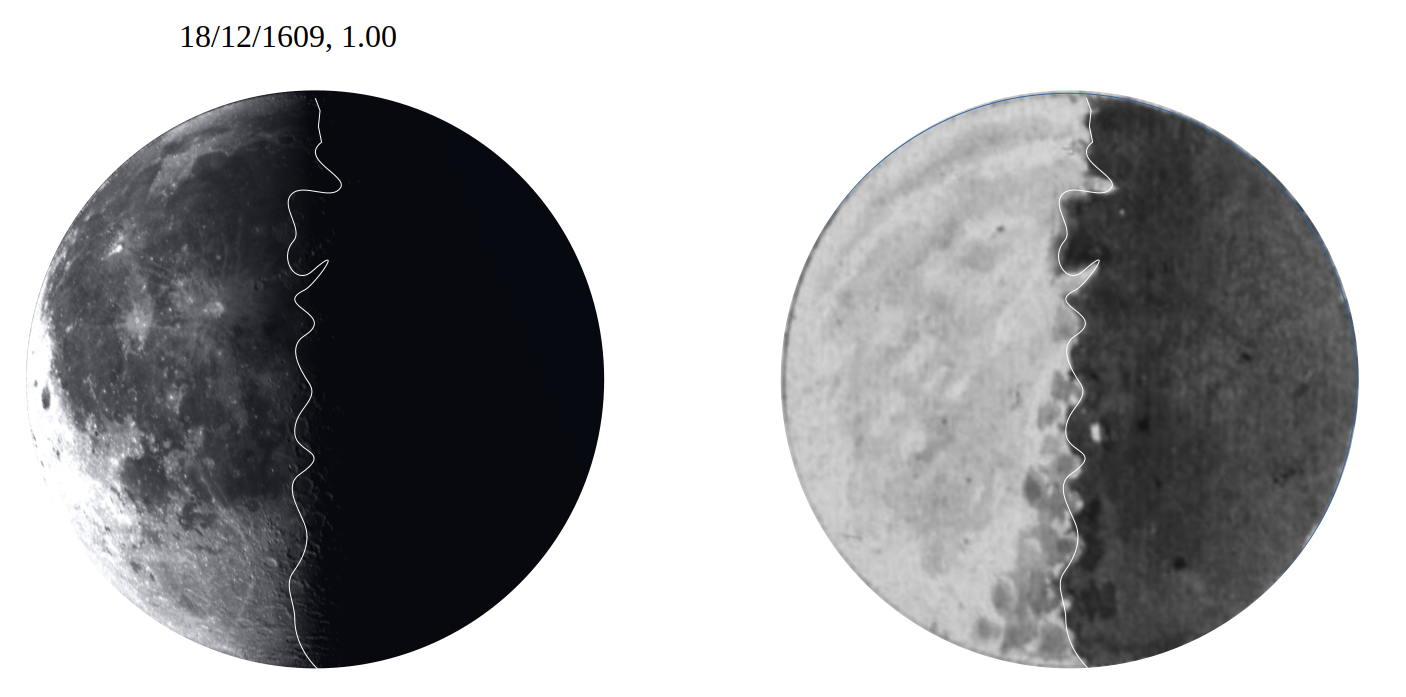}
\caption{\label{fig:moonB} Comparison of the Moon inkwashes with the prediction from the \textsc{Stellarium} simulator. The date is shown in the upper left side of the picture. The white line marks the terminator in the inkwash and it is superimposed to image from the simulation in the left to guide the eye in the comparison.}
\end{figure}

\begin{figure}[h!]
\centering
\includegraphics[width=0.9\linewidth]{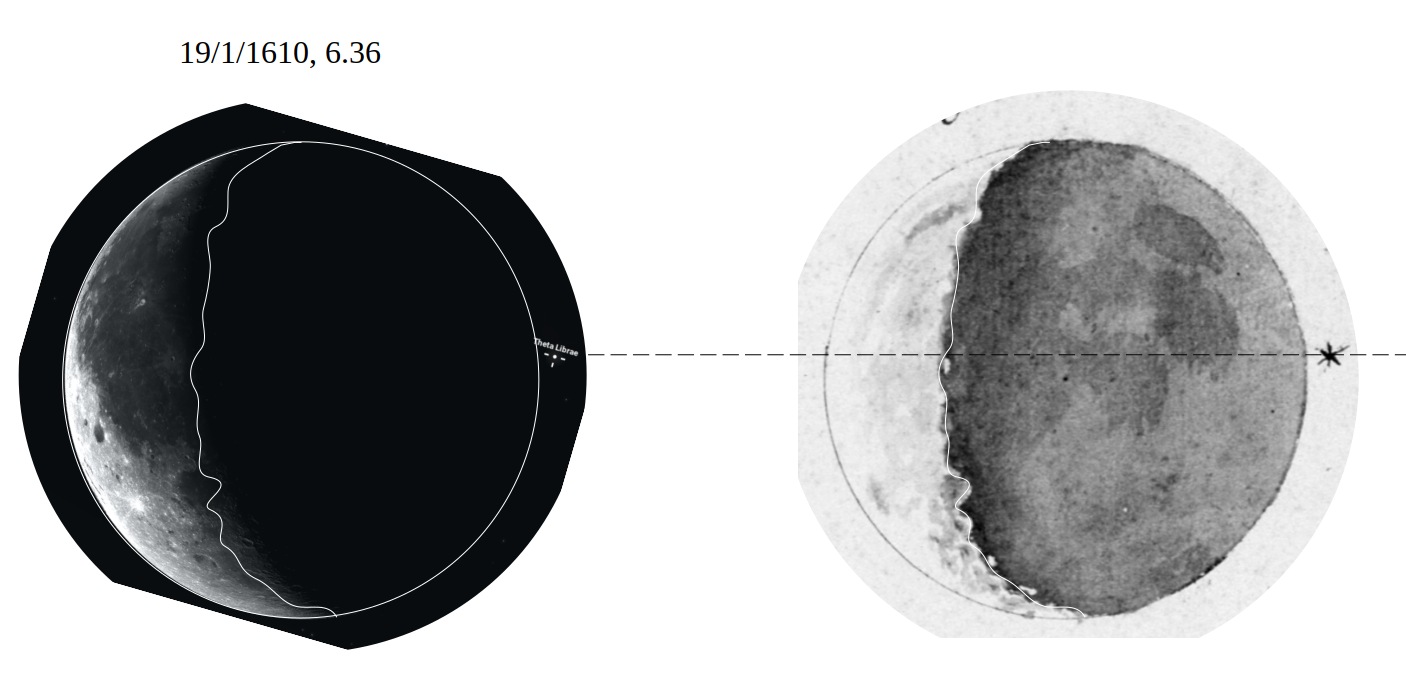}
\caption{\label{fig:moonC} Comparison of the Moon inkwashes with the prediction from the \textsc{Stellarium} simulator. The date is shown in the upper left side of the picture. The white line marks the terminator in the inkwash and it is superimposed to image from the simulation in the left to guide the eye in the comparison.}
\end{figure}



\section{Comparison of original hand-written notes and Sidereus Nuncius printed version}
\label{appendixE}
\begin{figure}[hbpt!]
\centering
  \includegraphics[width=\linewidth]{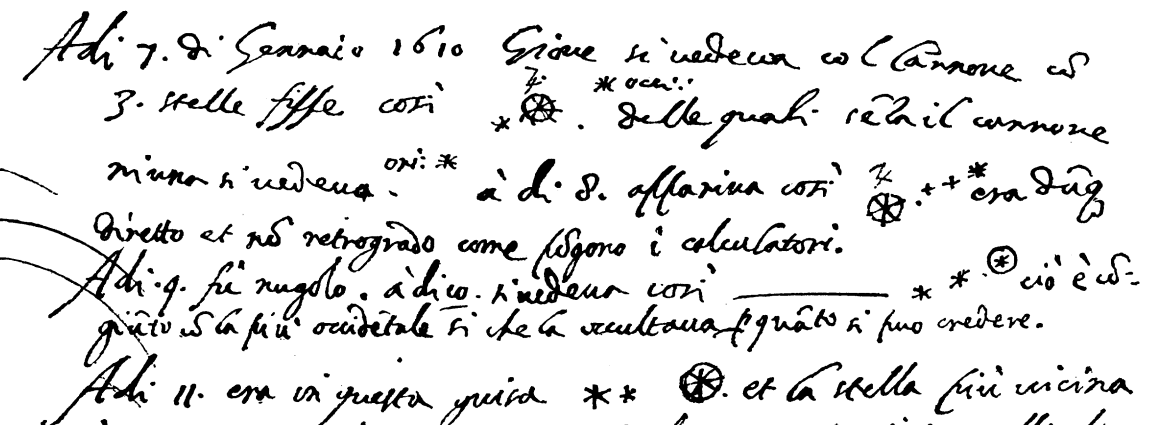}
  \caption{Example of the handwritten original notes.}
  \label{fig:exsketch}
\end{figure}

It is interesting to compare the relationship between the original
handwritten notes of Galilei and the printed version of the Sidereus
Nuncius. An example of the handwritten notes of the first observations
of January 1610 are shown in Fig.~\ref{fig:exsketch}. We took the
notes from \cite{GalileiOpere}, Volume 3 part II by Antonio Favaro in
pages 427-434, also available online for example at
\cite{GalileiOpereONLINE}.

It is interesting to notice that in the handwritten notes the Jupiter
system is drawn by reproducing its inclination with respect to the
horizon, i.e. in such a way that it is aligned with the bottom/top
sides of the page. This would allows to roughly guess when some
observations were taken even when the time was not reported in the
Sidereus Nuncius (we assumed 1 hour before sunset for those).  The
handwritten sketches of the satellites were cleaned from the
surrounding text and rotated to the horizontal plane as seen in
Tab.~\ref{tab:handw}.  To keep track of the original orientation, the
applied rotation angle is marked in the first column together with the
date.  The comparison is possible only until February 16$^{\rm{th}}$
(the last observation on page 434) as the observations until March 1st
are not available. From page 435 they restart from March 9.

As it can be seen, there is overall a good overlap between the notes
and the printed book. On the other hand it should be noted that the
information on the pattern of the satellites is only partially encoded
in the sketch. The full information is kept in the text where Galilei
either records in words the relative entity of the separation of the
satellites, or even their values in primes, starting from January
12$^{\rm{th}}$. This is also the reason why in the analysis we
directly considered (and compared) the angular recordings and the
diagrams from the printed version. In addition, considering the
handwritten sketches would be complicated by the fact the the size of
the Jupiter disk is rendered rather approximately and the scale of the
representation changes from case to case. In general there is a
one-to-one match between the sketches in the notes and the diagrams in
the Sidereus with the only exception of February 15$^{\rm{th}}$ when
the configurations corresponding to hour 5 and 6 are described in
words in the notes and represented in the Sidereus.

\setlength{\tabcolsep}{3pt} 

\renewcommand{\arraystretch}{1.05}

\begin{center}
\begin{longtable}{>{\raggedright\arraybackslash}p{3.1cm} c c}
\caption{Side-by-side images with progressive numbering}
\label{tab:handw}\\
\toprule
\textbf{No.} & \textbf{Notes} & \textbf{Sidereus Nuncius} \\
\midrule
\endfirsthead
\toprule
\textbf{No.} & \textbf{Notes} & \textbf{Sidereus Nuncius} \\
\midrule
\endhead
\midrule
\multicolumn{3}{r}{\small Continued on next page}\\
\bottomrule
\endfoot
\bottomrule
\endlastfoot
1) 1/7 1:00 (-24$^\circ$)   & \includegraphics[height=0.6cm]{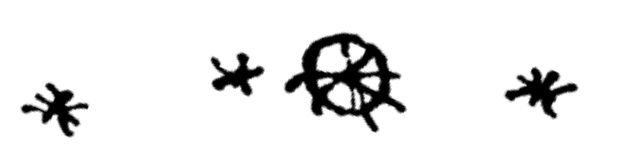} & \includegraphics[height=0.6cm]{figs/g00.png} \\
2*) 1/8 / (-20$^\circ$)     & \includegraphics[height=0.6cm]{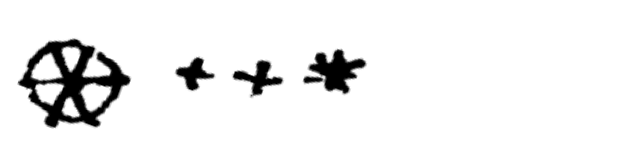} & \includegraphics[height=0.6cm]{figs/g01.png} \\
3*) 1/10 / (-20$^\circ$)    & \includegraphics[height=0.6cm]{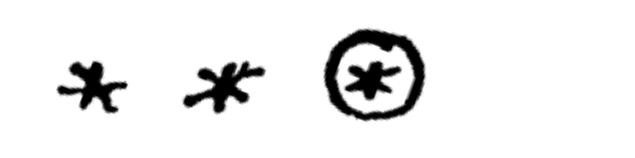} & \includegraphics[height=0.6cm]{figs/g03.png} \\
4*) 1/11 / (-5$^\circ$)     & \includegraphics[height=0.6cm]{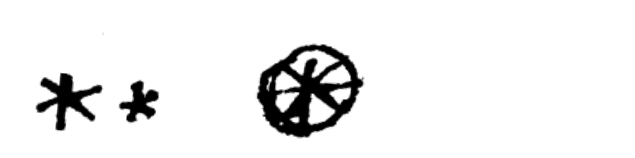} & \includegraphics[height=0.6cm]{figs/g04.png} \\
5) 1/12 1:00 (-12$^\circ$)  & \includegraphics[height=0.6cm]{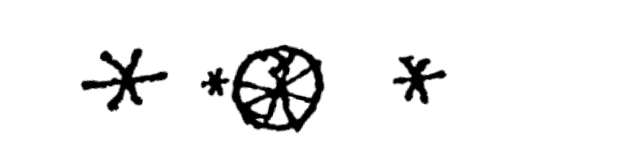} & \includegraphics[height=0.6cm]{figs/g05.png} \\
6*) 1/13 / (-12$^\circ$)    & \includegraphics[height=0.6cm]{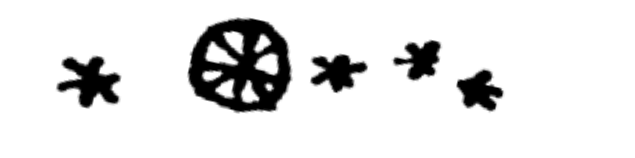} & \includegraphics[height=0.6cm]{figs/g06.png} \\
7) 1/15 3:00 (0$^\circ$)    & \includegraphics[height=0.6cm]{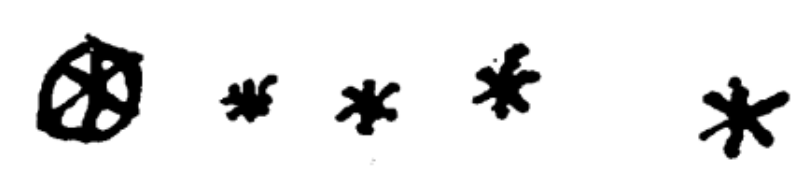} & \includegraphics[height=0.6cm]{figs/g07.png} \\
8) 1/15 7:00 (15$^\circ$)   & \includegraphics[height=0.6cm]{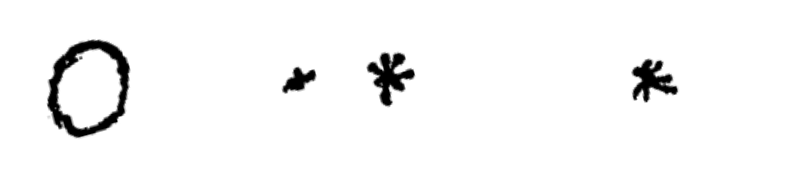} & \includegraphics[height=0.6cm]{figs/g08.png} \\
9) 1/16 1:00 (-13$^\circ$)  & \includegraphics[height=0.6cm]{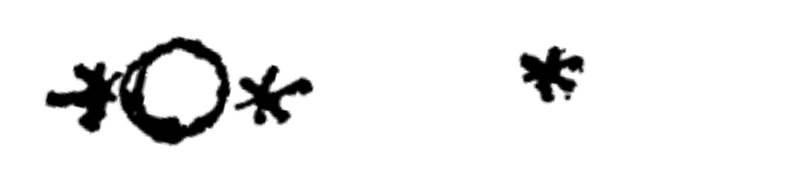} & \includegraphics[height=0.6cm]{figs/g09.png} \\
10) 1/17 0:30 (-15$^\circ$) & \includegraphics[height=0.6cm]{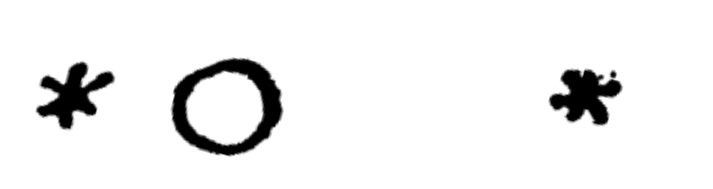} & \includegraphics[height=0.6cm]{figs/g10.png} \\
11) 1/17 5:00 (30$^\circ$)  & \includegraphics[height=0.6cm]{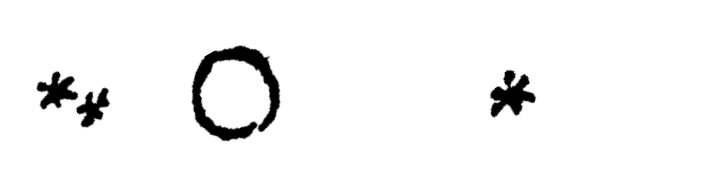} & \includegraphics[height=0.6cm]{figs/g11.png} \\
12) 1/18 0:20 (-22$^\circ$) & \includegraphics[height=0.6cm]{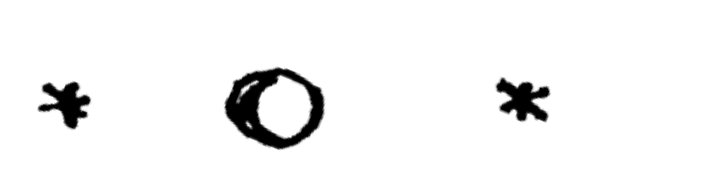} & \includegraphics[height=0.6cm]{figs/g12.png} \\
13) 1/19 2:00 (-10$^{\circ}$)& \includegraphics[height=0.7cm,angle=0]{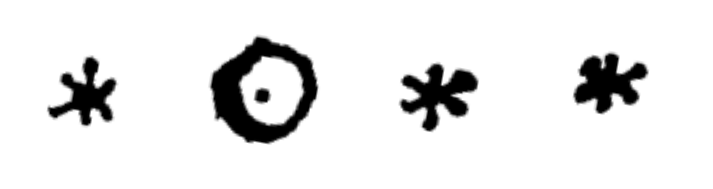} & \includegraphics[height=0.7cm]{figs/g13.png} \\
14) 1/19 5:00 (6$^{\circ}$)& \includegraphics[height=0.7cm,angle=0]{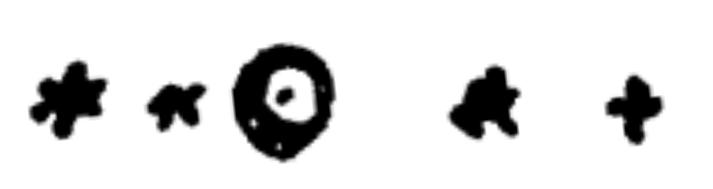} & \includegraphics[height=0.7cm]{figs/g14.png} \\
15) 1/20 1:15 (0$^{\circ}$)& \includegraphics[height=0.7cm]{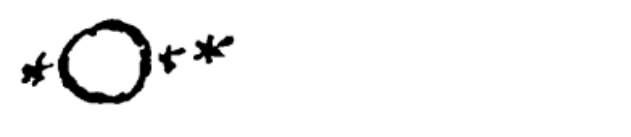} & \includegraphics[height=0.7cm]{figs/g15.png} \\
16) 1/20 6:00 (22$^{\circ}$)& \includegraphics[height=0.7cm,angle=0]{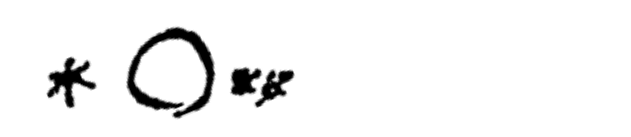} & \includegraphics[height=0.7cm]{figs/g16.png} \\
17) 1/20 7:00 (55$^{\circ}$)& \includegraphics[height=0.7cm,angle=0]{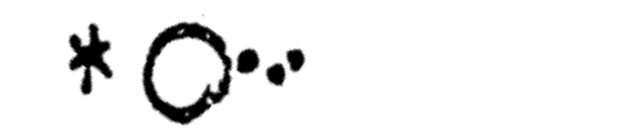} & \includegraphics[height=0.7cm]{figs/g17.png} \\
18) 1/21 0:30 (-15$^{\circ}$)& \includegraphics[height=0.7cm,angle=0]{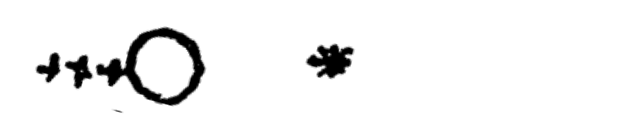} & \includegraphics[height=0.7cm]{figs/g18.png} \\
19) 1/22 2:00 (-5$^{\circ}$)& \includegraphics[height=0.7cm,angle=0]{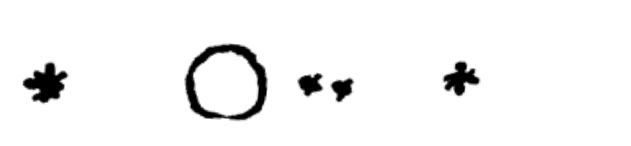} & \includegraphics[height=0.7cm]{figs/g19.png} \\
20) 1/22 6:00 (42$^{\circ}$)& \includegraphics[height=0.7cm,angle=0]{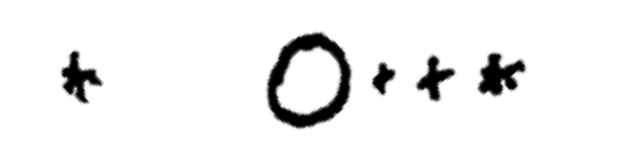} & \includegraphics[height=0.7cm]{figs/g20.png} \\
21) 1/23 0:40 (-19$^{\circ}$)& \includegraphics[height=0.7cm,angle=0]{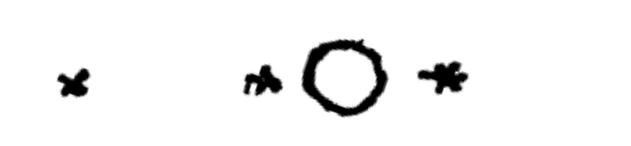} & \includegraphics[height=0.7cm]{figs/g21.png} \\
22) 1/23 5:00 (28$^\circ$)&\includegraphics[height=0.7cm,angle=0]{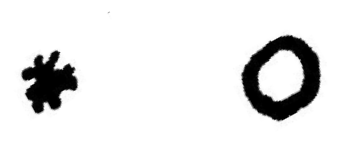} & \includegraphics[height=0.7cm]{figs/g22.png} \\
23*) 1/24 1:00 (21$^{\circ}$)& \includegraphics[height=0.7cm,angle=0]{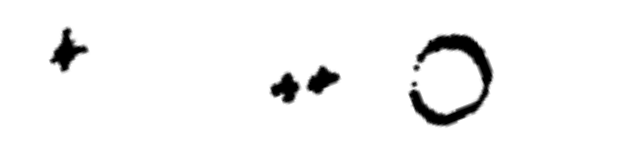} & \includegraphics[height=0.7cm]{figs/g23.png} \\
24) 1/24 6:00 (23$^{\circ}$)& \includegraphics[height=0.7cm,angle=0]{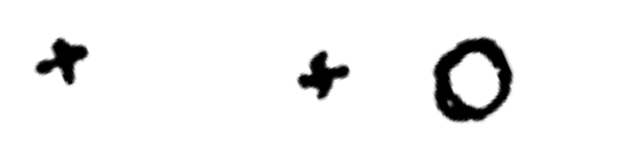} & \includegraphics[height=0.7cm]{figs/g24.png} \\
25) 1/25 1:40 (-25$^{\circ}$)& \includegraphics[height=0.7cm,angle=0]{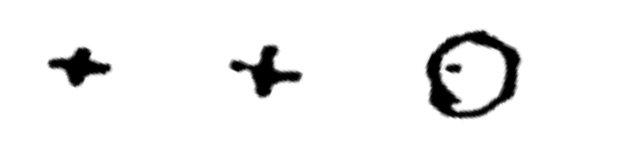} & \includegraphics[height=0.7cm]{figs/g25.png} \\
26) 1/26 0:40 (-30$^{\circ}$)& \includegraphics[height=0.7cm,angle=0]{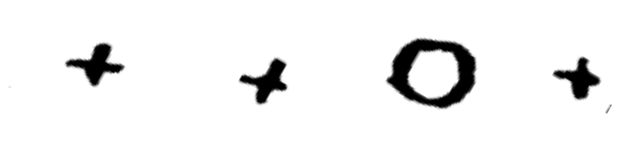} & \includegraphics[height=0.7cm]{figs/g26.png} \\
27) 1/26 5:00 (25$^{\circ}$)& \includegraphics[height=0.7cm,angle=0]{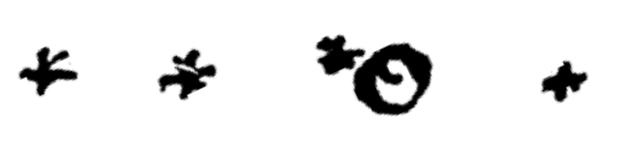} & \includegraphics[height=0.7cm]{figs/g27.png} \\
28) 1/27 1:00 (-15$^{\circ}$)& \includegraphics[height=0.7cm,angle=0]{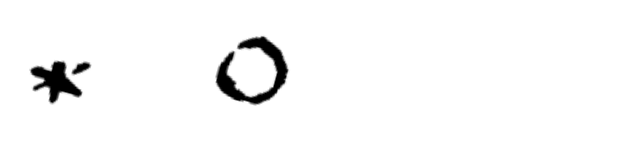} & \includegraphics[height=0.7cm]{figs/g28.png} \\
29) 1/30 1:00 (-8$^{\circ}$)& \includegraphics[height=0.7cm,angle=0]{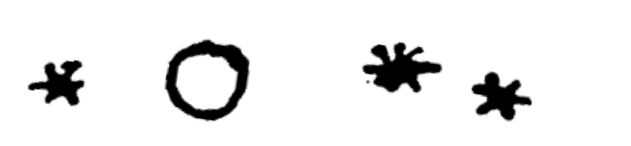} & \includegraphics[height=0.7cm]{figs/g29.png} \\
30) 1/31 2:00 (0$^{\circ}$)& \includegraphics[height=0.7cm,angle=0]{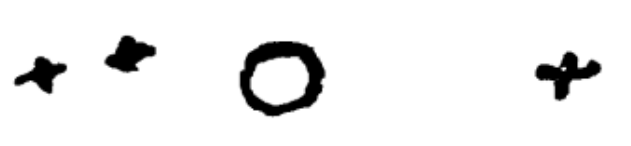} & \includegraphics[height=0.7cm]{figs/g30.png} \\
31) 1/31 4:00 (18$^{\circ}$)& \includegraphics[height=0.7cm,angle=0]{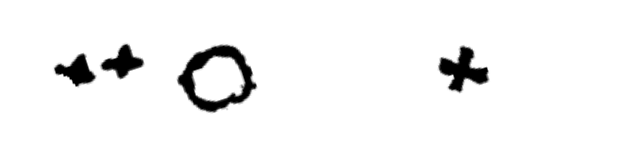} & \includegraphics[height=0.7cm]{figs/g31.png} \\
32) 2/1 2:00 (-30$^{\circ}$)& \includegraphics[height=0.7cm,angle=0]{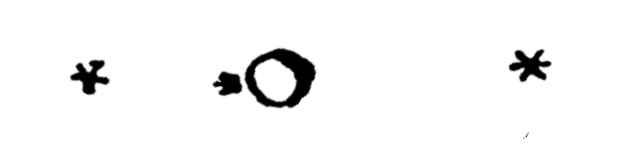} & \includegraphics[height=0.7cm]{figs/g32.png} \\
33*) 2/2 / (-13$^{\circ}$)& \includegraphics[height=0.7cm,angle=0]{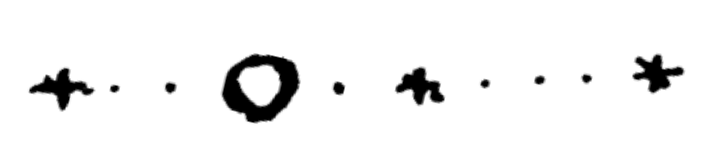} & \includegraphics[height=0.7cm]{figs/g33.png} \\
34) 2/2 7:00 (35$^{\circ}$)& \includegraphics[height=0.7cm,angle=0]{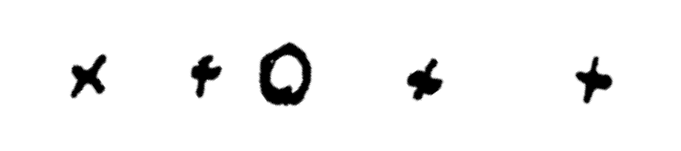} & \includegraphics[height=0.7cm]{figs/g34.png} \\
35) 2/3 7:00 (35$^{\circ}$)& \includegraphics[height=0.7cm,angle=0]{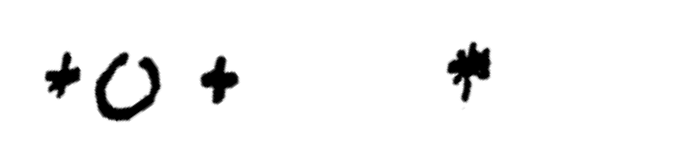} & \includegraphics[height=0.7cm]{figs/g35.png} \\
36) 2/4 2:00 (-2$^{\circ}$)& \includegraphics[height=0.7cm,angle=0]{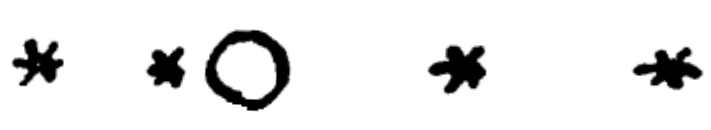} & \includegraphics[height=0.7cm]{figs/g36.png} \\
37) 2/4 7:00 (30$^{\circ}$)& \includegraphics[height=0.7cm,angle=0]{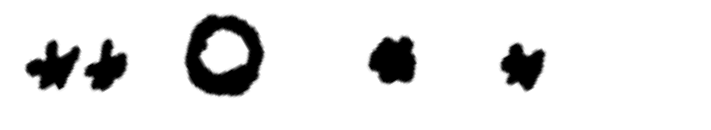} & \includegraphics[height=0.7cm]{figs/g37.png} \\
38*) 2/6 / (4$^{\circ}$)& \includegraphics[height=0.7cm,angle=0]{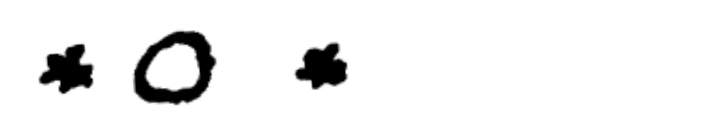} & \includegraphics[height=0.7cm]{figs/g38.png} \\
39*) 2/7 / (-5$^{\circ}$)& \includegraphics[height=0.7cm,angle=0]{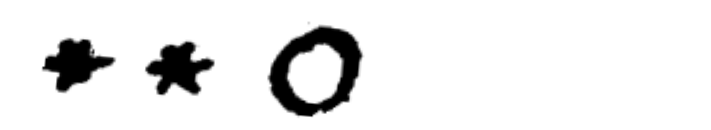} & \includegraphics[height=0.7cm]{figs/g39.png} \\
40) 2/8 1:00 (-22$^{\circ}$)& \includegraphics[height=0.7cm,angle=0]{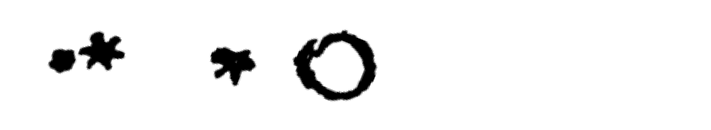} & \includegraphics[height=0.7cm]{figs/g40.png} \\
41) 2/9 0:30 (0$^{\circ}$)& \includegraphics[height=0.7cm,angle=0]{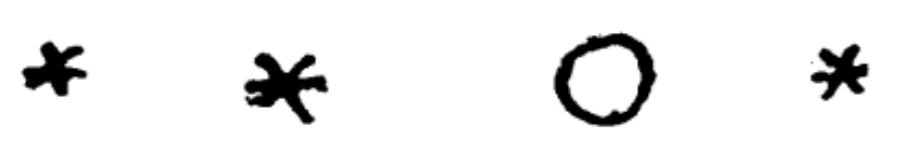} & \includegraphics[height=0.7cm]{figs/g41.png} \\
42) 2/10 1:30 (-5$^{\circ}$)& \includegraphics[height=0.7cm,angle=0]{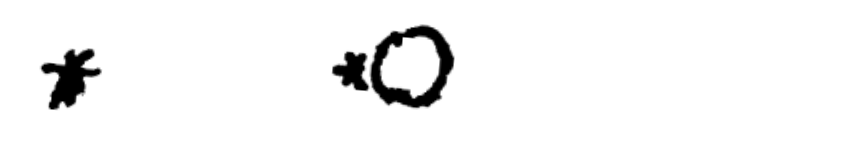} & \includegraphics[height=0.7cm]{figs/g42.png} \\
43) 2/11 1:00 (-10$^{\circ}$)& \includegraphics[height=0.7cm,angle=0]{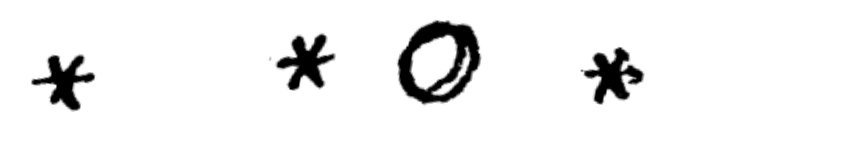} & \includegraphics[height=0.7cm]{figs/g43.png} \\
44) 2/11 3:00 (15$^{\circ}$)& \includegraphics[height=0.7cm,angle=0]{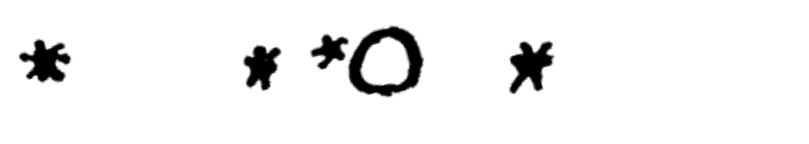} & \includegraphics[height=0.7cm]{figs/g44.png} \\
45) 2/11 5:30 (25$^{\circ}$)& \includegraphics[height=0.7cm,angle=0]{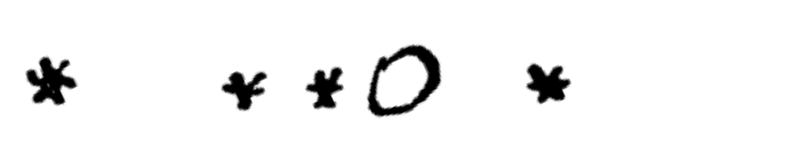} & \includegraphics[height=0.7cm]{figs/g45.png} \\
46) 2/12 0:40 (-9$^{\circ}$)& \includegraphics[height=0.7cm,angle=0]{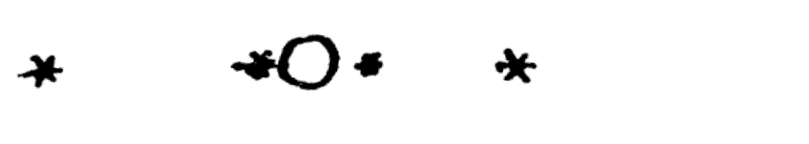} & \includegraphics[height=0.7cm]{figs/g46.png} \\
47) 2/13 0:30 (0$^{\circ}$)& \includegraphics[height=0.7cm,angle=0]{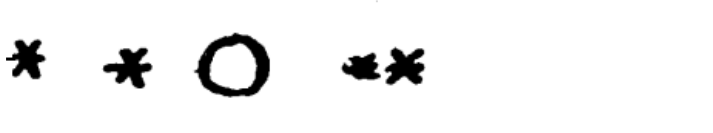} & \includegraphics[height=0.7cm]{figs/g47.png} \\
48*) 2/15 / (0$^{\circ}$)& \includegraphics[height=0.7cm,angle=0]{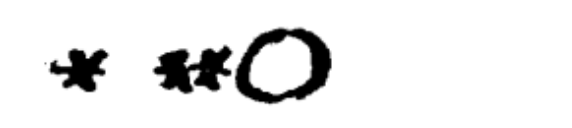} & \includegraphics[height=0.7cm]{figs/g48.png} \\
49) 2/15 5:00 (0$^{\circ}$)& in words& \includegraphics[height=0.7cm]{figs/g49.png} \\
50) 2/15 6:00 (0$^{\circ}$)& in words& \includegraphics[height=0.7cm]{figs/g50.png} \\
51) 2/16 6:00 (2$^{\circ}$)& \includegraphics[height=0.7cm,angle=0]{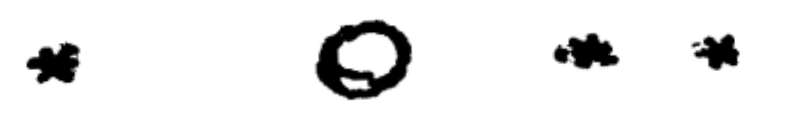} & \includegraphics[height=0.7cm]{figs/g51.png} \\
\end{longtable}
\end{center}
\section{Conversion from Italic hours}
\label{sec:conversion}
\begin{table}[b]
\centering
\resizebox{\textwidth}{!}{%
\begin{tabular}{r r r r r r | r r r r r r}
\toprule
\multicolumn{6}{c}{January 1610} & \multicolumn{6}{c}{February–March 1610} \\
\cmidrule(lr){1-6} \cmidrule(lr){7-12}
id & mo. & day & italic (h) & sunset & modern &
id & mo. & day & italic (h) & sunset & modern \\
\midrule
1  & 1 &  7 & 1:00 & 16.74 & 17:44 & 32 & 2 &  1 & 2:00 & 17.33 & 19:19 \\
2  & 1 &  8 & 1:00 & 16.77 & 17:46 & 33 & 2 &  2 & 1:00 & 17.36 & 18:21 \\
3  & 1 & 10 & 1:00 & 16.81 & 17:48 & 34 & 2 &  2 & 7:00 & 17.36 & 24:21 \\
4  & 1 & 11 & 1:00 & 16.84 & 17:50 & 35 & 2 &  3 & 7:00 & 17.38 & 24:22 \\
5  & 1 & 12 & 1:00 & 16.86 & 17:51 & 36 & 2 &  4 & 2:00 & 17.40 & 19:24 \\
6  & 1 & 13 & 1:00 & 16.89 & 17:53 & 37 & 2 &  4 & 7:00 & 17.40 & 24:24 \\
7  & 1 & 15 & 3:00 & 16.93 & 19:55 & 38 & 2 &  6 & 1:00 & 17.45 & 18:26 \\
8  & 1 & 15 & 7:00 & 16.93 & 23:55 & 39 & 2 &  7 & 1:00 & 17.47 & 18:28 \\
9  & 1 & 16 & 1:00 & 16.96 & 17:57 & 40 & 2 &  8 & 1:00 & 17.50 & 18:29 \\
10 & 1 & 17 & 0:30 & 16.98 & 17:28 & 41 & 2 &  9 & 0:30 & 17.52 & 18:01 \\
11 & 1 & 17 & 5:00 & 16.98 & 21:58 & 42 & 2 & 10 & 1:30 & 17.54 & 19:02 \\
12 & 1 & 18 & 0:20 & 17.00 & 17:20 & 43 & 2 & 11 & 1:00 & 17.57 & 18:34 \\
13 & 1 & 19 & 2:00 & 17.03 & 19:01 & 44 & 2 & 11 & 3:00 & 17.57 & 20:34 \\
14 & 1 & 19 & 5:00 & 17.03 & 22:01 & 45 & 2 & 11 & 5:30 & 17.57 & 23:04 \\
15 & 1 & 20 & 1:15 & 17.05 & 18:17 & 46 & 2 & 12 & 0:40 & 17.59 & 18:15 \\
16 & 1 & 20 & 6:00 & 17.05 & 23:02 & 47 & 2 & 13 & 0:30 & 17.61 & 18:06 \\
17 & 1 & 20 & 7:00 & 17.05 & 24:02 & 48 & 2 & 15 & 1:00 & 17.66 & 18:39 \\
18 & 1 & 21 & 0:30 & 17.07 & 17:34 & 49 & 2 & 15 & 5:00 & 17.66 & 22:39 \\
19 & 1 & 22 & 2:00 & 17.10 & 19:05 & 50 & 2 & 15 & 6:00 & 17.66 & 23:39 \\
20 & 1 & 22 & 6:00 & 17.10 & 23:05 & 51 & 2 & 16 & 6:00 & 17.68 & 23:41 \\
21 & 1 & 23 & 0:40 & 17.12 & 17:47 & 52 & 2 & 17 & 1:00 & 17.71 & 18:42 \\
22 & 1 & 23 & 5:00 & 17.12 & 22:07 & 53 & 2 & 18 & 1:00 & 17.73 & 18:43 \\
23 & 1 & 24 & 1:00 & 17.14 & 18:08 & 54 & 2 & 18 & 6:00 & 17.73 & 23:43 \\
24 & 1 & 24 & 6:00 & 17.14 & 23:08 & 55 & 2 & 19 & 0:40 & 17.76 & 18:25 \\
25 & 1 & 25 & 1:40 & 17.17 & 18:50 & 56 & 2 & 21 & 1:30 & 17.80 & 19:18 \\
26 & 1 & 26 & 0:40 & 17.19 & 17:51 & 57 & 2 & 25 & 1:30 & 17.90 & 19:23 \\
27 & 1 & 26 & 5:00 & 17.19 & 22:11 & 58 & 2 & 26 & 0:30 & 17.92 & 18:25 \\
28 & 1 & 27 & 1:00 & 17.21 & 18:12 & 59 & 2 & 26 & 5:00 & 17.92 & 22:55 \\
29 & 1 & 30 & 1:00 & 17.29 & 18:17 & 60 & 2 & 26 & 5:00 & 17.92 & 22:55 \\
30 & 1 & 31 & 2:00 & 17.31 & 19:18 & 61 & 2 & 27 & 1:04 & 17.94 & 19:00 \\
31 & 1 & 31 & 4:00 & 17.31 & 21:18 & 62 & 2 & 28 & 1:00 & 17.97 & 18:58 \\
   &   &    &      &       &       & 63 & 2 & 28 & 5:00 & 17.97 & 22:58 \\
   &   &    &      &       &       & 64 & 3 &  1 & 0:40 & 17.99 & 18:39 \\
\bottomrule
\end{tabular}
} 
\caption{Conversion from Italic to modern hours (January–March 1610).}
\label{tab:times}
\end{table}

\begin{figure}
    \centering
\includegraphics[width=1.1\linewidth]{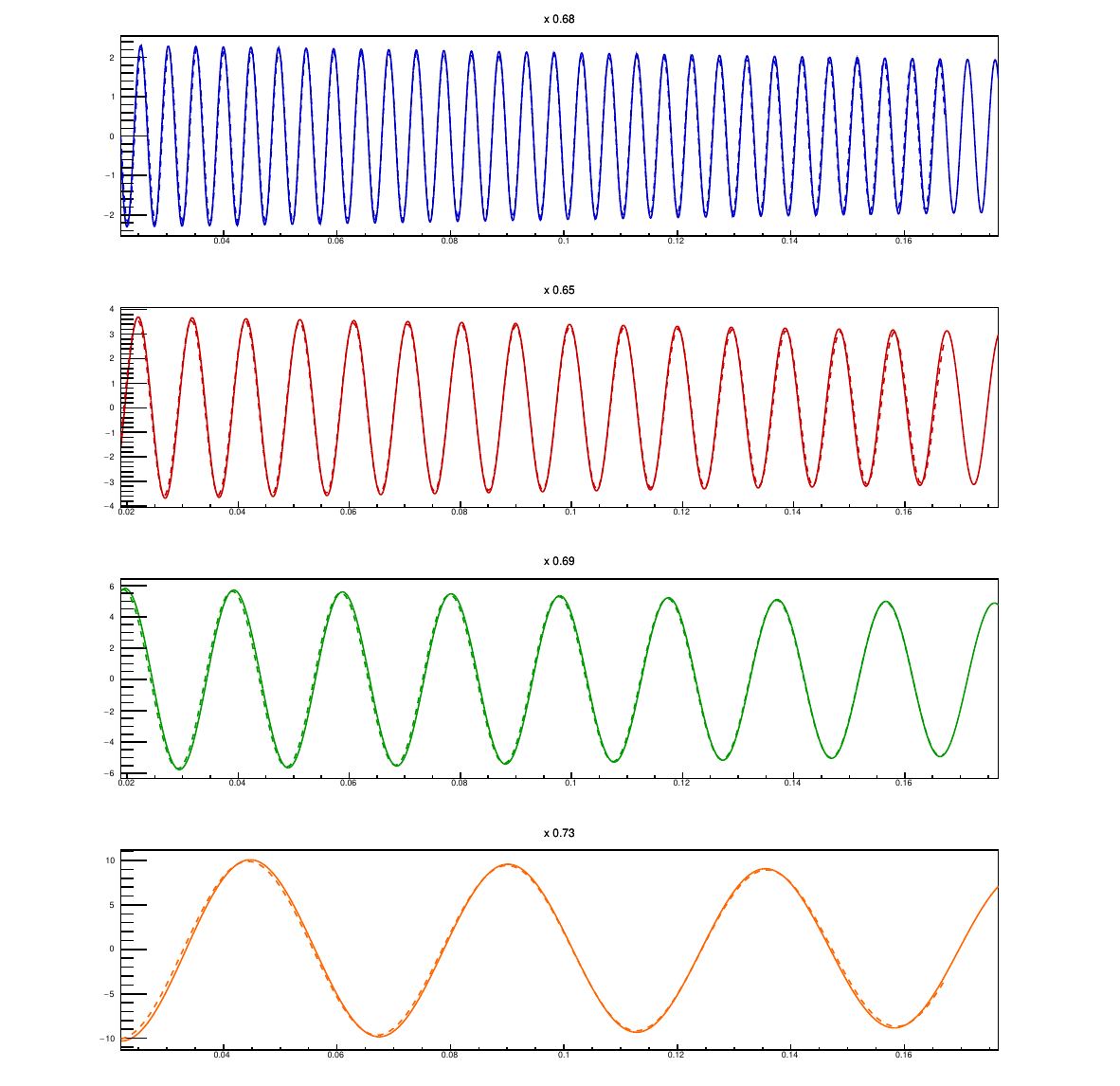}
    \caption{Comparison of the absolute predictions from the ephemerides and the fit for dataset-2 with an ad-hoc scaling factor on the amplitude of about 0.7.}
    \label{fig:fitS-eph}
\end{figure}

\end{appendices}

\bibliography{sn-bibliography}

\end{document}